\documentclass[fleqn,a4paper]{book}
\usepackage[font={small,it},labelfont=bf]{caption}
\usepackage{mdframed}
\definecolor{columbiablue}{rgb}{0.61, 0.87, 1.0}
\definecolor{ivory}{rgb}{1.0, 1.0, 0.94}
\definecolor{lightyellow}{rgb}{1.0, 1.0, 0.88}
\definecolor{pearl}{rgb}{0.94, 0.92, 0.84}
\definecolor{platinum}{rgb}{0.9, 0.89, 0.89}
\mdfdefinestyle{tpq}{%
    linecolor=blue,
    linewidth=1pt,
    roundcorner=20pt,
    innertopmargin=\baselineskip,
    innerbottommargin=\baselineskip,
    innerrightmargin=20pt,
    innerleftmargin=20pt,
    backgroundcolor=lightyellow}
\usepackage[utf8]{inputenc}
\usepackage{pdfpages}
\usepackage{eurosym}
\usepackage{amsmath}
\usepackage{amssymb}
\usepackage{makeidx}
    \makeindex
\usepackage{times}
\usepackage{ifthen}
\usepackage{wrapfig}
\usepackage{tabularx}
\usepackage{booktabs}
\usepackage[pdftex]{epsfig}
\DeclareGraphicsExtensions{.pdf}

\usepackage[pdftex,colorlinks=false]{hyperref}
\usepackage[figure]{hypcap}

\chardef\bslash=`\\ 

\newcommand{\bibtex}{\ifx\is@itshape\f@shape{\fontshape{scit}\selectfont
Bib}\else\textsc{Bib}\fi\kern-.1em\TeX}

 \newcommand{\ket}[1]{{| #1\rangle}}

\newcommand{\vecz}[2]{{\left(\begin{array}{c} #1 \\ #2 \end{array} \right)}}
\newcommand{\matz}[4]{{\left(\begin{array}{cc} #1 & #2 \\ #3 & #4 \end{array}
    \right)}}

\newcommand{\tre}[1]{\textcolor{red}{#1}}

\newboolean{offen}
\setboolean{offen}{false}
\newcommand{\geheim}[1]{\ifthenelse{\boolean{offen}}{#1}{}}
\newcommand{\zitateng}[1]{``#1''}

\begin{document}
\frontmatter   
\bibliographystyle{alpha}

\author{Prof. Dr. Joachim Stolze \\
(Technische Universität Dortmund)\\}
\title{A Short Guide to Quantum Mechanics\\
---\\
 Some Basic Principles}
\maketitle

\setcounter{page}{4}

\Huge

\tre{Important note:}
\\

\normalsize

This is the internet version of the handbook for the \textit{Treffpunkt Quantenmechanik} laboratory. It contains only the
overview part in six chapters.

For copyright reasons, we cannot make the detailed instructions for the individual
experiments in \textit{Treffpunkt Quantenmechanik} available on the internet.

Those instructions are only available in the printed version of the handbook or as
individual printouts in the rooms of the \textit{Treffpunkt Quantenmechanik}.

\chapter{Foreword to the English edition}\label{new_vorwort}
This is a translation into into (probably broken) English of the first part of a \zitateng{Handbook} intended as a companion to the \textit{Treffpunkt Quantenmechanik}, a laboratory at TU Dortmund University, where high-school students can get acquainted with the wonderful world of quantum physics.

The original printed version of this handbook consists of two parts: an overview connecting different aspects of quantum physics, written by me, and detailed instructions for the individual experiments available in the lab, written by Dr. Sebastian Duffe. The overview part (in German) was also made available on the internet. The detailed instructions contain copyrighted material and thus are only available to individuals interested in pursuing experiments in person.

Due to interest by colleagues from abroad, or involved in international activities, I decided to translate the overview part of the handbook, starting from a software-created raw version. I am not a professional translator of literature, but let me note that  scientific and literary texts put very different demands concerning precision, using different words (literary) vs. always precisely the same technical terms (scientific), etc. So the process of smoothening the raw zeroth version of the translation brought with it some surprises and funny moments, and if the english text has some definitely german flavor, this is due to my insufficient linguistic abilities. I hope the text will be useful anyway.

Dortmund, March 2024:

\phantom{x}

Joachim Stolze

\chapter{Foreword to the German edition}\label{vorwort}
\textit{Treffpunkt Quantenmechanik}  is a laboratory at the Faculty of Physics at  TU Dortmund University that offers high-school students an insight into the fundamentals of modern physics. Basic experiments can be carried out there that make the essential properties of quantum mechanics clear. Both detailed instructions for the individual experiments, written by Dr. Sebastian Duffe, and an overview of the background and the connections between the various surprising findings of quantum physics are (hopefully) helpful. 

Before continuing with the table of contents and the \zitateng{instructions for use} for this book, I would like to thank several people and institutions. 
I would like to thank Prof. Dr. Metin Tolan for suggesting this project, for his encouragement, and for the great cooperation in many jointly taught courses over many years. The Wilhelm and Else Heraeus Foundation has supported me as a senior professor, and
the Faculty of Physics has also provided me with space and work equipment as a retiree.
I am very grateful to both institutions for this.

I would like to thank my old schoolmate Dr. Udo Zielinski, a retired math and physics teacher, who read the entire text sympathetically but with relentless criticism, sometimes several times. Udo, you are a hero!

I would also like to thank Dr. Marlene Doert and Priv.-Doz. Dr. Dominik Elsässer for their comments and suggestions on my text, and of course Dr. Sebastian Duffe for the trusting and smooth collaboration.

Finally, I would like to thank all readers in advance for pointing out errors and
ambiguities, which can be sent by e-mail to joachim.stolze@tu-dortmund.de.\\

Dortmund,  June 2023:  

Joachim Stolze

\tableofcontents

\mainmatter

\chapter*{Instructions for use}\label{manual}
\section*{What this book contains}

This book consists of an introduction to quantum mechanics in six chapters and instructions for the experiments in \textit{Treffpunkt Quantenmechanik}. Chapter \ref{I} gives a rough idea of whether quantum physics is important, strange, useful or nothing
special at all.

Waves and particles form the content of Chapter \ref{II} . Both notions are important for quantum mechanics and sometimes it is not entirely clear whether something is a wave or a particle. Various experiments in \textit{Treffpunkt Quantenmechanik} deal with
this aspect. These experiments are described, but without going into the details of the experimental setup and execution, because that is what the experimental instructions in the second part of the book are for.

Chapter \ref{III} deals with the fact that quantum mechanics can often only make statements about probabilities. To understand this, we need to understand how quantum mechanics describes the motions of particles in general and what exactly happens when a measurement is carried out. It becomes clear why an atom or molecule can only have very specific (\textit{quantized}) energies, and how this can be
understood in the experiments at \textit{Treffpunkt Quantenmechanik}.

Chapter \ref{V} then describes the experiments that confused the scientific world 100 or more years ago and led to the development of quantum mechanics in its present form.
Most of these experiments can also be carried out at \textit{Treffpunkt Quantenmechanik}.

The two remaining chapters are special in that they \textit{do not} refer to the experiments in the \textit{Treffpunkt Quantenmechanik}\footnote{With the exception of the \textit{Quantum cryptography} experiment (under construction), which can only be
meaningfully discussed in chapter \ref{VI}.
} but offer material for particularly interested people who would like to know a little more precisely (typical physicist characteristic!) and are not entirely satisfied with the roughly qualitative explanations in the other chapters. It turns out that some mathematical tools are necessary for the somewhat more precise investigation, which are introduced in chapter \ref{IV} . This makes it possible to better understand how such strange effects as the Heisenberg uncertainty principle occur.

The last chapter (\ref{VI}) introduces an important additional quantity, the spin. Why it is important in medical diagnostics as well as in chemistry, and what it has to do with the
recently discussed \textit{quantum computers} will (hopefully) become clear. This, too, will not be possible without the mathematics from chapter \ref{IV} . Finally, the strange fact of quantum mechanical
\textit{entanglement} will be discussed. An experiment with a cat will be discussed, which fortunately
has never been carried out. We will also learn about teleportation (\zitateng{beaming}) and the Nobel Prizes in Physics for 2012 and 2022.

At the end of the book there is a list of the experiments, with references to the chapters in which they are discussed, a glossary of technical terms, and finally the individual  instructions to the experiments.

\section*{How to use this book}

There are printed copies of this book in the rooms of the \textit{Treffpunkt Quantenmechanik} and it is available as a PDF file via its homepage. There are thus three different ways of reading: (i) on paper, (ii) online in a browser and (iii) offline after
downloading the PDF file to your own device. Option (i) is also possible while lying back on the sofa, but first you have to get your hands on a printed copy (or print one out yourself). With
(ii) you ensure that you always look at the current version, as the text is of course constantly corrected and updated. Nevertheless, I would personally prefer the variant (iii) because there will be no updates and corrections \textit{that} frequently.

Reading on screen has the advantage that you can easily search for a word in all PDF display programs (\textit{Acrobat Reader, Foxit}, etc.) if you no longer know exactly what it means or where you have seen it before. Jumping back and forth to illustrations
mentioned in the text, other sections, etc. is also easy, provided the author has taken the trouble to set appropriate jump marks. (He has.) Unfortunately, however, jumping back and forth is often not well implemented, especially in PDF viewer programs for tablets or smartphones. You can usually jump somewhere, but not back again. This
even applies to the tablet versions of \textit{Acrobat Reader} and \textit{Foxit}, at least on my tablet.
However, these two programs work perfectly on the laptop (at least on mine). A PDF file of your own also has the important advantage that you can leave annotations, underlines, notes etc. in it.

The experiment instructions are linked to the text of book chapters \ref{II} - \ref{V} by forward and backward references: In the individual chapters, the experiments are described in the
physical context in which they belong, and the list of experiments indicates in which chapter the respective experiment occurs.

There are many terms and contexts that are not 100\% familiar to every reader: \textit{Amplitude, spectrum, momentum}, etc. There is a glossary (i.e. a collection of explanations) at the end of the book. There is an info box for each topic, which is also inserted in the running text where, for example, the word \textit{spectrum} appears for the first time. The info boxes are highlighted in color and can also appear several times if a topic appears repeatedly. If you think you already know enough about the keyword in the heading, you can of course skip the box. If you then become unsure a little later, you can still read the box.

\section*{Who can read this book}

Short answer: everyone! The text was written with interested high-school students in mind who have a basic knowledge of classical physics and would like to learn more about the
strange world of quanta. Even if the experiments in \textit{Treffpunkt Quantenmechanik} are mentioned from time to time, you don't have to have performed them to get something out of the book. Of course, teachers who would like to find out about the possibilities of the \textit{Treffpunkt Quantenmechanik}
are also very welcome. For them, as for everyone
else, curiosity is enough!

\begin{mdframed}[style=tpq]

\begin{center} 
\textsf{Classical physics}
\end{center}

The state of knowledge of physics around 1900 is termed classical: mechanics obeys the principles found by Newton, extended by Einstein's special theory of relativity. Electrical and magnetic phenomena are fully described by Maxwell's electrodynamics. Even today, classical physics still describes very well all physical phenomena that occur in everyday life and in many areas of technology. Around 1900, only a few phenomena were not
fully described by classical physics.
The spectral lines of the chemical elements and the spectrum of thermal radiation, for example, could \textit{not} be explained in terms of classical physics.
Only the development of quantum mechanics in 
 the 1920s provided  the theory of these \zitateng{non-classical} phenomena.
\end{mdframed}
\Huge

\tre{Important note:}
\\

\normalsize

This is the internet version of the handbook for the \textit{Treffpunkt Quantenmechanik} laboratory. It contains only the
overview part in six chapters.

For copyright reasons, we cannot make the detailed instructions for the individual
experiments in \textit{Treffpunkt Quantenmechanik} available on the internet.

Those instructions are only available in the printed version of the handbook or as
individual printouts in the rooms of the \textit{Treffpunkt Quantenmechanik}.

\vfill

\chapter{{Curious} Questions}
\label{I}

\zitateng{\textit{Things on a very small scale behave like nothing that you have any direct experience about. They do not behave like waves, they do not behave like particles, they do not behave like clouds, or billiard balls, or weights on springs, or like anything that you have ever seen.
}

\textit{(.......)}

\textit{
Because atomic behavior is so unlike ordinary experience, it is very difficult to get used to, and it appears peculiar and mysterious to everyone --- both to the novice and to the experienced physicist. Even the experts do not understand it the way they would like to, and it is perfectly reasonable that they should not, because all of direct, human experience and of human intuition applies to large objects. We know how large objects will act, but things on a small scale just do not act that way. So we have to learn about them in a sort of abstract or imaginative fashion and not by connection with our direct experience.
}}\\

\textit{Richard P. Feynman, Nobel laureate for Physics 1965}\\

These remarks can be found at the beginning of the third part of the
\zitateng{Feynman Lectures on Physics}. Feynman wanted his
students in the second year of their physics studies to be prepared for the fact
that quantum physics is different from what they had learned in their studies up to that point.
These words of warning are now
about 60 years old, but still true. Therefore, let us begin by discussing a few
questions that will help us to contextualize
quantum physics a little better: First of all, let's
see \textit{what} it's all about before we try to understand
try to understand \textit{how} it works.

{In this introductory chapter, we can quickly answer a few curious questions. However, we can only briefly touch on other questions; more detailed explanations can be found in the later chapters}.

\section{Is quantum physics important?}

In the last 100 years, our lifestyle has changed more 
than in the millennia before; this was largely due to the
the findings of the natural sciences and the advances in technology
that have resulted from them. \textbf{Quantum physics} is one of these
sciences. Comparably great influence was probably only exerted by
\textbf{electrodynamics} (with the universal availability of light,
energy and communication possibilities) and \textbf{chemistry} (with 
materials, medicines, chemical fertilizers, etc.)
{It is important to note that the
chemical processes can only really be understood in the context of quantum physics.}

A quantum physical device that each of us carries is
the \textbf{smartphone}. The age of modern electronics began in 1947 with
the invention of the \textbf{transistor}, which cannot be understood without quantum physics.
 On a chip, as it is found today not only in all
smartphones and TV sets, but also in cars and washing machines
a huge number of transistors is packed
(\zitateng{integrated}). A modern smartphone contains more
transistors than a professional computer a few
years ago. The \textbf{camera} 
of your smartphone also uses findings from quantum physics, as does the
\textbf{photovoltaic system}, which may be installed on the roof of the house
 you live in.

The \textbf{laser} was invented around 60 years ago. The word was originally an
abbreviation for \zitateng{Light Amplification by Stimulated Emission of
 Radiation}{and the stimulated emission is a purely quantum mechanical effect}.
 Initially, the laser was ridiculed as \zitateng{a solution
 in search of a problem}, but that changed
quickly. Today, every floor layer or painter and decorator measures your home with a laser device.
 There are laser pointers instead of pointer sticks, CD, DVD and 
BluRay players contain lasers, as do laser printers. Lasers are
used to cut sheet metal, to operate on eyes and by physicists and chemists to
 examine materials by spectroscopy.

\begin{mdframed}[style=tpq]

\begin{center} 
\textsf{Spectrum (plural: spectra), spectroscopy
}
\end{center}
A \textbf{spectrum} is the set of the wavelengths (or frequencies) of
electromagnetic radiation that can be observed in a range, together with
their respective intensities. 
If all wavelengths can occur within some range, the spectrum is called
\textbf{continuous}, e.g. in the case of thermal radiation from a hot object. If only
very specific wavelengths can occur, the spectrum is called \textbf{discrete}. This is
the case for pure chemical elements, for example. Discrete spectra are also
called \textbf{line spectra}, named after the image of colored lines on a dark
background that appears in an optical spectrometer. By examining such line
spectra in \textbf{spectral analysis}, it is possible to determine which chemical
elements the substance under investigation contains. The study of
spectra is called \textbf{spectroscopy}.

In more general terms, the possible values of the energy of a quantum system
are also referred to as the \textbf{energy spectrum}.

\end{mdframed}

\textbf{GPS navigation} via satellites is based on extremely accurate atomic clocks 
{and 
thus uses the findings of quantum physics} (and incidentally also Einstein's
general theory of relativity). 
{Also the functioning of the energy-saving
\textbf{LED lamps}, which are now standard, cannot be explained without quantum physics}.

The quantum physical 
tunnel effect, for which there is no explanation in classical physics
enables the release of energy through \textbf{nuclear fusion} in the
sun (or even in a hydrogen bomb), 
{and is also the basis for} the fact that we can work with the
\textbf{scanning tunneling microscope} to study the properties of individual atoms on
surfaces (also at TU Dortmund University).

\begin{mdframed}[style=tpq]

\begin{center} 
\textsf{Classical physics}
\end{center}

The state of knowledge of physics around 1900 is termed classical: mechanics obeys the principles found by Newton, extended by Einstein's special theory of relativity. Electrical and magnetic phenomena are fully described by Maxwell's electrodynamics. Even today, classical physics still describes very well all physical phenomena that occur in everyday life and in many areas of technology. Around 1900, only a few phenomena were not
fully described by classical physics.
The spectral lines of the chemical elements and the spectrum of thermal radiation, for example, could \textit{not} be explained in terms of classical physics.
Only the development of quantum mechanics in 
 the 1920s provided  the theory of these \zitateng{non-classical} phenomena.
\end{mdframed}

With \textbf{magnetic resonance imaging} (MRI), you can see things that are 
difficult to see on an X-ray, e.g. the damaged cruciate ligament of a
soccer player. Improvements to this
quantum physics technology are also being worked on at TU Dortmund University.

If the \textbf{geckos} knew when hunting insects at night that
they use a quantum effect (van der Waals forces) to run along window panes or under the 
ceiling, they would probably be terrified and 
drop down in shock.

Over 100 years ago the explanation of the spectrum of \textbf{heat radiation} was 
the reason for Max Planck to introduce quanta into physics.
 Today, thanks to the heat radiation from outer space, we know
that the universe has a temperature of around 2.7 K
(Brrrrr!), and from very small temperature differences between different directions of the
sky we learn something about the structure of the early universe.
For good reason, the European Space Agency ESA
gave the satellite that made the most accurate measurements to date the name
name \zitateng{Planck}.

A relatively new development that has not yet arrived on a broad front in
commercial technology concerns 
\textbf{quantum information processing} and \textbf{quantum computers}. The
basic question here is: What {information} does quantum mechanics contain
about the state of a system, and how can this information be
used, protected and processed. Here there are
a wealth of possibilities, but also great difficulties that must 
have to be overcome before we can think about broad practical application.
 The enormous interest in this area is witnessed by the fact that
 in addition to large companies (Google, IBM, Microsoft,...),
countries (USA, China,...) and communities (EU) are also investing a lot of money in this field. Several groups at the Faculty of Physics at TU Dortmund University are researching the fundamentals of this future technology.

{An information that is quoted again and again and allegedly comes from
from a report to the British Parliament: 25\% 
of the gross domestic product of highly developed societies 
are based on products that contain quantum physics. Based on this
 statement, even economists will accept that quantum physics is important. }

\section{Is quantum physics weird?}

\label{seltsam}

This is partly a matter of taste and (as always) also has to do with how much you know about it. The more you know about something, the less mystical it seems. Quantum physics is unusual, but it's not esoteric. Esotericists only use the somewhat mystical ideas that are in circulation to sell their wares.

If someone offers you quantum healing (often also on the Internet), be careful and better not transfer any money. You can also easily find books that combine quantum physics with thought transference, love, consciousness and
destiny. Not all of this is certain knowledge according to scientific standards, to put it mildly.

On the other hand, it is quite possible to seriously deal scientifically with problems of understanding and interpreting quantum physics. However, it is important to remember the remarks made by Richard Feynman at the beginning of this chapter: All our ideas and concepts and terms, and the entire language in which we discuss are shaped by our experiences in everyday life. The fact that in this \zitateng{everyday language} questions can be formulated that quantum mechanics {cannot} answer unambiguously must be accepted; after all, quantum mechanics is not an everyday thing. You have to seriously engage with the formal (mathematical) language of quantum physics; in \textit{this} language you can ask those questions that quantum physics  can answer {unambiguously}. 


To put it more succinctly, some questions that you can ask quite naturally in everyday life are simply not allowed when it comes to very small things. You then often have to calculate and compare numbers to see whether the oddities and inaccuracies that arise should really worry you. It is true that quantum physics cannot answer some questions precisely, but can only make statements about the probability of processes,
instead of providing clear and unambiguous results{; more on this in the later chapters}. But even this should not worry us in our everyday lives: When we press the light switch in the bedroom when going to bed, we can be sure that the light will go off. 

However, if we leave the everyday level and go to the level of individual atoms or other small quantum systems, surprising, non-intuitive situations can be constructed time and again. The founders of quantum physics also used such constructions when discussing their science, but only ever in the form of thought experiments. Today, experimental technology has advanced to such an extent that many of the original thought experiments can actually be carried out in the laboratory and thus be used to  investigate the behavior of quantum systems in individual cases. 

\section{Is quantum physics only for geniuses?}

\label{genies}

Short answer: No. Anyone studying physics at TU Dortmund University learns quantum mechanics in the fourth semester, i.e. in the second year of their degree course. It is similar at all other universities. However, the three semesters of \zitateng{warm-up} are also necessary, because the ways of thinking and concepts developed in classical physics are the foundations of quantum physics, and above all a mathematical formalism must be learned, which unfortunately cannot be avoided if you really want to formulate the laws of the quantum world precisely.

\section{Is quantum physics a remote specialty?}

\label{abseitig}

If that were the case, then quantum physics would certainly not be a compulsory course in \textit{all} physics degree programs in Germany. Quantum mechanics is the key to understanding all recent developments in the physics of elementary particles, solids, and semiconductors, molecules, atoms, and atomic nuclei, electronics and the physics {of stars and the universe}. By virtue of their training, all physicists are quantum physicists.

Even the fact that we and the world around us exist in the form we know {can only be understood through quantum mechanics}. Without the \textbf{uncertainty principle}, the negatively charged electrons would crash into the positively charged atomic nuclei and the entire earth would be a lump of incredibly dense matter with a diameter of around 400 meters. The stability of matter is based on quantum physics.

\begin{mdframed}[style=tpq]

\begin{center} 
\textsf{Uncertainty relation}
\end{center}
Heisenberg's uncertainty relation, often also called the indeterminacy
relation, states the accuracy $\Delta x$  with which the position $x$ of a particle
moving along the $x$ axis can be determined if the momentum $p$ of the
particle is also determined with accuracy $\Delta p$ in the same state:
$$
\Delta x \Delta p \geq \frac{\hbar}2 .
$$
The \textit{uncertainties} $\Delta x$ and $\Delta p$ are therefore related: If you want to determine $x$ very precisely, $p$ becomes very imprecise, and vice versa.

Not only the position and momentum, but also other pairs of \zitateng{incompatible} quantities of quantum mechanics are linked by indeterminacy relations.

$\hbar$ (read: \zitateng{h-bar})   is the Planck constant divided by $2\pi$; $$\hbar = 1,055 \cdot 10^{-34} \textrm{Js}.$$
Since this number is so tiny, we do not notice the uncertainty principle in
everyday life, but the atoms feel it very clearly.

\end{mdframed}

\begin{mdframed}[style=tpq]

\begin{center} 
\textsf{Momentum}
\end{center}
The momentum $p$ of a classical particle is simply the product of mass
$m$ and speed $v$:
$$
p=m v .
$$
The momentum of a force-free quantum mechanical particle is related to the
wavelength $\lambda$ of the wave function:
$$
p=\frac h{\lambda},
$$
where $h$ is the Planck constant.

If forces act on the quantum mechanical particle, $p$ is somewhat more
complicated to calculate.

A photon (light quantum) also has momentum; this is defined by the
wavelength $\lambda$
of light: $$p=\frac h{\lambda}.$$ 

The momentum of the photon is related to its energy $E$
by $$p=\frac Ec,$$ with $c$ the speed of light.

\end{mdframed}

\begin{mdframed}[style=tpq]

\begin{center} 
\textsf{State of a particle}

\end{center}
The state of a classical particle moving in the $x$ direction is determined at time $t$ by its position $x$ and its momentum $p$ (or alternatively its velocity $v$). The 
values of $x$ and $p$ are precisely determined at every instant.

The state of a quantum mechanical particle, on the other hand, is
determined by a function that depends on $x$ and $t$, the wave function. The wave function allows statements to be made about the position and momentum of the particle, but only within the limits of the Heisenberg uncertainty principle.

Vectors are an alternative way of representing a quantum mechanical state
mathematically. More about this can be found in section \ref{vektoren} .

\end{mdframed}

\section{What is so special about quantum physics?}

\label{besonders}

This is exactly the topic of the next chapters and the experiments that you can experience for yourself in \textit{Treffpunkt Quantenmechanik}; here we will only give a few keywords, together with the experiments for which these keywords are important.

\textbf{Waves} can overlap. This is the basis for \textbf{interference} and \textbf{diffraction}. First of all, this has nothing to do with quantum physics, but has long been known in classical physics. Light is a wave process. (Experiments: \textit{Wave trough, Diffraction of laser light, Polarization
of light}). In quantum physics, however, waves appear in places where you would rather expect particles. The fact that these waves can overlap and interfere with each other is the cause of many of the surprising effects in quantum physics. It is therefore worth taking a closer look at the properties of waves, especially interference and diffraction, even if this requires some effort, as we will see in Chapter \ref{II}.

In some situations, \textbf{electrons} behave like \textbf{waves}. (Experiment: \textit{Electron diffraction}) But sometimes they also behave like particles, for example all electrons have the same charge $-e$. The charge of each material object is quantized, i.e. it consists of all the same smallest units, and the unit of charge is the {elementary charge} $e$. (Experiment:
\textit{Millikan experiment})

\textbf{Light} sometimes behaves like a stream of \textbf{particles} (photons). (Experiment: \textit{Detection of single photons}) However, if a large number of photons are observed, the overall behavior is again that of waves. (Experiment: \textit{Double slit with single photons})

The \textbf{energy of atoms} is quantized, so only certain values of energy are permitted. These energy values identify different atoms like a fingerprint. (Experiments: \textit{Franck-Hertz experiment, Spectroscopy})

The \textbf{energy of light} is quantized and the permissible values of the energy depend on the wavelength of light. (Experiments: \textit{Photoelectric effect, Blackbody radiation})

The result of a \textbf{measurement} can be random, even if the measured system is in a very specific, clearly defined state. The result of an \textit{individual} measurement cannot usually be calculated by 
quantum mechanics. Very \textit{many} measurements on identical systems in the same state provide a certain distribution of possible measurement values; quantum mechanics can calculate this distribution. A measurement influences the system: the state after a measurement is different from the state before and depends on the result of the measurement. (Experiments: \textit{{Quantum cryptography}} (under construction), \textit{Polarization of light})

\chapter{Wave or particle, or what else?}
\label{II}
In this chapter, we first clarify in Section \ref{teilchen} what the essential properties of a particle are in classical physics. In Section \ref{welle} we do the same for classical waves. This will be the first time we talk about \textbf{uncertainty}, but initially only in a rather uncertain way. Waves can be diffracted, and if several waves are present at the same time, they can interfere with each other. In  Section \ref{interferenz} we will see what this means and what consequences it has. The phenomena of interference and diffraction can be observed quite clearly and with the naked eye in the experiments with the \textit{Wave Trough} in \textit{Treffpunkt Quantenmechanik}. 
{The experiment \textit{Diffraction of laser light} shows the same phenomena at a much shorter wavelength. Section \ref{wellenfunktion} then explains how the waves come into play as \textbf{wave functions} in quantum physics. These} can also exhibit interference and diffraction, just like other waves{, and this is the cause of many of the \zitateng{strange} properties of quantum physics.  Section \ref{interferenz_quanten} explains the details and the experiment \textit{Electron diffraction} provides an example of diffraction of waves in quantum physics.} In  Section \ref{unschaerfe} {we take a closer look at what uncertainty has to do with waves } and realize that it is by no means just annoying, but actually fundamentally important for understanding the matter that surrounds us.

\section{What is a particle?}
\label{teilchen} 

\begin{figure}[h] 
\includegraphics[width=0.8\textwidth,keepaspectratio=true]{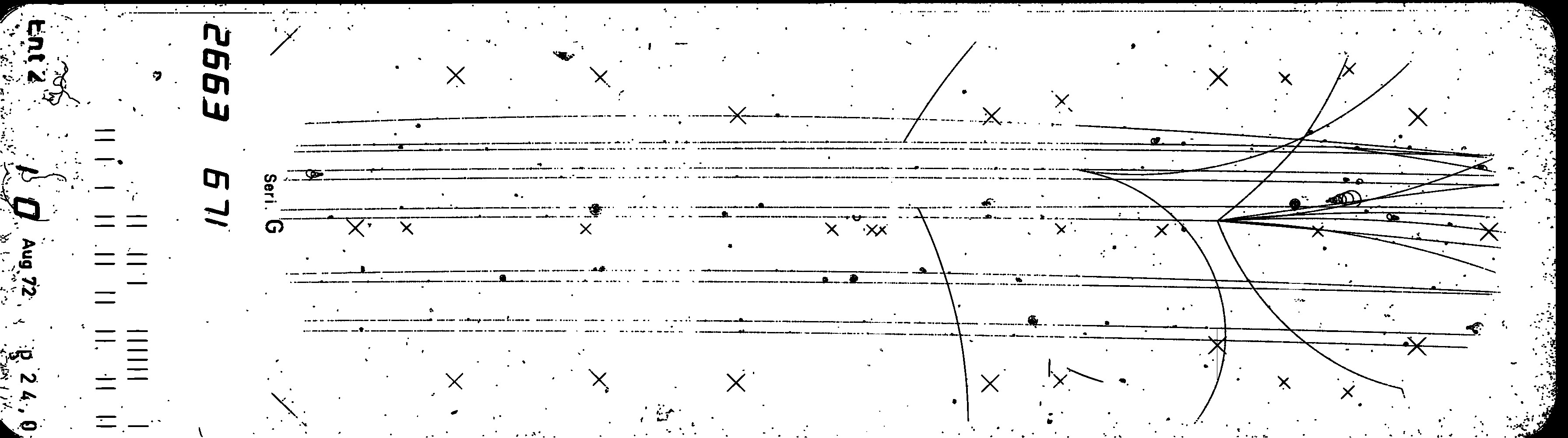}
\caption{Traces of elementary particles in a bubble chamber. The picture dates from 1972; today bubble chambers are obsolete and have been replaced by other kinds of detectors. \tiny{(Photo: CERN}) 
}
\label{teilchenspuren}
\end{figure}

A particle is a small object with a certain mass. It is so small that it can be imagined as a point; it is often also referred to as a \zitateng{mass point}. It moves on a path through space and has a specific position and a specific speed or a specific momentum (product of mass and speed) at all times. In classical physics, even very large objects can be treated as particles, e.g. the earth when calculating its orbit around the sun, because the earth is very small compared to the distance to the sun. However, in the context of quantum physics, we always mean atoms or their components when we talk about particles.

The elementary particles that make up matter often also have an electrical charge; they do not have any other properties, such as color or smell. As it is point-shaped, a particle cannot rotate \zitateng{around its axis}. In quantum physics, however, some particles have a \zitateng{spin}, a quantity that reflects the properties of rotation; we will come back to this later.

If no force acts on a particle (if it is \zitateng{force-free}), it continues to move in a straight line at the speed it had at the beginning. If it had no speed at the beginning, it simply stays in the same place. If a force acts, the speed of the particle changes, and with it its momentum. One of the laws of motion discovered by Isaac Newton is that the change in momentum of a particle over time is equal to the force acting on it. This law makes it possible to calculate movements within the framework of classical physics, from the throwing of a stone to the orbit of a satellite and the motions of the planets.

For our considerations 
we want to make things as simple as possible. That is why we only consider movements in one dimension. Our particle therefore moves along a straight line, the $x$-axis. We define a zero point $x=0$ on the line and the number $x$ then tells us where the particle is.

If the particle is at rest at the beginning and a force acts in the positive $x$-direction (\zitateng{to the right}), then the speed of the particle increases and so does its kinetic energy (energy of motion). If the force acts to the left, the particle has a negative velocity (\zitateng{backward}), but the kinetic energy also increases because it does not depend on the sign (direction) of the velocity.

\begin{mdframed}[style=tpq]

\begin{center} 
\textsf{Kinetic and potential energy, classical}
\end{center}

Kinetic energy is the energy of motion. If a particle of mass $m$ moves at velocity $v$, it has the energy
$$
E_{\textrm{kin}} = \frac 12 m v^2.
$$
If the particle transfers its energy to a spring, for example, which is tensioned by the impact, then the energy is stored in the spring as potential energy $E_\textrm{pot}$. The total energy, i.e. the sum of kinetic and potential energies, is constant. This law of conservation of energy  is one of the most important principles of physics.

$E_\textrm{pot}$ often depends on a spatial coordinate $x$, in our example on the position at
which the particle is stopped by the spring. The faster it was at the
beginning, the further the spring is compressed and the greater the potential
energy. In physics, such a potential energy dependent on the location $x$ is
usually referred to as $V (x)$ for short and is also briefly called potential.
(A somewhat misleading term, as in electricity theory the potential energy
\textit{per charge} is referred to as electric potential).

The kinetic energy is often expressed by the momentum $p = mv$; then one has the relationship
$$
E_{\textrm{kin}} = \frac{p^2}{2m}.
$$
However, this relationship only applies if the particle is moving at speeds that are very small compared to the speed of light $c$. If this is not the case, the theory of
relativity must be employed, with the result
$$
E_{\textrm{kin}} = \sqrt{m^2 c^4 + p^2 c^2} - m c^2.
$$
In this text, however, the theory of relativity only becomes important at one
point, namely in the Compton effect.

\end{mdframed}

If we exclude frictional forces, there is only one other form of energy apart from kinetic energy: potential energy, sometimes also called position energy. As an example, imagine two particles with the same electrical charge, one of which is stationary at the point $x=0$ and the other is moving towards this point. As the electrical repulsion slows down the moving particle, its kinetic energy decreases. At some point, the kinetic energy will be zero. (Remember that the electrical repulsion can become infinitely large if the distance between the two particles becomes smaller and smaller). All the kinetic energy has been converted into potential energy. In our simple case, the law of conservation of energy states that the sum of kinetic and potential energy remains constant. The conservation of energy is one of the most important principles of physics.

What happens next with the two charged particles? The electrical repulsive force continues to act, and this pushes the moving particle back in the direction from which it came. The momentum of the particle changes according to Newton's law, and its kinetic energy increases again, while its potential energy decreases. Every one-dimensional movement can be calculated completely (and mathematically simply) using the law of conservation of energy.

If you have a fixed force field, i.e. a certain force, constant in time, acting on the moving particle depending on its position $x$ (as in our example), then instead of the force at position $x$ you can also specify the potential energy that the moving particle has when it is at position $x$; we call this potential energy $V(x)$. Often we also speak briefly and simply of the potential. (This is widespread, but actually somewhat imprecise).

We summarize:
{A particle moves along a path that is determined solely by the forces acting on it}. All the properties of a particle are concentrated in a very small area of space. This applies in particular to its energy, which is quite different from the energy of a wave.

\section{What is a wave?}
\label{welle}

\begin{figure}[h] 
\includegraphics[width=0.5\textwidth,keepaspectratio=true]{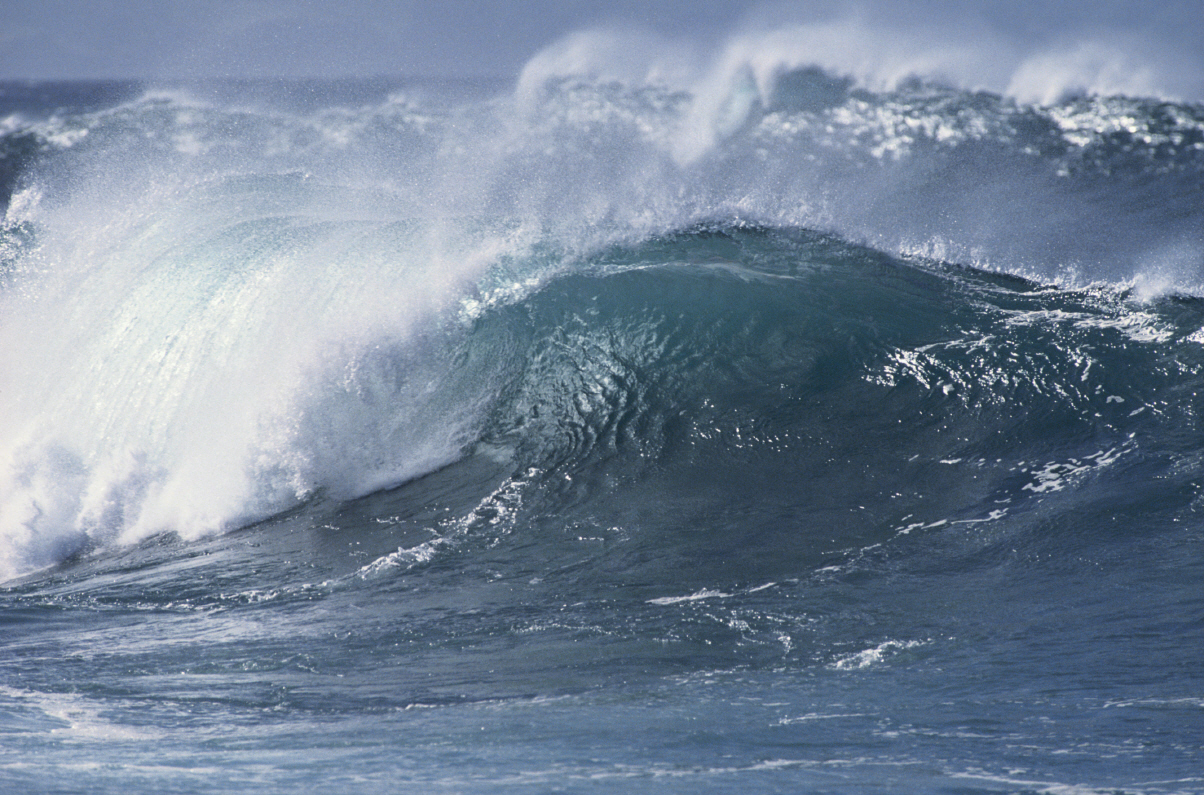}
\includegraphics[width=0.5\textwidth,keepaspectratio=true]{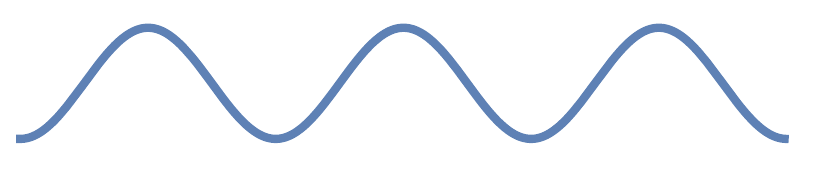}

\caption{Waves in the surf and idealized \zitateng{physicist wave}. \tiny{(Photo and graphic: Stolze) }
}

\label{brandung} 

\end{figure}

Everyone has seen waves on the surface of water. They usually look much messier than the perfect waves that physics likes to deal with. More important than water waves are the waves that we deal with every day: electromagnetic waves that our smartphones send and receive, light waves that illuminate our homes, sound waves from our loudspeakers, etc. They are usually rather as neat as the right-hand image in Fig. \ref{brandung}, but {unfortunately you can't see them as well as water waves}.

Let's make life easy for ourselves again and first look at waves that move in one dimension, like the simple wave on the right in Figure \ref{brandung}. Such a wave can be described by a few quantities. The \textbf{wavelength} $\lambda$ (Greek letter: lambda) is the distance between two neighboring wave crests (or troughs, of course). The \textbf{displacement} indicates how much the wave deviates from the absolutely calm water surface, for example; it can be positive or negative and changes in space and time. The \textbf{amplitude} is the maximum displacement, i.e. the height of the wave crest above the level of the calm water surface.

Waves move; imagine, for example, that the wave pattern in Figure \ref{brandung} (right) moves to the right at a constant speed. Then, at regular intervals, wave crests and wave troughs pass by alternately at a fixed point. The time between two wave crests is called the \textbf{period} $T$ of the wave, and $\frac 1T = f$ is called the \textbf{frequency} of the wave. The frequency is just the number of wave crests that pass a fixed point per second. Then $v=\lambda f$ is the speed at which the wave moves. The displacement of a wave is periodic in space for a fixed point in time (as a \zitateng{momentary snapshot}). The spatial pattern repeats itself exactly after a wavelength $\lambda$. At a fixed position, the displacement of a wave is periodic in time. The temporal behavior repeats itself exactly after a period $T$.

Here you can already see that ideal waves cannot actually exist in the real world, because something that always repeats itself in space and time must be infinitely extended in space and last infinitely long in time. This seemingly very subtle statement is the cause of the infamous uncertainty principle. We will come back to this later.

Waves contain and transport energy. This can be seen, for example, in the fact that water waves can move ships and destroy shore reinforcements. The fact that your smartphone battery runs out sooner than you would like is also at least partly due to the waves it sends into the mobile network (but possibly also to the music you listen to or the games you play to pass the time). The energy of a wave is distributed over larger areas of space, in contrast to the sharply localized energy of a particle.

\begin{figure}[h] 
\includegraphics[width=0.8\textwidth,keepaspectratio=true]{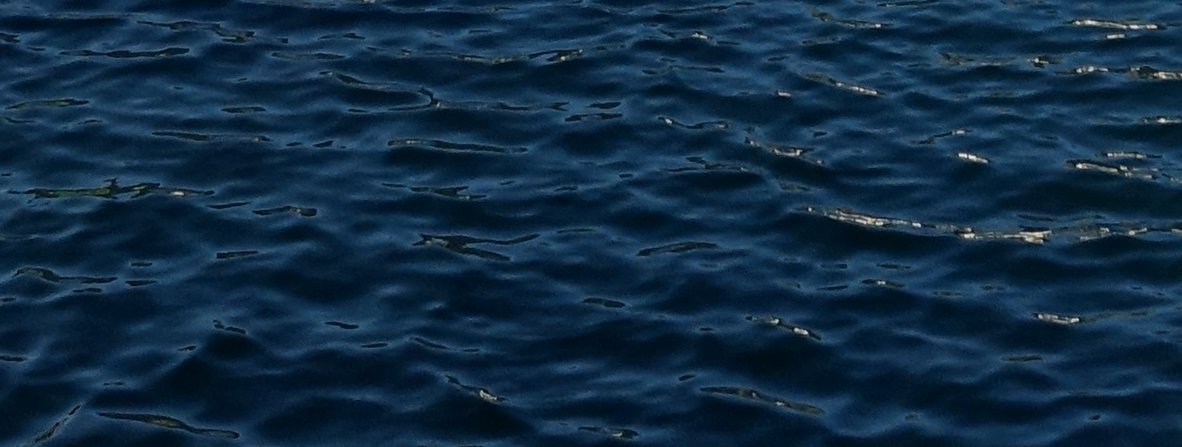}

\caption{Small waves on a calm water surface \tiny{(Photo: Stolze)} 
}
\label{wellen1}
\end{figure}

In  Figure \ref{wellen1} you can see waves on a water surface, i.e. in two dimensions, which are somewhat regular but not completely \zitateng{orderly}. The crests and troughs of the waves are roughly equally spaced. They are roughly straight, but not quite, and they end somewhere. The simplified version considered in physics has \textit{exactly} equal distances between the \textit{perfectly straight} wave crests, which are \textit{infinitely wide}. Such an entity is called a \textit{plane wave} because the wave crests lie on planes in three dimensions or on straight lines in two dimensions.

\begin{figure}[h] 
\includegraphics[width=0.8\textwidth,keepaspectratio=true]{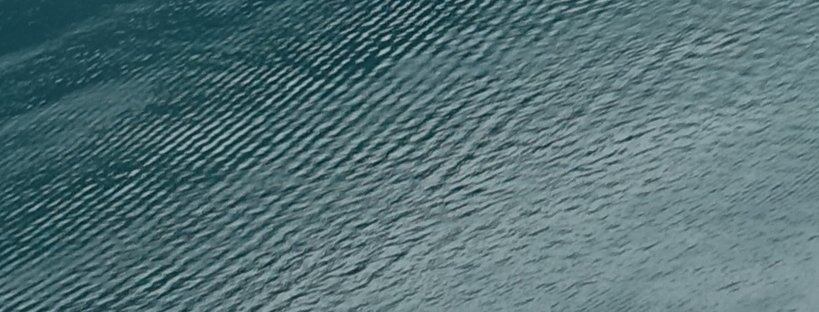}

\caption{Almost ideal plane water waves. 
{Due to the camera perspective, the wavelength on the left appears shorter than on the right}. \tiny{(Photo: Stolze)}
}
\label{wellen3}
\end{figure}

The water waves in Figure \ref{wellen3} are already pretty close to the ideal case of the plane wave. On closer inspection, however, you can see that there is not just one type of wave in play here. {In some areas of the image, you can still see waves with a different wavelength, whose crests run in a different direction. This is an example of how waves can overlap without disturbing each other. In physics, employing the Latin word for overlapping, this is often referred to as the \textbf{superposition principle}. This sounds of course much more scientific, but note that this principle is also extremely important. If waves could in fact influence each other, reliable GPS navigation, for example, would not be possible, because at the same time as the signals from the GPS satellites, the waves from mobile communications, weather radar, and thousands of television programs which are also moving through space would distort the GPS signal. An example of undisturbed\footnote{{Electromagnetic waves in free space overlap completely undisturbed; this only applies approximately to water waves.}} superposition of water waves can be seen in Figure \ref{wellen4}: The bow wave of the motorboat runs over the water surface without affecting the other waves. The superposition of waves leads to the phenomenon of \textbf{interference}, which we will come back to in Section \ref{interferenz}.

\begin{figure}[h] 
\includegraphics[width=0.8\textwidth,keepaspectratio=true]{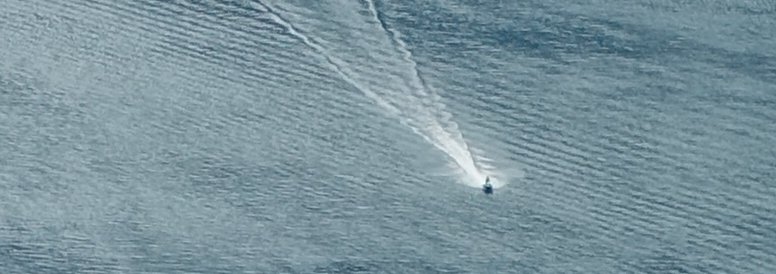}

\caption{Several water waves overlapping.  \tiny{(Photo: Stolze)} 
}
\label{wellen4}
\end{figure}

We will briefly return to the concept of wavelength and the measurement of wavelengths. For a wave like the one in Figure \ref{wellenzug_lang}, the wavelength is easy to measure: Since the wave is so regular, you can simply measure the distance between two neighboring wave crests. It is even more precise to measure the distance between two distant crests and divide it by the number of valleys in between. On the other hand, it is difficult to answer the question of the \zitateng{position of the wave}. The wave is obviously \zitateng{everywhere} in the figure and the question of the position is actually pointless.

\begin{figure}[h] 
\includegraphics[width=0.8\textwidth,keepaspectratio=true]{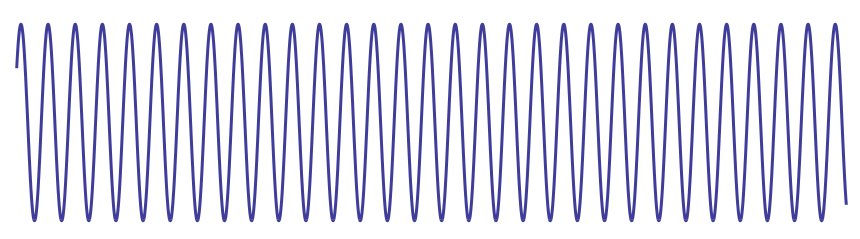}
\caption{A wave with easily measurable wavelength.  \tiny{(Graphic: Stolze)} 
}
\label{wellenzug_lang}
\end{figure}

This is completely different with a wave like the one in Figure \ref{wellenzug_kurz}: Here you can  tell where (approximately) the wave is. However, it is difficult to determine an exact wavelength: Where does this wave train start, where does it end, so how long is it? And which maxima or minima should be counted as wave crests or wave troughs; only the large ones or also the very small ones?

\begin{figure}[h] 
\includegraphics[width=0.8\textwidth,keepaspectratio=true]{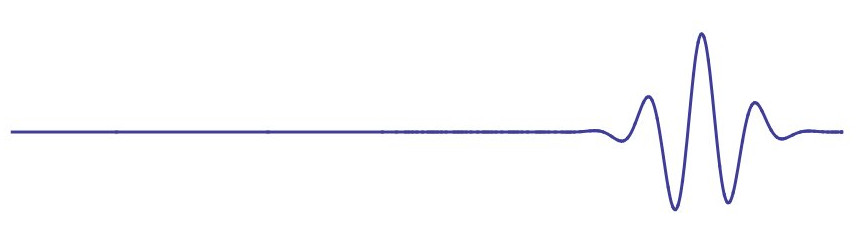}
\caption{A wave with easily measurable position.  \tiny{(Graphic: Stolze)} 
}
\label{wellenzug_kurz}
\end{figure}

For one type of wave, the wavelength is \zitateng{sharp} (exactly known) and the position \zitateng{unsharp} and for the other type it is exactly the other way round. Mathematical methods can be used to prove that such a relationship is generally valid and does not only apply to our specific example. Since in quantum mechanics the wavelength of the wave function is linked to the momentum 
{(more details in Section \ref{wellenfunktion})}, there is the uncertainty relation between position and momentum. The uncertainty relation outlined above for \zitateng{completely normal} waves, e.g. radio waves, is everyday knowledge for communications engineers. If you want to transmit a particularly short signal, you cannot do this with a single wavelength, but need many wavelengths (or frequencies), i.e. a \zitateng{large bandwidth}, as experts say.

The fact that position and wavelength (or time and frequency) are connected to each other in an uncertainty relation is neither unusual nor surprising. What is unusual (and initially very surprising) about quantum mechanics is the fact that waves can describe the properties of particles and that wavelength is related to momentum. More about uncertainty, { the uncertainty relation and the} consequences can be found in  Section \ref{unschaerfe}.

{Before we get to that, however, we need to go into more detail about the meaning of interference. Since interference is so characteristic and important for quantum mechanics, we will need several sections to explain the not-so-simple details bit by bit. So perseverance is required here; there is a reason why it takes three semesters in a physics course before you can seriously get started with quantum physics}.

\section{Interference of classical waves}

\label{interferenz}

\subsubsection*{Interference in one dimension}

We have already seen that several waves can overlap. It is then also said that these waves \textbf{interfere} with each other or that \textbf{interference} exists. The situation is simplest again in one dimension and when the waves involved have the same wavelengths. Then it only depends on how the waves are positioned relative to each other. If the crests of one wave coincide exactly with the crests of the second wave, then the waves are amplified; this is called \textbf{constructive interference}. This is shown in Figure \ref{konstruktiv}.

\begin{figure}[h] 
\includegraphics[width=0.8\textwidth,keepaspectratio=true]{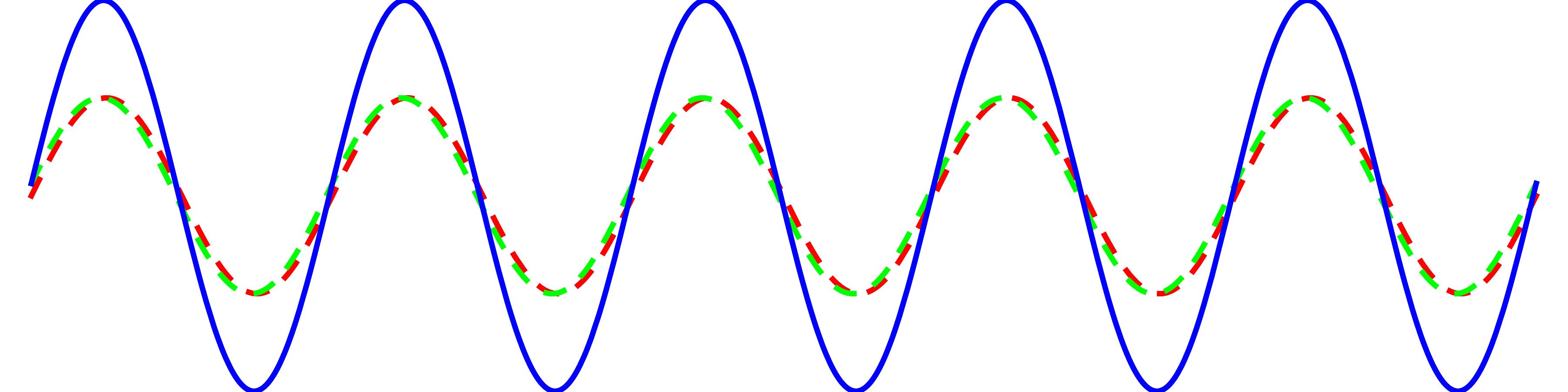}

\caption{Two waves (green and red, dashed) interfere constructively and form the blue, solid wave. The two dashed waves are shown slightly shifted against each other in the drawing so that they can be distinguished. \tiny{(Graphic: Stolze)} 
}
\label{konstruktiv}
\end{figure}

The opposite case is shown in Figure \ref{destruktiv}: The crests of the first wave coincide exactly with the troughs of the second wave and the two waves cancel each other out; this is called \textbf{destructive interference}.

\begin{figure}[h] 
\includegraphics[width=0.8\textwidth,keepaspectratio=true]{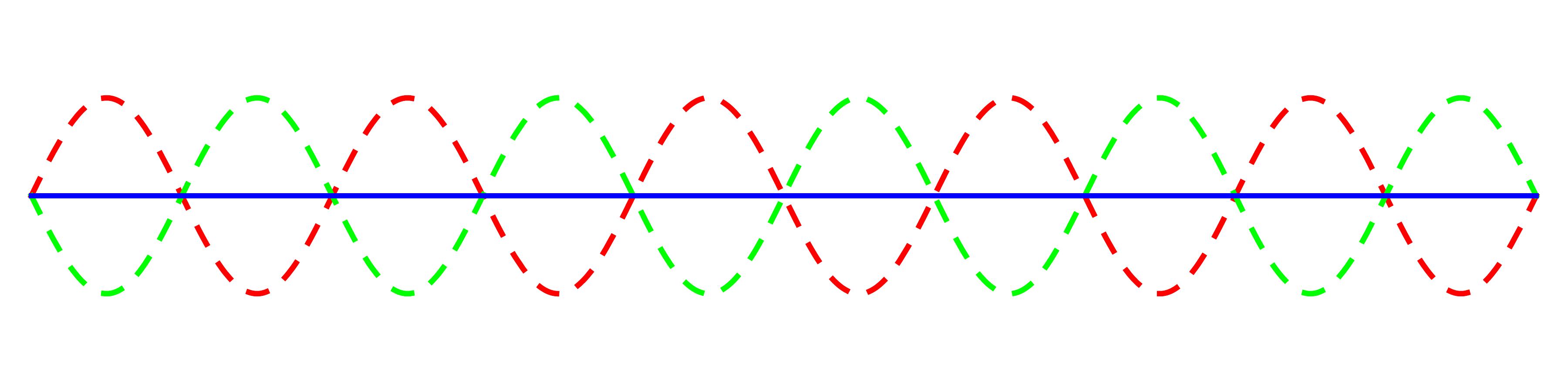}
\caption{Two waves (green and red, dashed) interfere destructively and form the blue, solid \zitateng{wave}; they thus cancel each other out completely. \tiny{(Graphic: Stolze)}
}
\label{destruktiv}
\end{figure}

Of course, all intermediate situations can also occur if the displacement between the wave crests of both waves is between zero (constructive interference) and half a wavelength (destructive interference).

If waves of \textit{different} wavelengths interfere with each other, the situation becomes more complicated (Figure \ref{zweilambdas}); however, we do not need to consider such situations.

\begin{figure}[h] 
\includegraphics[width=0.8\textwidth,keepaspectratio=true]{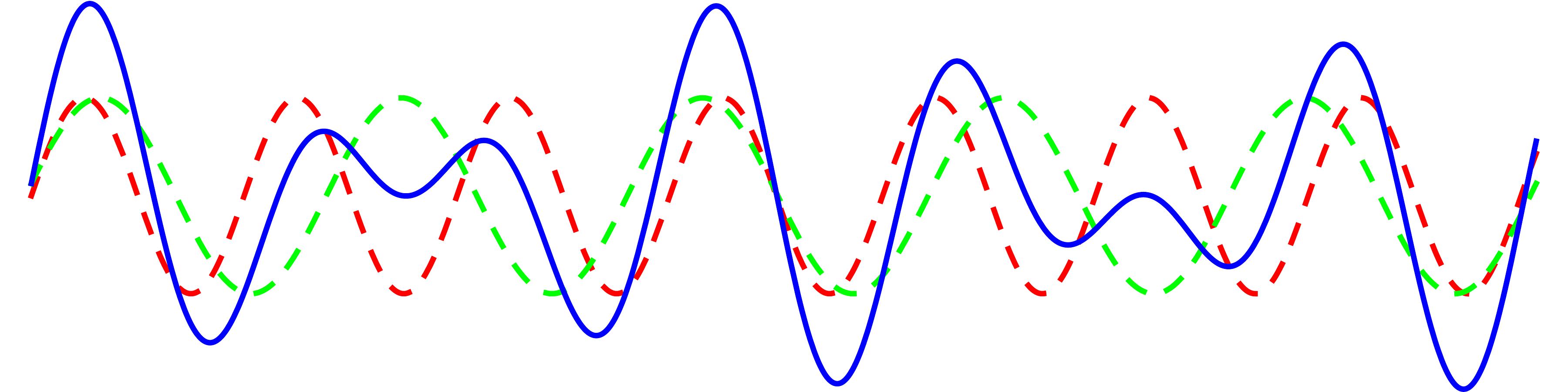}

\caption{Two waves (green and red, dashed) with different wavelengths (in the ratio $1:\sqrt 2$) interfere and form the blue, solid, complicated {wave}. \tiny{(Graphic: Stolze)}
}
\label{zweilambdas}
\end{figure}

\subsubsection*{Interference in two dimensions and diffraction}

We now consider {a more complicated situation: the interference between the waves of two transmitters in \textit{two} dimensions} . This situation is also investigated in the experiment \textit{Wave trough} and is shown in Figure \ref{zweiquellen}. The two transmitters $S1$ and $S2$ (often also referred to as sources) generate waves in time with each other (\zitateng{in phase}), the superposition of which is observed at points $A$ and $B$. Point $A$ is at exactly the same distance from the two sources, so that the wave crests of $S1$ and $S2$ arrive there at exactly the same time. There is therefore constructive interference and waves with a particularly large amplitude are formed at $A$.

\begin{figure}[h] 
\includegraphics[width=0.8\textwidth,keepaspectratio=true]{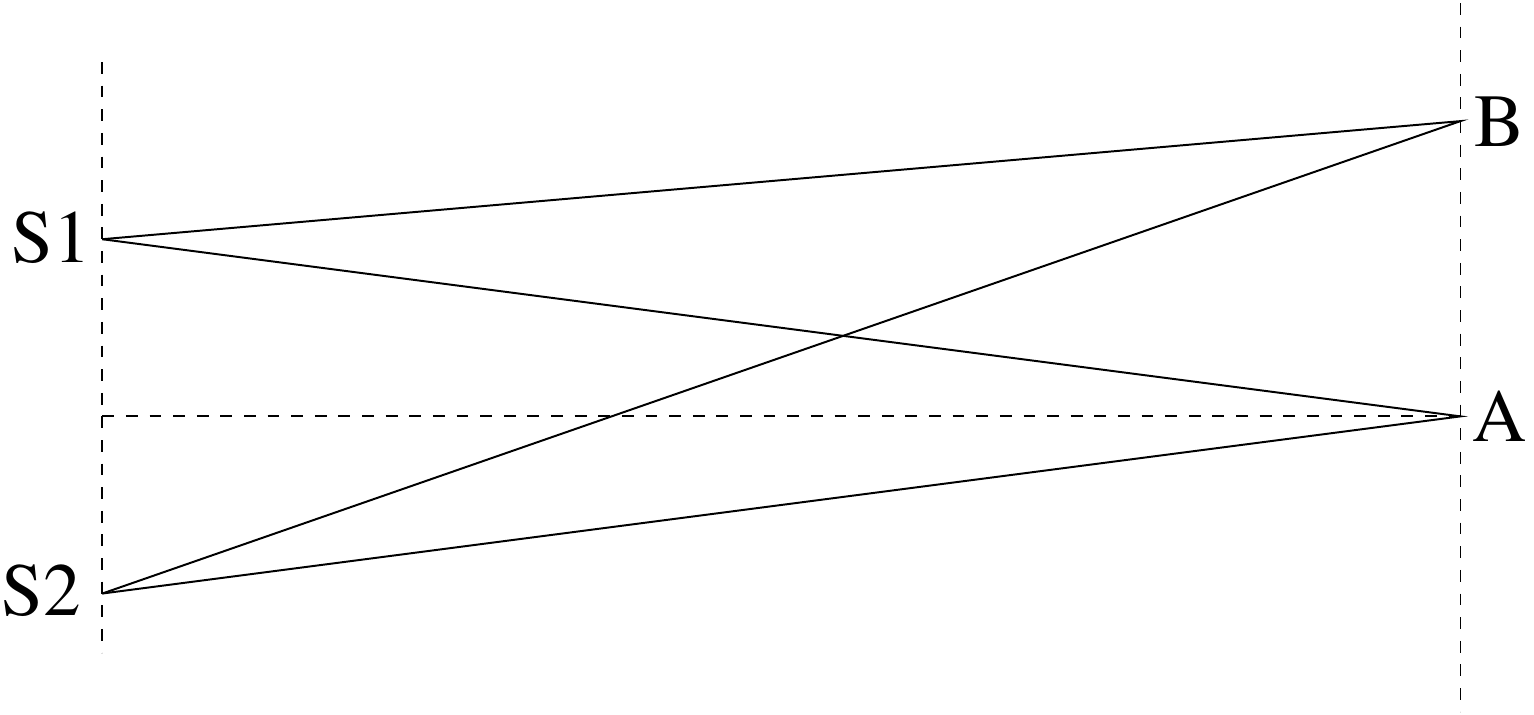}
\caption{On the interference between the waves of two transmitters. \tiny{(Graphic: Stolze)}
}
\label{zweiquellen}
\end{figure} 

At $B$ things looks different. Since the source $S2$ is further away from $B$ than $S1$, the waves from the two sources have to travel a different distance; this is called a \textbf{path difference}. If the path difference is just half a wavelength, then the crests of the waves from one source coincide exactly with the troughs of the waves from the other source and destructive interference occurs. However, if it is an entire wavelength (or exactly two wavelengths, or three, etc.), then wave crests meet wave crests and constructive interference occurs. Between these points, with a path difference of 1.5 (or 2.5 or 3.5 etc.) wavelengths, there is again destructive interference. So if you gradually move the observation point $B$ away from $A$ in your mind, you will alternately observe constructive and destructive interference, i.e. maximum and minimum amplitude of the resulting wave pattern. The distance between these extreme values of the amplitude becomes smaller if the transmitters (sources) emit waves with a shorter wavelength, because then it only takes a shorter distance between $A$ and $B$ to achieve a path difference of $\lambda/2$ (half the wavelength). The situation is similar if the distance between the two sources is increased; the path difference then also changes more quickly. The easiest way to understand this is to imagine the opposite extreme situation: If you \textit{reduce} the distance between $S1$ and $S2$ so far that it is less than $\lambda/2$, you can move the point $B$ as much as you like; you will not reach a path difference of $\lambda/2$ anywhere. The point where destructive interference occurs for the first time has moved to infinity.

If you experiment at the wave trough with one or two sources (or if you simply throw a stone into the water), you will see that the waves spread out in a circle on the surface of the water. The waves emitted by your smartphone spread out in a spherical shape in space. (If they didn't, you would always have to align the device so that the nearest mobile phone station receives your signal; that would be quite annoying). But how do you get from these circular waves or spherical waves to the uniform plane waves that you see in Figure \ref{wellen3}, for example, and how do such waves propagate? This question is answered by the \textbf{Huygens' principle}: Elementary waves emanate circularly (or spherically, depending on the dimension) from each point of a wave front, and their superposition forms the new wave front at the next moment. If the original wave front was a straight line (or plane, depending on the dimension), then the new wave front is also a straight line, provided that the propagation of the wave is not influenced by obstacles.

If you let plane waves run against an obstacle, for example by setting up a \zitateng{wave breaker} in the wave trough, then you can see that the water surface behind the wave breaker does not remain completely calm, but that the wave fronts bend near the end of the wave breaker so that the waves \zitateng{run around the corner}. This is due to the Huygens elementary waves that arise near the end of the breakwater and which no longer have \zitateng{partner waves} with which they can form a plane wave again. If you set up two breakwaters with a narrow gap between them, only very few elementary waves will get through and the wave pattern behind the obstacle will correspond to a single circular elementary wave. This is the case when the width of the gap is approximately as wide as the wavelength.

If the propagation of plane waves is almost completely stopped by a breakwater that leaves only two narrow gaps (a \textbf{double slit}), then you have exactly the situation shown in Figure \ref{zweiquellen}. The two slits are two sources from which waves with the same phase emanate (wave crests start moving at exactly the same time). You can then check the interference between two sources in the experiment \textit{wave trough}, which was discussed theoretically using Figure \ref{zweiquellen}.

All these phenomena can be observed not only with water waves, but also with light waves. 
The fact that the same phenomena occur for waves of very different types (mechanical and electromagnetic) and very different wavelengths (some 10$^{-2}$m and some 10$^{-7}$m) is shown by the experiments \textit{Wave trough} and \textit{Diffraction of laser light}.

The fact that light not only goes \zitateng{straight ahead} through a hole, but also a little \zitateng{around the corner}, is called \textbf{diffraction} of light. Diffraction is a strong indication that light is not a stream of particles moving along the light rays, but clearly has wave properties. Interestingly, however, in the context of quantum physics, light also has particle properties; we will come back to this later.

The wave nature of light is investigated in the experiment \textit{Diffraction of laser light}. In this experiment, a double slit is again used to obtain two sources for waves. Could you also simply take two lasers or any other light sources and recreate the exact situation from Figure \ref{zweiquellen}? --- This would not work with two light bulbs, gas flames or candles, because these sources are not \textbf{coherent}; this means that the wave patterns they produce are not as regular as in Figure \ref{wellen3}, but look more like Figure \ref{wellen1}. Only a few wave crests are emitted one after the other at exactly the same intervals, followed by an interruption, a \zitateng{fault} in the wave pattern. This means that a constant and precise phase relationship between the two waves would not be possible at any time or place, i.e. the shift between the crests of the two waves would vary so much in space and time that no interference pattern would be visible. Even if two lasers were used, it would not be possible, because each laser can emit coherent waves over a certain period of time, but the phase relationship between the two lasers changes again and again, and so quickly that at least the human eye no longer perceives any interference. Interference can only be observed if the coherent waves from \textit{one} light source are sent through two slits and brought together again.

\begin{mdframed}[style=tpq]

\begin{center} 
\textsf{Phase relation, phase difference}
\end{center}

Two waves with the same wavelength $\lambda$ can have different amplitudes and they can be shifted relative to each other so that the wave crests are at different positions. For mathematical reasons, this shift is expressed as an angle. Half a wave length corresponds to an angle of 180$^\circ$, for example. This angle is
called the \textbf{phase difference}, and if the phase difference between two
waves remains constant, they are referred to as having a fixed \textbf{phase relation}.
Important special cases are 0$^\circ$ phase difference (in-phase) and 180$^\circ$ (out-of-phase). An angle of 360$^\circ$ is, as always, completely equivalent to an angle of 0$^\circ$, because a shift of one entire wavelength has no effect, as does a shift
of two wavelengths, i.e. 720$^\circ$, etc.

\end{mdframed}

An interesting situation arises when not only two, but many slits are arranged next to each other at the same distance and the interference is investigated at a \textit{very} large distance, i.e. completely different from that shown in Figure \ref{zweiquellen}. (Realistically, for example, in the experiment \textit{Diffraction of laser light} between the laser and the double slit and also between the double slit and the observation point there are distances of a few {decimetres}, but the two slits are only fractions of a millimetre apart). Then the {light paths} from the two or more slits to one of the observation points (\zitateng{$A$ or $B$}) are practically parallel. But then there is always the same path difference between the waves from two neighboring slits. If this path difference is now exactly one wavelength {(or exactly two wavelengths, or exactly three...)}, \textit{all} waves from the \textit{many} slits interfere constructively and the light becomes very bright at this point. Where this happens depends of course on the wavelength, and therefore such an arrangement of many parallel slits (a \textbf{diffraction grating}) can be used to study which wavelengths the light from a source (e.g. a star) contains. Such a diffraction grating is also used in the \textit{Spectroscopy} experiment. The effect works not only with a screen with many slits, but also with a reflective surface with many regularly arranged lines; with a CD, for example, you can see which colors are contained in the sunlight. (If you don't know what a CD is, ask your parents.) Using a (very special) diffraction grating, Davisson and Germer in the early days of quantum mechanics 
{(1927)} were able to prove that electrons have wave properties. We will come back to this in  Section \ref{interferenz_quanten}, where the experiment \textit{Electron diffraction} is described.

\section{What is a wave function?}
\label{wellenfunktion}

\subsubsection*{De Broglie's wave}

How can one come up with the crazy idea that a particle could have the properties of a wave, for example a wavelength? --- Without going deep into the history of science, we can say that the reverse idea came first: Planck (1900, thermal radiation) and Einstein (1905, photoelectric effect) saw signs that electromagnetic waves could only absorb or emit energy in very specific \textbf{portions}. The size of these \textbf{energy quanta}, i.e. their energy, is firmly linked to the frequency of the waves:
$$
E=h f.
$$
Where $E$ is the energy of the quantum, $f$ is the frequency of the electromagnetic wave and $h$ is Planck's constant, also known as the quantum of action. The energy of the electromagnetic wave therefore consists of a type of \zitateng{energy particles}. In 1922, Compton discovered that the momentum of electromagnetic waves is also quantized. In addition to energy, a light particle or \textbf{photon} also has a momentum $p$, which is linked to the wavelength $\lambda$:
$$
p= \frac h{\lambda}.
$$
It turned out that waves also exhibit particle properties whenever they exchange energy or momentum with their surroundings.

In his doctoral thesis in 1924, de Broglie then pursued the idea that, conversely, particles could also have wave properties, in particular that the momentum $p$ of a particle could be linked to a wavelength $$\lambda=\frac hp .$$ However, neither de Broglie himself nor anyone else knew what actually forms waves, like the surface of water in the case of water waves or the electric and magnetic fields in the case of light waves. Also, in physics, waves normally arise as solutions to equations of motion; however, the de Broglie waves were initially \zitateng{free invention}.

\begin{mdframed}[style=tpq]

\begin{center} 
\textsf{Equations of motion}
\end{center}

In physics, equations of motion are always differential equations,
i.e. they indicate at a certain point in time $t$ how the state changes over
time (mathematically: derivative with respect to time). From this, the state
at a later point in time can be calculated.

For a \textit{classical} particle, the equation of motion is given by Newton's laws, e.g. in the form \zitateng{force = mass times acceleration} If the force is known, the acceleration can be used to
calculate how the velocity changes, and from this, how the position
changes. Then you can follow the path of the particle in space and time, e.g. for a
satellite orbiting the earth.

The equation of motion for a \textit{quantum mechanical} particle is the Schrödinger
equation, a differential equation for the wave function. Due to Heisenberg's
uncertainty principle, the concept of orbit does not exist for a quantum
mechanical particle (e.g. an electron in an atom).

\end{mdframed}

\subsubsection*{Schrödinger's equation}

Schrödinger then began the search for the equation of motion for the de Broglie waves and found it. He investigated how the \textbf{wave function} of an electron behaves when it is \zitateng{held in place} by the electric attraction of a proton; in other words, he calculated the properties of a hydrogen atom and in 1925/26 found exactly the energy spectrum that had previously only been explained in a makeshift manner. This made it clear that the de Broglie waves and the Schrödinger equation represented a breakthrough. However, it was still not clear what physical significance Schrödinger's wave function had, as can be seen from the mocking verses written by participants at a conference in Zurich in the summer of 1926:
\begin{verse}
\textit{Erwin with his Psi can do\\
Calculations quite a few.\\
But one thing has not been seen:\\
Just what Psi does really mean.}\\
\end{verse}
This actually only became clear over time, when the Schrödinger equation was applied to many different physical situations. Today it is quite clear \zitateng{what Psi does really mean}, even if there are a small number of physicists who interpret the wave function differently in detail.

\subsubsection*{Some wave mathematics}

What do physical waves have to do with mathematical functions? --- A more or less complicated wave that moves in one dimension along the $x$-axis, as in Figures \ref{wellenzug_lang} or \ref{wellenzug_kurz}, is described at each instant by a mathematical function $f(x)$ of the coordinate $x$. Of course, the wave moves over time, i.e. $f(x)$ changes over time $t$. You then write $f(x,t)$ and read \zitateng{$f$ of $x$ and $t$}. To describe a two-dimensional surface, you need two coordinates, usually denoted by $x$ and $y$. Waves on a water surface can then be described by the displacement $h(x,y,t)$, which indicates the height $h$ of the moving water surface (compared to the water surface at rest) at any point $(x,y)$ at time $t$. Of course, $h(x,y,t)$ can also be negative in the wave troughs. Things get a little more confusing with electromagnetic waves, e.g. light waves, because they involve electric and magnetic fields that move in three-dimensional space, which is described by the coordinates $x,y$ and $z$. As the two fields themselves are vectors, i.e. they point in a certain direction in space, they each have three components (parts) in the three spatial directions, which are designated $E_x$, $E_y$ and $E_z$ for the electric field $\vec E$, for example. Each of the components depends on the three spatial coordinates and time, so you have functions $E_x(x,y,z,t)$, $E_y(x,y,z,t)$ and $E_z(x,y,z,t)$. To simplify the writing, some abbreviations have been invented, but we do not need to learn them as we will not use them.

We all know that the higher the sea waves are, the more damage they can cause. This has to do with the fact that higher waves contain and transport more energy. This is not so easy to calculate in detail for water waves, but it is comparatively simple for electromagnetic waves. It turns out that the energy content is related to the \textit{square} of the fields involved. The energy density (energy per volume) at the point $(x,y,z)$ at time $t$ is made up of a part proportional to
$$
 |\vec E|^2=(E_x(x,y,z,t))^2 + (E_y(x,y,z,t))^2 + (E_z(x,y,z,t))^2 
$$
and the analoguous expression formed from the three components of the magnetic field. The quantity $|\vec E|$ (read: modulus of the vector $\vec E$) is the length of the electric field vector, i.e. the electric force on a unit charge.

The wave function of quantum mechanics needs less paperwork than the fields of electromagnetic waves, especially if we want to limit ourselves to one dimension $x$ in space, as we have done so far. Traditionally, the wave function is denoted by the Greek letter $\Psi$ (read: Psi), and in one dimension it is a function of position $x$ and time $t$:
$$
\Psi(x,t).
$$
There is, however, a little complication, since $\Psi(x,t)$ is a \textbf{complex} function, i.e. it consists of two parts, $ \textsf{Re}\Psi(x,t)$ and $\textsf{Im}\Psi(x,t)$ (read: \zitateng{real part of Psi and imaginary part of Psi}). These two parts play roles similar to the components $E_x$, $E_y$ und $E_z$ of the electric field or the coordinates $x$ and $y$ of a point in  a plane where, for example, the distance $r$ of the point $(x,y)$ to the origin $(0,0)$ can be calculated following Pythagoras:
$$
r^2=x^2+y^2 .
$$
For now, that's all we need to know about complex numbers and complex functions. 

\begin{mdframed}[style=tpq]

\begin{center} 
\textsf{Complex number}
\end{center}

A \textbf{complex number} $z$ is made up from two real numbers $x$ and $y$ according to the prescription $$ z= x+iy.$$ Here $i$  is the \textbf{imaginary unit} with the property $$i^2=-1.$$
$x$ is called the \textbf{real part} and $y$ the \textbf{imaginary part} of $z$. Complex numbers
can be represented in a plane with $x$ and $y$ as coordinates in the usual way (Gaussian number plane). The quantity $$|z|^2=x^2+y^2$$ is called \textbf{absolute value squared} of $z$. The absolute value $|z|$ of the number $z$ then indicates the distance between $z$ and the zero point of the Gaussian number plane.

To warm up, simply calculate the product of $3 + 4i$ and
$3 - 4i$. Hint: The result is real and a square.

\end{mdframed}

Similar to $|\vec E|^2$ above, the wave function yields the important quantity
$$
| \Psi(x,t)|^2 = ( \textsf{Re}\Psi(x,t))^2 + (\textsf{Im}\Psi(x,t))^2
$$
(read: \zitateng{Psi modulus squared}). In contrast to $|\vec E|^2$ for electromagnetic waves, $| \Psi(x,t)|^2$ for the wave function has nothing to do with the energy, which is determined in a completely different way, namely as \textbf{eigenvalue} from the time-independent Schrödinger equation. (More details can be found in the article \zitateng{Schrödinger equation} in the glossary but will not be needed here.)

\subsubsection*{Wave function and probability}

It turns out that $| \Psi(x,t)|^2$ is related to   the probability of finding the particle described by the wave function at time $t$ at position $x$\footnote{{What the role of $\Psi(x,t)$ \zitateng{itself} (i.e. without $|...|^2$) is, we will look at in more detail in Section \ref{interferenz_quanten} and in Chapter \ref{IV}}}. This needs to be made a little more precise. Since both $\textsf{Re}\Psi(x,t)$ and $\textsf{Im}\Psi(x,t)$ are \zitateng{entirely normal} (i.e. real) functions of $x$ and $t$, $| \Psi(x,t)|^2$ as the sum of two squares is positive in any case, at worst zero, but never negative. Since negative probabilities are meaningless, this is reassuring.

To establish the exact connection to probability, we look at Figure \ref{aufenthalt}. It shows the magnitude squared of a possible wave function for a particle moving in a dimension $x$ at a certain time $t$. The probability of finding the particle between $x=0.5$ and $x=1$ by a measurement is the area under the curve of $| \Psi(x,t)|^2$ on the interval between $x=0.5$ and $x=1$. (For know-it-alls: The definite integral of $| \Psi(x,t)|^2$ from $x=0.5$ to $x=1$). The probability of finding the particle \textit{somewhere} on the $x$-axis must be equal to one, because it must be somewhere. Therefore, the area that the curve of $| \Psi(x,t)|^2$ encloses with the \textit{whole} $x$-axis must be equal to one. A wave function $\Psi(x,t)$ that does not have this property is physically meaningless and therefore inadmissible. (Experts say that decent wave functions must be \textbf{normalized}.) $| \Psi(x,t)|^2$ is an example of a \textbf{probability density} that indicates how the probability is distributed in space (here: on the $x$-axis).

\begin{figure}[h] 
\includegraphics[width=\textwidth,keepaspectratio=true]{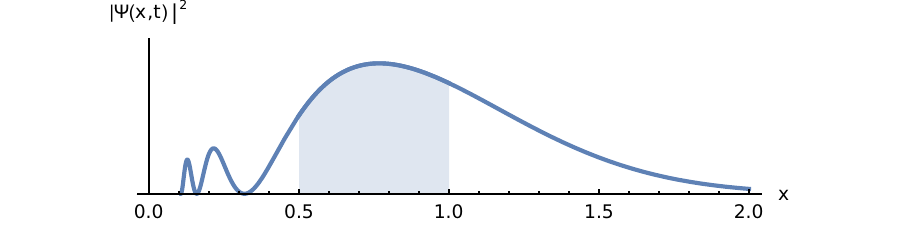}
\caption{$| \Psi(x,t)|^2$ for a typical wave function. The shaded area indicates the probability of finding the particle between $x=0.5$ and $x=1$. \tiny{(Graphic: Stolze)}
}
\label{aufenthalt}
\end{figure}

The wave function contains \textit{all} information that can be obtained about a particle. Quantum mechanics cannot make any statements about the \textit{exact} position of the particle; this concept is ultimately meaningless. You can only state the probability with which a particle can be found in a certain area {if you look}. Many physicists in the early days of quantum mechanics (including Einstein) believed that this could not yet be the whole truth, that there must still be \zitateng{hidden variables}, i.e. quantities that determine exactly \zitateng{where the particle really is}, and that we simply do not know. It is now clear that such hidden mechanisms cannot exist.

\begin{mdframed}[style=tpq]

\begin{center} 
\textsf{Measurements and probabilities}
\end{center}

The state of a particle in one dimension at time $t$ is \textit{completely} determined by the wave function $\Psi(x,t)$. Nevertheless, the result of a \textbf{single measurement} of the position $x$ (or another measurable quantity) is not predictable. However, if the following operations are carried out very often one after the other:
\begin{enumerate}
\item Prepare the state $\Psi(x,t)$,
\item Measure the position $x$ and
\item Register the result of the position measurement,
\end{enumerate}
then the distribution of the measurement results of the position, which is given by the probability density $|\Psi(x,t)|^2$, is gradually obtained.

In contrast to a single measurement, this is a statistical measurement; experts often speak of \textbf{ensemble measurements}. It is important that the system is always prepared anew, i.e. the initial state $\Psi(x,t)$ is created.
Alternatively, one could also imagine to prepare a  large number of particles (an ensemble of particles) all in the desired state $\Psi(x,t)$ and measure each particle once.

After a single measurement, the state of the particle is usually \textit{no longer} $\Psi(x,t)$, but a different one, depending on which physical quantity was measured and which value this measurement produced.

\textit{A measurement changes the state of a system.}

\end{mdframed}

\begin{mdframed}[style=tpq]

\begin{center} 
\textsf{Measurements change states}
\end{center}

If a particle is in a state $\Psi$ and a (single) measurement is carried out on this particle, then after the measurement the state of the particle is usually \textit{no longer} $\Psi$, but a different one, depending on which physical quantity was measured and which value this measurement produced. \textit{A measurement
changes the state of a system.}

A simple example is a measurement that determines \zitateng{on which side} a particle is, i.e. if it is found in the region $x > 0$ (\zitateng{right}) or at $x \le 0$ (\zitateng{left}). Before the
measurement, in the state $\Psi$, the particle is on the left with a certain probability and on the right with a certain probability. (The sum of the two probabilities is, of course, one.) The measurement places the particle \textit{either} in a state $\Psi_{\textrm{right}}$ \textit{or} in a state $\Psi_{\textrm{left}}$. If the measured value was \zitateng{right} and if a second measurement is carried out immediately after the first (without influencing the particle in any other way), the result will certainly be \zitateng{right} again.

The first measurement has determined the state here. The initial state $\Psi$ is said to have \textbf{collapsed} to the state $\Psi_{\textrm{right}}$  or to have been \textbf{reduced} to this
state. The states $\Psi_{\textrm{right}}$ and $\Psi_{\textrm{left}}$  are referred to as \textbf{eigenstates} of the measurement quantity \zitateng{side}. An eigenstate is a state in which the measurement of the associated quantity provides a very specific value (the \textbf{eigenvalue}) with certainty.

\textit{After} the measurement, the state of the particle continues to develop
according to the Schrödinger equation, and thus the particle can possibly also move to the left again, so that the eigenstate property determined by the measurement is lost again. For this reason, the same results are only obtained for the two measurements described if they are carried out \textit{immediately} one after the other.

\end{mdframed}

We have only ever talked about the measurement of the position $x$ and read off the corresponding probabilities directly from the probability density $| \Psi(x,t)|^2$. However, quantum mechanics also provides rules for calculating probabilities from the wave function $\Psi(x,t)$ for other measurable quantities (momentum, energy, angular momentum, ...). For those quantities, too, the result of a single measurement can only be predicted in exceptional cases.

Here we have become familiar with a feature of quantum mechanics that takes some getting used to, namely the fact that it does not provide clear answers to many questions, but only probabilities for the various possibilities. Nevertheless, we note (once again, because it is so important) that the wave function $\Psi(x,t)$ describes the state of a particle \textit{completely}; more information than is contained in $\Psi(x,t)$, \textit{does not exist}. Nevertheless, this does not open the door to chance, as some people believe. If $\Psi(x,t)$ is known at time $t=0$, the Schrödinger equation yields $\Psi(x,t)$ for all futures. Chance only ever has a chance (and uses it) when a measurement is carried out. Even if the wave function is uniquely determined for all times by the Schrödinger equation, it is not possible to predict the result of a single position measurement at a future time. Here you could protest and say something like: \zitateng{But surely I can accurately calculate the orbit of a satellite or the fall of a stone; what's all this quantum nonsense about}? To deal with such objections, you have to take a closer look and do something that many people don't like at all, namely calculate and compare a few numbers. We will deal with this in  Section \ref{wie_bewegt}.

\subsubsection*{Two states at the same time}

There is a second property of quantum mechanics, or more precisely, of the wave function, that takes at least as much getting used to. We have already seen this with water waves: Waves can be superimposed. The sum of two possible waves is another possible wave. This is not really exciting at all. It only becomes exciting when you identify the term \zitateng{wave(function)} with the term \zitateng{state of a particle}, as quantum mechanics does. If $\Psi_a(x,t)$ and $\Psi_b(x,t)$ are two admissible states of a particle, then $\Psi_a(x,t) + \Psi_b(x,t)$ is also an admissible state. \footnote{If you have been paying attention all this time, you will see a problem: The sum of two normalized wave functions is probably no longer normalized. --- Bravo! Good attention! You do indeed have to add a suitable pre-factor so that the sum is normalized again. But this is not a serious problem.}. \textit{The particle can therefore be in two different states at the same time!} The most extreme example, which we certainly cannot fully discuss, is Schrödinger's infamous cat, which is supposedly alive and dead at the same time. We discuss some less worrying aspects in Chapter \ref{III}.

To make it clear once again: The superposition of two possible states is not about two particles, one of which is in the state $\Psi_a(x,t)$ and the other in the state $\Psi_b(x,t)$, but about \textit{a single} particle whose state is a \zitateng{weighted sum} 
$\alpha \Psi_a(x,t) + \beta \Psi_b(x,t)$ with (complex) numbers $\alpha$ and $\beta$; technically this is called a \textbf{linear combination}.

\begin{mdframed}[style=tpq]

\begin{center} 
\textsf{Complex number}
\end{center}

A \textbf{complex number} $z$ is made up from two real numbers $x$ and $y$ according to the prescription $$ z= x+iy.$$ Here $i$  is the \textbf{imaginary unit} with the property $$i^2=-1.$$
$x$ is called the \textbf{real part} and $y$ the \textbf{imaginary part} of $z$. Complex numbers
can be represented in a plane with $x$ and $y$ as coordinates in the usual way (Gaussian number plane). The quantity $$|z|^2=x^2+y^2$$ is called \textbf{absolute value squared} of $z$. The absolute value $|z|$ of the number $z$ then indicates the distance between $z$ and the zero point of the Gaussian number plane.

To warm up, simply calculate the product of $3 + 4i$ and
$3 - 4i$. Hint: The result is real and a square.

\end{mdframed}

An example would be a state whose probability density looks like Figure \ref{zweibuckel}. If the two parts of the wave function move towards each other, the question arises as to what happens when the two parts collide. Answer: Nothing actually happens, because the partial wave functions simply overlap. However, the combination of the superposition principle for $\Psi(x,t)$ with the role of $|\Psi(x,t)|^2$ as a probability density gives rise to situations that require explanation, which are dealt with in the next section \ref{interferenz_quanten}.

\begin{figure}[h] 
\includegraphics[width=\textwidth,keepaspectratio=true]{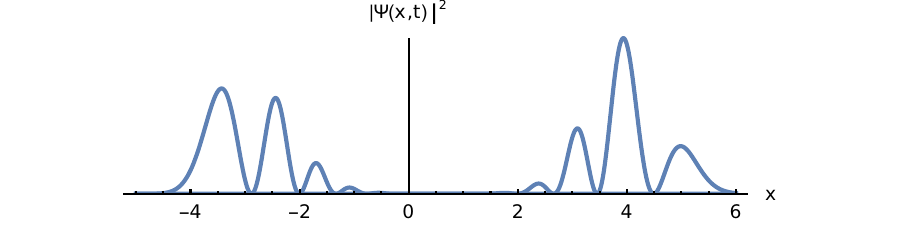}
\caption{$| \Psi(x,t)|^2$ for a wave function consisting of two spatially separated parts. \tiny{(Graphic: Stolze)}
}
\label{zweibuckel}
\end{figure}

\section{Interference and quantum physics}
\label{interferenz_quanten}

{Let's recap the most important facts about quantum mechanics from the last section. The state of a particle is described by the wave function $\Psi(x,t)$. A sum of two (or more) admissible wave functions is also an admissible wave function. The probability of finding a particle at time $t$ at position $x$ is given by $| \Psi(x,t)|^2$. Two (or more) wave functions can therefore interfere with each other and amplify or attenuate each other, just like classical light waves, which are also studied in \textit{Treffpunkt Quantenmechanik}.

{In order to understand all these experiments, including those with classical light waves, we will take a closer look at the interference between two transmitters, as shown in Figure \ref{zweiquellen}. In the experiment \textit{Wave trough}, the waves can be observed directly on the surface of the water. This is no longer possible in experiments with light waves, as their wavelength is much too short and the light travels through space without leaving any traces. Only when the light hits a surface can it be observed, on a screen or with a suitable detector.

\subsubsection*{Interference between two point sources}

{We are therefore investigating} what the brightness pattern on a screen looks like when the light waves from two point sources (transmitters) interfere with each other. In  Section \ref{interferenz} we saw that it depends on the path difference between the waves that meet on the screen: If the path difference is zero, there is constructive interference, if it is half a wavelength, there is destructive interference, and if the path difference is an entire wavelength, the interference is constructive again. What does this mean in concrete terms? What is the distance between two bright spots on the screen?

\begin{figure}[h] 
\includegraphics[width=\textwidth,keepaspectratio=true]{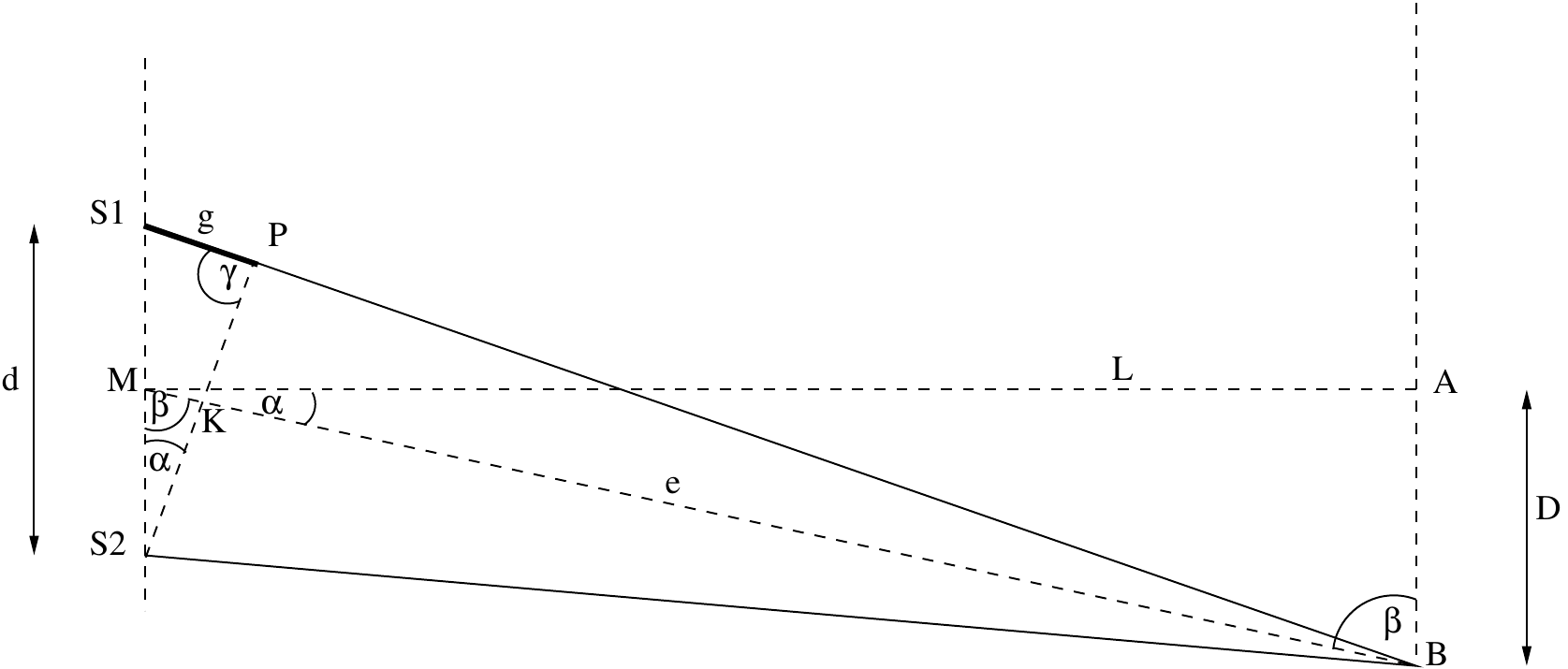}
\caption{Interference between two point sources $S1$ and $S2$ on an observation screen at a large distance $L$. At point $A$ there is constructive interference, at point $B$ the elementary waves have a path difference $g$. The drawing is not to scale; the distance $L$ is \textit{very} much larger than the distances $d$ and $D$. \tiny{(Graphic: Stolze)} }
\label{punktquellen}
\end{figure}

{Let us consider the situation in Figure \ref{punktquellen}. Two sources $S1$ and $S2$ are at a distance $d$ from each other. At a large distance $L \gg d$ from the sources, the screen is parallel to the plane in which the sources lie. At point $A$ there is constructive interference, as it is exactly the same distance from the two sources $S1$ and $S2$. At what distance $D$ must the point $B$ be so that constructive interference prevails there again? For constructive interference, the path difference $g$ between the two waves from $S1$ and $S2$ arriving at $B$ must be just equal to the wavelength: $g=\lambda$. Then the distance $D$ between the two bright spots $A$ and $B$ on the screen is
$$
D=\frac{\lambda L}d.
$$}

{This relationship can be justified geometrically. If you are not interested in the details, you can skip this paragraph. The sine theorem in the triangle $S1$ $S2$ $P$ returns}
$$
\frac g{\sin \alpha} =\frac d{\sin \gamma}.
$$
{The acute angle at $M$ is also $\alpha$. To see this, we compare the small triangle $M$ $S2$ $K$ with the large triangle $M$ $B$ $A$. Both triangles are right-angled: $A$ has a right angle due to its construction, and the angle at $K$ in the small triangle is also $90^{\circ}$, since $K$ $B$ is the height in the isosceles triangle $P$ $S2$ $B$. The angle $\beta$ at $B$ in the large triangle $M$ $B$ $A$ appears again as an alternate angle at $M$ in the small triangle $M$ $S2$ $K$. Both triangles (small and large) are therefore right-angled and contain the angle $\beta$. This means that the third angle, i.e. $\alpha$, must also be identical}.

{In the large triangle $M$ $B$ $A$ the definition of the sine then provides} 
$$
\sin \alpha = \frac De.
$$  
{Now we exploit the proportions. For a realistic experiment with visible light, the distances $d$ and $D$ are of the order of millimetres or smaller, but the distance $L$ is half a meter or more. Then the angle $\alpha$ is very small and the angle $\gamma \approx 90^{\circ}$, so that $\sin \gamma \approx 1$. Inserting $\sin \alpha$ and $\sin \gamma$ into the sine theorem then yields 
$$\frac {ge}D=d$$ and since $e \approx L$ is due to the size ratios, the result is $$ \frac {gL}D=d.$$ Since the path difference $g$ should be equal to the wavelength $\lambda$, the equation for $D$ is obtained by inserting and rearranging. The next points with constructive interference, i.e. with path difference $2\lambda$, $3\lambda$ etc. are then at the distance $2D$, $3D$ etc. from point $A$, because the corresponding angles (e.g. $2\alpha$, $3\alpha$ etc.) are still very small and we can argue in the same way as above.

For a realistic experimental setup, we use red light with $\lambda=600$nm$=600 \cdot 10^{-9}$m, a distance between the sources of $d=0.3$mm$=0.3 \cdot 10^{-3}$m and a distance to the screen $L=0.5$m. Then the distance between the points $A$ and $B$ results in
$$
D=1 \mathrm{mm}.
$$
The ratios are therefore much more extreme than shown in Figure \ref{punktquellen}: $B$ is much closer to $A$. The formula for $D$, which we have just considered, shows that
\begin{itemize}
\item $D$ becomes larger as $L$ increases, i.e. the screen is pushed further away.
\item $D$ becomes larger if $\lambda$ increases, i.e. light with a longer wavelength is used.
\item $D$ increases if $d$ decreases, i.e. the distance between the sources \textit{reduces}.
\end{itemize}
In the experiment \textit{Diffraction of laser light}, two slits are used as light sources, 
{which are illuminated by a single laser (from behind)}. We have already mentioned the reasons for this (keyword: coherence) in  Section \ref{interferenz}. As interference patterns on the screen behind the slits, we then expect many stripes at a distance of $D$, because if we move on from point $B$ by $D, 2D, 3D...$, we get a path difference of $2\lambda, 3\lambda, 4\lambda...$ and thus constructive interference again and again.

\subsubsection*{An extended source: diffraction at the slit}

If we actually carry out the experiment \textit{diffraction of laser light}, a different picture emerges: Besides the bright fringe at position $A$ in Figure \ref{punktquellen}, you can see a few more bright fringes at distances $D$, $2D$,... but then it gets dark; {examples of this can be seen in Figure \ref{doppelspalt}}. This is because the two slits that serve as light sources (transmitters) have a finite width, and this must be the case for enough light to penetrate to be able to see or measure anything.

\begin{figure}[h] 
\includegraphics[width=0.3\textwidth,keepaspectratio=true]{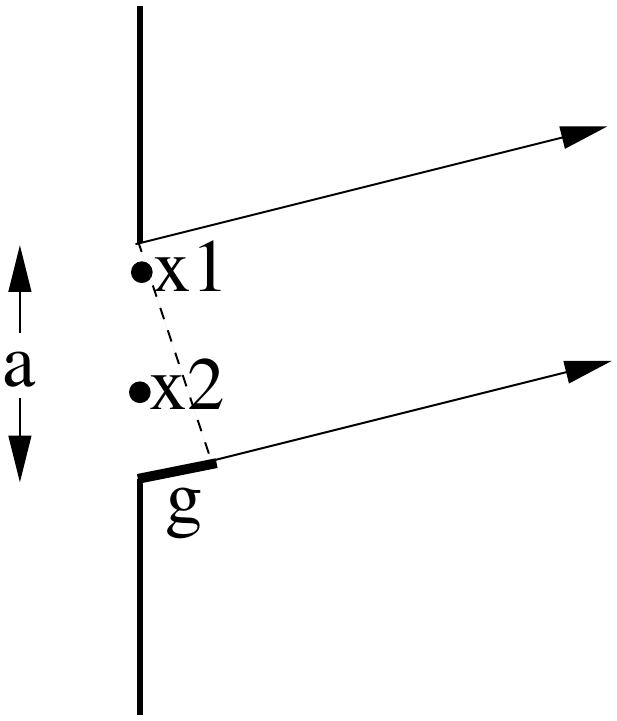}
\includegraphics[width=0.7\textwidth,keepaspectratio=true]{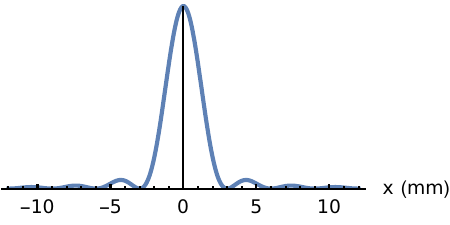}
\caption{\textit{Left}: A single slit of width $a$; the path difference $g$ between elementary waves from the two edges of the slit is one wavelength; between the two points $x1$ and $x2$ it is half a wavelength. \textit{Right}: Light distribution (intensity) on a screen at a distance of 0.5m behind a slit with $a=0.1$mm, for light with $\lambda=600$nm. $x$ is the distance from the point of maximum intensity. {The two zeros of the intensity next to the maximum just correspond to the situation in the left partial image: The path difference between the elementary waves from the two edges of the slit is exactly one wavelength}. \tiny{(Graphic: Stolze)} }
\label{einzelspalt}
\end{figure}

{To understand what happens, we first consider a single slit of width $a$ (Figure \ref{einzelspalt}). According to Huygens' principle (Section \ref{interferenz}), elementary waves emanate from each point of the slit and interfere on the screen. In the direction exactly \zitateng{straight ahead} (i.e. on the screen opposite the center of the slit), all these elementary waves interfere constructively; we therefore expect maximum brightness there}. Now, however, we consider a \zitateng{slanted} direction, which causes a path difference between the elementary waves. We choose the direction so that the path difference $g$ in the image \ref{einzelspalt} between the elementary waves from the upper edge of the slit and the lower edge of the slit is exactly one wavelength, i.e. $g=\lambda$ for the two points with distance $a$. Then for each point in the slit (e.g. $x1$) there is a point (e.g. $x2$) that is exactly at a distance of $a/2$. The elementary waves from these two positions then have a path difference $\lambda/2$ and therefore interfere destructively. In this direction, we therefore expect \textit{no light} on the screen. If we turn the direction further away from the straight-ahead direction, the path difference $g$ between the two edges of the slit becomes greater than $\lambda$ and not all elementary waves interfere destructively, but most of them still do. Therefore, there is a little more light on the screen again, but nowhere near as much as in the straight-ahead direction. The distribution of the light intensity on the screen is shown in Figure \ref{einzelspalt} (right). This distribution then limits the brightness of the fringe pattern discussed earlier, which is expected for two {ideal point sources or two \zitateng{infinitely narrow} slits}, respectively.

{The intensity distribution for a single slit depends on the wavelength $\lambda$ and the slit width $a$. The larger $a$ is, the less we have to twist the arrows in Figure \ref{einzelspalt} (left) from the straight-ahead direction to reach the path difference $\lambda$, i.e. the first zero of the intensity distribution. The width of the central maximum in Figure \ref{einzelspalt} (right) (and thus of the entire intensity distribution) therefore decreases as $a$ increases: The wider the slit, the narrower the diffraction pattern. Instead of increasing $a$, we can also reduce $\lambda$ according to this reasoning: the smaller the wavelength, the narrower the diffraction pattern. Ultimately, it is only the ratio $\frac{\lambda}a$ that matters.}

\subsubsection*{Two extended sources: Diffraction of light at the double slit}

\begin{figure}[h] 
\includegraphics[width=0.5\textwidth,keepaspectratio=true]{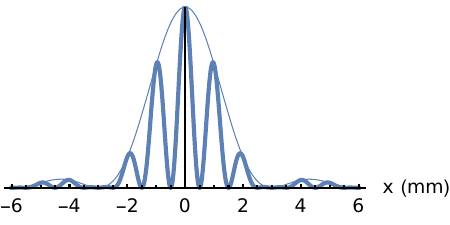}
\includegraphics[width=0.5\textwidth,keepaspectratio=true]{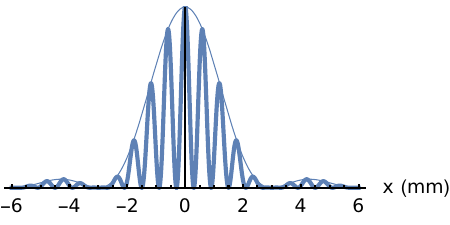}
\caption{Intensity distribution on a screen 0.5m behind a double slit (thick line). 
{Each slit is $a=0.1$mm wide, the centers of the slits are $d=0.3$mm 
(left) and $d=0.5$mm apart (right).} $\lambda=600$nm. The thin line is the intensity distribution for a single slit of 0.1mm width. $x$ is the distance from the point of maximum intensity. \tiny{(Graphic: Stolze)} }
\label{doppelspalt}
\end{figure}

Figure \ref{doppelspalt} shows two examples of the interference pattern behind a double slit. The two slits are each $a=0.1$mm wide, and the distance between the centers of the slits is $d=3a=0.3$mm in the left image and $d=5a=0.5$mm in the right image. {The equation $D=\frac{\lambda L}d$ derived above for the distance between the brightness maxima is confirmed in the figure: The larger distance $d$ between the slits (right image) provides a smaller distance $D$ of the brightness maxima.} {The finite width of the individual slits ensures that the brightness maxima are strongly attenuated towards the outside; the thin line in the figure shows the intensity distribution for \textit{one} slit. The ratio 3:5 of the slit spacing is observed again if the brightness maxima are counted from $x=0$ to $x=3$mm, where the brightness pattern of the single slit has its first zero point.

{You can measure a similar brightness distribution as in Figure \ref{doppelspalt} in the experiment \textit{Diffraction of laser light}. Depending on the width and distance of the two slits, the brightness pattern will be different; two possibilities are shown in the figure. In the experiment \textit{Double slit with single photons} you can choose to block one of the two slits. You can then observe the interference pattern for a single slit (Figure \ref{einzelspalt} or thin line in Figure \ref{doppelspalt}) and compare it with the interference pattern of the double slit (thick line in Figure \ref{doppelspalt}). Note that the intensity distributions for one or two slits (thin or thick lines in Figure \ref{doppelspalt}) are drawn at different scales to make them easier to compare. The intensity at $x=0$ is of course greater for a double slit than for a single slit. Due to destructive interference, less light reaches the screen in certain areas for the double slit than for the single slit; however, the \textit{total} light intensity is twice as high for two identical slits as for one slit, because twice as much energy reaches the screen through two slits as through one}.

The diffraction and interference of classical light waves at two slits is already an astonishing phenomenon: If you \textit{cover} one of the two slits, the diffraction pattern becomes \textit{brighter} in some places, and if you open the second slit again, it becomes dark again in these places. \zitateng{light + light = darkness} is therefore sometimes a correct equation. However, intuition already refuses to recognize this in the case of classical light waves. The fact that quantum mechanics blurs the boundaries between waves and particles does not make things any easier. Interference phenomena are at the heart of quantum mechanics and contribute significantly to the unease that it causes for many people.

The double-slit experiment is very well suited to discussing {similarities and differences between classical physics and quantum physics} and will therefore be used repeatedly in this section.

\subsubsection*{Double slit experiment with classical particles}

\begin{figure}[h] 
\includegraphics[height=5.5cm]{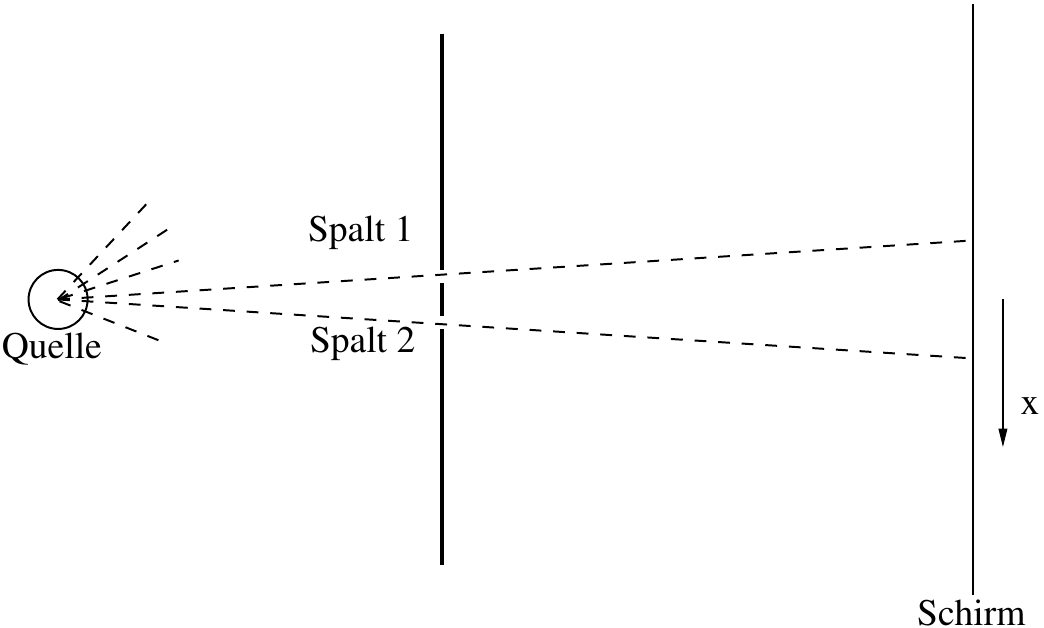}
\includegraphics[width=5.5cm,angle=-90,origin=rb,keepaspectratio=true]{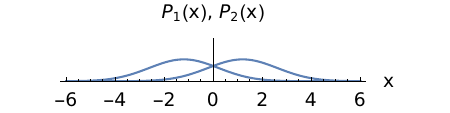}
\includegraphics[width=5.5cm,angle=-90,origin=rb,keepaspectratio=true]{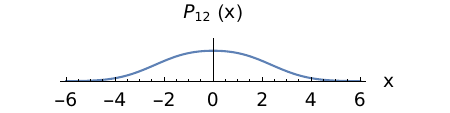}
\caption{Double-slit experiment with classical particles. The source (Quelle) sends untargeted particles in the direction of the double slit. If only slit 1 (Spalt 1) is open, the particles on the screen (Schirm) form the distribution $P_1(x)$, if only slit 2 (Spalt 2) is open, the result is $P_2(x)$. If both slits are open, the distribution is $P_{12}(x)$.
\tiny{(Graphic: Stolze)}}
\label{doppelspalt_teilchen}
\end{figure}

We have discussed the double-slit experiment with classical light waves in detail. We now compare it with a corresponding experiment with classical particles, such as small spheres, which are shot by an apparatus against the wall with the double slit. Most of the particles will bounce off the wall, some will fly through the slit, some will touch the edges of the slit and be deflected. Behind the wall there is a screen that registers where and how often it was hit by a particle. The result is a distribution $P(x)$, which tells us how many particles arrived at the position $x$ of the screen after the experiment (Figure \ref{doppelspalt_teilchen}). We now carry out the experiment three times in succession. First we open only slit 1, then only slit 2 and finally both slits. We call the measured distributions $P_1(x)$, $P_2(x)$ and $P_{12}(x)$. In the first experiment, all balls go through slit 1, in the second experiment all balls go through slit 2. What distribution will we get if both slits are open? --- Very simple: In addition to the balls that passed through slit 2 in our second experiment, we also get those that would have passed through slit 1 if it had been open, i.e. exactly those that we measured in the first experiment. The overall distribution is therefore as follows
$$
P_{12}(x)=P_1(x)+P_2(x),
$$
or in words: \zitateng{particles + particles = more particles}.

In the first two experiments, we know exactly which slit the particles passed through (because only one was open in each case), and in the third experiment we could have determined this because we can follow the paths of the particles in space if we want to, e.g. with a video camera.

{This is not at all surprising and corresponds exactly to our ideas; however, the experiment with light waves at the double slit is completely different}. If only one slit is open, we can of course say through which slit the wave has passed, and the brightness pattern for a single slit results, as in Figure \ref{einzelspalt}. If both slits are open, however, we can  \textit{not} say, \zitateng{through which slit the wave has passed} and the result is a brightness pattern that is \textit{not} simply the sum of the results for two single slits, see Figure \ref{doppelspalt}. In some places, \zitateng{light + light = darkness} then applies.

\subsubsection*{Double-slit experiment with quantum mechanical particles, electron diffraction and an exciting movie from Japan}

If we were to carry out the double-slit experiment with electrons (or other quantum mechanical particles), the picture would be exactly the same as with light waves. \textit{Electrons therefore move through the double slit like waves}.

In practice, it is very difficult to carry out the double-slit experiment with electrons as described\footnote{{Electrons are charged, and if electrons accumulate in the apparatus that have not passed through the slits, they exert electrical repulsive forces on the electrons that come later, which influence the experiment. In extreme cases, the wall with the slits is then so strongly charged that all further electrons are repelled.}}; this is why another experiment is used in \textit{Treffpunkt Quantenmechanik} to demonstrate the wave properties of electrons, namely the experiment \textit{Electron diffraction}. This experiment demonstrates the diffraction of electron waves by a \textit{grating}. We already briefly mentioned this situation at the end of  Section \ref{interferenz}: When waves from many equally spaced slits interfere constructively with each other, very high and sharp interference maxima arise. If the path difference between the waves of two neighboring slits is just one wavelength, this is referred to as the first-order maximum; if the path difference is two wavelengths, this is referred to as the second order, and so on.

\begin{mdframed}[style=tpq]

\begin{center} 
\textsf{De-Broglie wavelength}
\end{center}

A particle with mass $m$ moving at speed $v$ has momentum $p = mv$. In
quantum mechanics, this particle then has a wavelength
$$
\lambda = \frac hp,
$$
where $h$ is Planck's constant. The heavier the particle and the faster it
moves, the shorter the wave length.

\end{mdframed}

The grating in the experiment \textit{Electron diffraction} consists of atoms arranged regularly in planes in a thin graphite crystal. To allow the electrons to move freely, the experiment takes place in a glass flask that is pumped empty of air. After the electrons {hit the graphite crystal}, they fall onto a screen on which light is produced. (In the past, similar picture tubes were used in televisions and computer monitors; your parents may still remember them). Only when the de Broglie waves of electrons fall at a very specific angle on the planes of {regularly arranged carbon atoms in the graphite crystal do they interfere constructively with each other, like light hitting an array of many slits at equal distances or of many lines on a reflective surface (a diffraction grating). In this case, they are reflected at the same angle; otherwise, they interfere destructively with each other and provide no signal}. As the graphite crystal used consists of many small, differently aligned sub-crystals, the electrons are always deflected by the same angle, but in a different direction at each sub-crystal, so that the deflected electrons form a ring around the passing beam on the screen. {Since a graphite crystal has two different distances between different planes}, two such first-order diffraction rings are observed. The higher orders are too weak to be seen in this setup.

Of course, all of this only works if all electrons have the same wavelength. This is ensured by setting the electrons in motion with a precisely defined acceleration voltage. This gives them a very specific kinetic energy, a very specific momentum $p$ and, according to de Broglie's formula $\lambda=h/p$, also a very specific wavelength. When generating \zitateng{electron waves}, we therefore strangely fall back on the image of the electron \textit{as a particle}, which has a very specific charge $-e$ and is therefore provided with a kinetic energy $E_{\textsf{kin}}=eU$ by an accelerating voltage $U$, which, due to the very specific mass $m_e$, corresponds to a very specific momentum $p$ ($E_{\textsf{kin}}=p^2/2m_e$), which then provides the de Broglie wavelength $\lambda=h/p$. As a wave, the electron then interferes happily and is registered on the screen -- but how and as what?

In the experiment \textit{Electron diffraction} you can see bright and dark areas on the fluorescent screen {which prove the wave nature of the electrons. From observation with the naked eye, however, it is not possible to decide exactly how these arise, because at} every moment a very large number of electrons (or a very strong electron wave?) fall onto the screen and cause it to glow. However, there are more precise experiments that work with very small numbers of electrons and allow an insight into the details. The detectors used must be much more sensitive (and therefore unfortunately also much more expensive and difficult to handle) than the fluorescent screen in our experiment.

\begin{wrapfigure}{r}{3cm}
  \includegraphics[width=3cm]{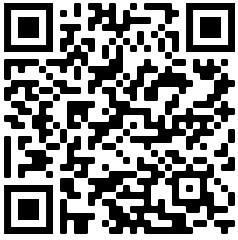}
  \phantom{x}
  \includegraphics[width=3cm]{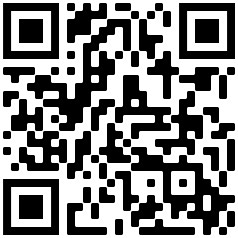}
\end{wrapfigure}
An interference pattern of bright and dark areas can then be seen again, but these are composed of more or less closely spaced bright dots. Each such dot corresponds to a single electron hitting the detector. The number of incoming electrons can be reduced to such an extent that the bright points of light can be seen appearing individually on the detector. At first they appear in seemingly random places and only after a long time does a stripe pattern slowly emerge. 

There are several short videos on the internet of such experiments with electrons at a double slit\footnote{The apparatus is {considerably} trickier than just a wall with two slits, but the result corresponds exactly to a double slit for electrons. {If you want to repeat the experiment, you first need an electron microscope, which then has to be modified.}}. One video comes from the research laboratories of a Japanese electronics company (\url{https://rdg.ext.hitachi.co.jp/rd/moviee/doubleslite.mpeg}), another can be found at \url{https://www.youtube.com/watch?v=ZqS8Jjkk1HI} If these links no longer work, use a suitable search engine to search for \zitateng{electron double slit experiment}, for example. The Japanese video shows a 30-minute experiment; the beginning is recorded at the original speed, followed later by time-lapse recordings.

Electrons that move \textit{as waves} are therefore obviously registered \textit{as particles}. Their properties as particles (mass, charge) are also important for setting a certain wavelength, as we saw above. The two aspects, waves and particles, are \textit{inextricably linked}. This fact is called \textbf{dualism of wave and particle}.

\subsubsection*{When is there interference and when is there not?}

You can see how unusual the situation is if you watch one of the above videos again. (Feel free to do so now.) When the electrons arrive at the detector one after the other, there is \textit{at most one electron in the apparatus} at any given time. Nevertheless, an interference pattern is created at the end. It is therefore not the case that the electrons \zitateng{interfere with each other}, as each electron only flies off when the previous electron has long since completed its journey in the detector. Nevertheless, an interference pattern is created! Does the electron interfere \zitateng{with itself} by splitting and passing through both gaps? -- No one has ever observed a half electron. Since the electron scatters light because of its charge, you can use a light source near the slit to determine which of the two slits it has passed through. But if you do this, you don't get an interference pattern. You have, so to speak, forced the electron to move \zitateng{like a particle} and thus switched off its wave properties. More generally: \textit{Interference only occurs in a quantum mechanical process between an initial state and a final state if the process can take different ways and if there is no information about which way the process has taken}. As soon as such \zitateng{which-way-information} is available, there is no more interference.

So what interferes in quantum mechanics? -- From a technical-mathematical point of view, it's simple: the wave functions for the different possibilities overlap (add up), just like the electromagnetic fields in the interference of light waves, and this leads to the electromagnetic field being zero (darkness) or the wave function being zero (no probability of finding the particle) in some places. To put it more philosophically, you could say that the different possibilities that the particle has interfere with each other, but only as long as you allow the particle all these possibilities, i.e. you don't try to measure \zitateng{which-way-information}.

We will see below how such information can be determined using the example of the interference of ({classical}) light in the \textit{Mach-Zehnder interferometer} experiment. Interference of individual quantum objects is investigated in the experiment \textit{Double-slit with single photons}. Although photons are somewhat more abstract than electrons at first glance, they are much easier to handle experimentally because they are not charged; this is why the double slit with single electrons does not exist in \textit{Treffpunkt Quantenmechanik}.

Very similar to our experiment \textit{Electron diffraction}, Davisson and Germer proved as early as 1927 that electrons can be described by waves. They directed a beam of electrons at a nickel crystal and found that the electrons were reflected particularly strongly from the crystal surface in very specific directions. Their analysis showed that the electrons behaved like waves whose wavelength was, as they wrote as cautious scientists, \zitateng{in acceptable agreement with the values $h/mv$ of the undulatory mechanics}.

The interference phenomena of quantum mechanics clearly show that the concept of \textit{trajectory of a particle}, which is so important in classical mechanics, is meaningless in quantum mechanics. This is most obvious when there are several classically clearly distinguishable alternatives, such as the two slits in the double-slit experiment. We will see why classical mechanics nevertheless offers a (very) meaningful description for \zitateng{classical situations}, such as the motion of macroscopic objects, when we take a closer look at the important concept of uncertainty in  Section \ref{unschaerfe}. The fuzziness and meaninglessness of the concept of orbit also mean that individual quantum mechanical particles of the same type (such as electrons or neutrons or protons) cannot be distinguished from one another. (You can distinguish between the different types of particles, but not between two electrons.) This is not a question of measurement accuracy, but something fundamental and has very important consequences for the structure of matter.
\subsubsection*{Double slit with single photons}

The experiment \textit{Double slit with single photons} in \textit{Treffpunkt Quantenmechanik} allows you to measure the interference patterns for single and double slits (Figures \ref{einzelspalt} and \ref{doppelspalt}) yourself, in two different ways.

The double slit is illuminated with a \textit{strong} light source with a fixed wavelength (a laser) and either slit can be closed or both can be left open. 
{The diffraction pattern is then not only observed qualitatively on a screen, but also measured quantitatively. For this purpose, a detector is used behind a movable narrow slit at the point where you would set up an observation screen. If you move the slit in small steps over the interference pattern, you can measure the light intensity behind this slit and record it in a table}. In this way, you can obtain the intensity distributions for \zitateng{slit 1 open}, \zitateng{slit 2 open} and \zitateng{both slits open}.

The whole experiment can then be repeated with a \textit{weak} light source (an incandescent light bulb
operated at low voltage). In order to obtain an interference pattern, a certain
wavelength is filtered out of the continuous spectrum of the lamp. At the end, a
highly sensitive detector is used to count how many photons per second pass through
the movable observation slit. An acoustic display (\zitateng{cricket}) can be used to make it
audible when the detector registers a photon\footnote{Due to the properties of a thermal source such as the light bulb used here, photons are occasionally generated in pairs. This is therefore not a single photon experiment in the strictest sense. Light sources that \textit{only} generate single, well-separated photons are difficult to produce.} and thus easily determine where
many or few photons arrive. The spatial distribution of the photon numbers measured
and recorded in a table should ultimately be similar to the intensity distributions from
the first part of the experiment, with the strong light source.

\begin{mdframed}[style=tpq]

\begin{center} 
\textsf{Spectrum (plural: spectra), spectroscopy
}
\end{center}
A \textbf{spectrum} is the set of the wavelengths (or frequencies) of
electromagnetic radiation that can be observed in a range, together with
their respective intensities. 
If all wavelengths can occur within some range, the spectrum is called
\textbf{continuous}, e.g. in the case of thermal radiation from a hot object. If only
very specific wavelengths can occur, the spectrum is called \textbf{discrete}. This is
the case for pure chemical elements, for example. Discrete spectra are also
called \textbf{line spectra}, named after the image of colored lines on a dark
background that appears in an optical spectrometer. By examining such line
spectra in \textbf{spectral analysis}, it is possible to determine which chemical
elements the substance under investigation contains. The study of
spectra is called \textbf{spectroscopy}.

In more general terms, the possible values of the energy of a quantum system
are also referred to as the \textbf{energy spectrum}.

\end{mdframed}

This experiment shows that light, too, on the one hand behaves like a wave by showing interference, and on the other hand, as a photon, is registered at a certain point on the screen like a particle.  Only the distribution of a very
large number of registered photons generates the interference pattern of the light
waves again. Just like electrons, photons also exhibit \textit{wave-particle dualism}.

\subsubsection*{The Mach-Zehnder interferometer: How to prevent interference}

We will now discuss another interference experiment in which the different
 \zitateng{paths} of the light are much further apart than in the double slit, and where it is possible
to observe how the \zitateng{which path information} influences the interference. This is the
\textit{Mach-Zehnder interferometer} experiment.

\begin{figure}[h] 
\includegraphics[width=0.8\textwidth,keepaspectratio=true]{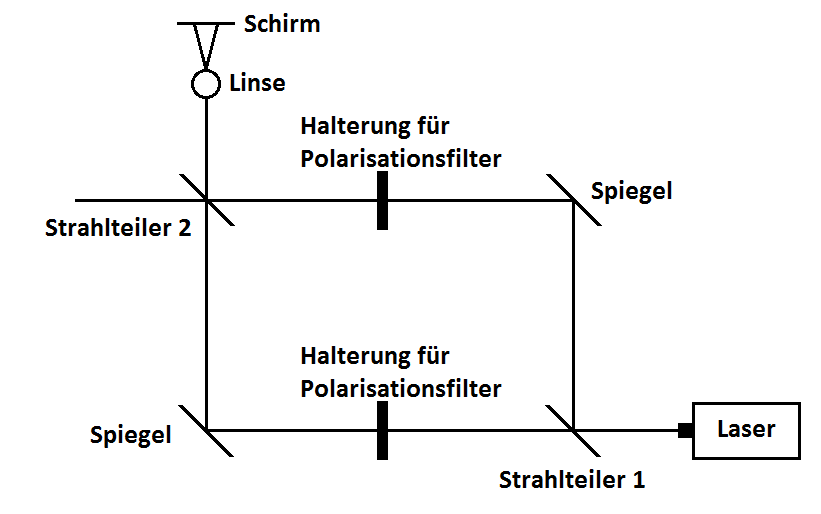}
\caption{Structure of a Mach-Zehnder interferometer. The light from the laser is split by
beam splitter 1 (Strahlteiler 1) and the two partial beams are recombined via two mirrors (Spiegel) at beam splitter 2 (Strahlteiler 2).
Part of the light then passes through a lens (Linse) (upwards) to the observation screen (Schirm), the remaining
light leaves the apparatus (to the left); Halterung für Polarisationsfilter = mount for polarization filter.
\tiny{(Graphic: Duffe)} 
}
\label{mach_zehnder}
\end{figure}

In principle, a Mach-Zehnder interferometer is a simple device. It consists of a light
source, two good mirrors (which fully reflect the light), two bad mirrors (which partly
reflect and partly transmit the light) and an observation screen. The schematic structure is shown in Figure  \ref{mach_zehnder}. In physics, the
\zitateng{bad mirrors} have the nicer name of \zitateng{beam splitters}. The polarization filters
mentioned in the figure only come into play later. On the observation
screen, interference occurs between the two light waves, which were separated at the first
beam splitter. Whether the interference is constructive or destructive
depends on the path difference between the two light paths, as usual. If the path
difference changes by half a wavelength, e.g. by moving a mirror, constructive
interference becomes destructive and vice versa. As visible light only has a wavelength
$\lambda$ of a few 100 nm, the interferometer is a very sensitive device that can be used to
determine the optical properties (refractive index) of a substance that is introduced
into one of the two light paths very precisely. At \textit{Treffpunkt Quantenmechanik},
however, we are interested in other questions that we would now like to discuss.

As already mentioned, on the observation screen you can see the interference
between the two wave trains that pass through the apparatus on different paths. If all the mirrors and lenses are positioned correctly, a pattern of light and dark
rings forms, like on a target. If it is bright in the center of the rings, there is
constructive interference. A little further out, the waves involved have traveled slightly
different paths and destructive interference occurs, and so on. The special feature of
the Mach-Zehnder interferometer compared to the double
slit is that it is now possible to \zitateng{mark} the two light paths in a simple way, and thus
distinguish on which path the light arriving on the screen is traveling.
 The \textit{polarization of light} is used for this purpose, which can be
examined in detail in the experiment of the same name in the \textit{Treffpunkt Quantenmechanik}.

\begin{mdframed}[style=tpq]

\begin{center} 
\textsf{Polarization of light}
\end{center}
Electromagnetic waves consist of coupled electric and magnetic fields that change in space and time. In the case of a monochromatic (only one frequency or wavelength) plane wave in free space, the electric and magnetic fields are perpendicular to each other and both are perpendicular to the direction in which the light propagates. If the electric field only oscillates back and forth in one direction, it is referred to as a \textbf{linearly polarized} wave. If you look in the direction of wave propagation, you can
distinguish between two fundamentally different polarizations (directions of oscillation): \zitateng{right and left} ($x$ polarization) and \zitateng{up and down} ($y$ polarization). All
more complicated waveforms can be composed from these basic forms.

\end{mdframed}

We now imagine two $y$-polarized waves that meet with half a wavelength difference.
Then we have exactly the situation shown in Figure \ref{destruktiv}. The electric fields of the two
waves are exactly the same (in magnitude, but opposite in direction) everywhere and cancel each other out; there is therefore
destructive interference. Of course, the same happens if both waves are $x$-polarized; again
there is destructive interference. The situation is different if one wave is $x$-polarized
and the other $y$-polarized. As the electric fields of the two waves then point in
mutually perpendicular directions, they can never cancel each other out and there can
be no interference.

This behaviour of equally or differently polarized waves can be easily observed with
the Mach-Zehnder interferometer. For this purpose, polarization filters are inserted
into the setup shown in Figure \ref{mach_zehnder}. A polarization filter consists of a substance
(usually a special plastic film) that only allows light waves with a certain polarization
to pass through. As a rule, polarization filters are rotatably mounted in a ring so that
the transmitted polarization direction can be adjusted. If you install polarization
filters in both light paths and set them both in the $x$-direction, you will observe
interference; the same applies if you set them both in the $y$-direction. However, if you
turn one of the two filters from the y-direction to the $x$-direction, the interference
pattern will become blurred and the screen will be (more or less) uniformly
illuminated.

If one polarization filter is set in the $x$-direction and the other in the $y$-direction, we
have marked the two light paths by the polarization directions and can tell which light has come via which path. If we turn the two filters back to the same orientation, we gradually delete the \zitateng{which path} information and restore the interference.

Up to this point, we have explained the behavior of light in the Mach-Zehnder
interferometer completely with the means of classical physics by talking about opposing
or mutually perpendicular electric fields. This is of course completely correct, but what
does all this have to do with quantum mechanics? - This becomes clear when we
imagine\footnote{Unfortunately, this is not possible in practice with a simple observation screen, as our eyes are not sensitive enough to register individual photons.} reducing the light intensity to such an extent that there are only a few
photons in the apparatus at any time, in extreme cases only one or sometimes none at
all. Then, depending on the position of the polarization filters, we have
either the \zitateng{which path} information and thus (after a longer observation time) no
interference pattern arises, or (if both polarization filters are oriented in the same
direction) we have no way of distinguishing the light paths and can therefore observe
the gradual emergence of an interference pattern, just as in the Japanese experiment
with electrons discussed above.


\subsubsection*{A photon takes no path}
At the end of this section, a word of warning about the term \zitateng{which path information} is appropriate.
It is
very misleading to think of a photon as \textit{something similar to a particle} that moves
\textit{along a path}. A photon is a certain amount of energy, which is contained in a certain
oscillation form of the electromagnetic field, and this electromagnetic field extends over all space.
In contrast, an electron or another material particle is not perfectly localized in space
due to the uncertainty principle, as we will discuss in Section \ref{unschaerfe}, but the wave
function can be used to define its mean position with a certain degree of (im)precision.
For a photon, no meaningful wave function can be defined and therefore no position.
A photon only shows its particle nature when it is actually no longer there because a
detector has registered it at a certain position.

\section{Uncertainty and the consequences}
\label{unschaerfe}

In order to be able to discuss uncertainty, we should first define the terms with more certainty and determine what is actually meant by uncertainty. It is often impossible to determine a quantity
precisely. There can be various reasons for this: The exact measurement can be difficult,
or the quantity is not clearly defined from the outset, such as the age or body size of a
whole group of people, or the height of the trees in a forest. In such cases, a repeated
measurement of a quantity $x$ results in a distribution of measured values that are more or
less widely scattered.

\subsubsection{Some statistics: mean values and standard deviations}

An average value can then be calculated in the usual way; in science, this is referred
to as the \textbf{mean value} and is denoted by $\langle x \rangle$.\footnote{There are also other ways of writing the mean value, but we use the notation with the angle brackets.} Often one would also like to know the
accuracy with which the mean value is determined, i.e. whether the many measured
values are all very close to the mean value or widely scattered. To do this, you
determine the deviation from the mean value for each individual measured value $ x $,
i.e. $x-\langle x \rangle$, and square this value. The mean value of this quadratic deviation is then
calculated, i.e.$\langle (x -\langle x \rangle )^2 \rangle$ . Squaring means that upward and downward deviations cannot
cancel each other out. The calculated
\zitateng{mean square deviation} is also called \textbf{variance}, and the larger it is, the more widely
the individual measured values scatter around the mean value.

However, the variance cannot be directly compared with the mean value: For
example, if $x$ is a length, then the variance of $x$ due to squaring is an area, i.e. something
completely different. To obtain a length again, simply take the square root of the
variance and call this quantity the \textbf{standard deviation} of $x$, in formulas:
$$
\sigma(x) = \sqrt{\langle \; (x-\langle x \rangle)^2 \; \rangle}.
$$
This quantity is exactly what is called the \textbf{uncertainty} of $x$ in physics. The usual
notation for the uncertainty in physics uses the Greek capital letter delta; we then
have
$$
\Delta x = \sqrt{\langle \; (x-\langle x \rangle)^2 \; \rangle}= 
\sqrt{\langle x^2 \rangle -\langle x \rangle^2}.
$$
(The two forms $\langle \; (x-\langle x \rangle)^2 \; \rangle$ und $\langle x^2 \rangle -\langle x \rangle^2$ of the variance are mathematically equivalent.)

\begin{mdframed}[style=tpq]

\begin{center} 
\textsf{Mean values and fluctuations}
\end{center}

The mean value is what is colloquially known as the \zitateng{average}
or \zitateng{average value}, and everyone intuitively knows how to calculate the average grade of an exam. As an 
example, the average age of the 20 students in an advanced physics course could be 16.8 years, just like the mean age of 20 people who enter a department store one after the other on a Saturday morning. If we call the age of the first person $A_1$ ,
that of the second person $A_2$ , etc., then the mean age of 20 people is simply
$$
\langle A \rangle = \frac{A_1+A_2+...+A_{20}}{20}.
$$
The angle brackets $\langle ...\rangle$ are a common way of denoting mean values.

The two groups of people (physics class and department store) will
certainly show different age distributions, even if  $\langle A \rangle$ for both groups is the same. There will also be older people and small children in the department store, i.e. people with large deviations from the mean. The
\textbf{variance} (the mean square deviation from the mean) is a measure that describes the extent of deviations from the mean:
$$
\textsf{Var}(A) = \langle \; (A-\langle A\rangle)^2 \;\rangle =  \langle A^2 \rangle - \langle A \rangle ^2 .
$$
(The square of the deviation from the mean is used so that the upward and downward deviations do not cancel each other out. The two specified forms of variance are equivalent. This is not immediately obvious, but can be calculated).

In order to be able to compare the deviations with the mean value of the age in a meaningful way, the square root of the variance is used; this is the \textbf{standard deviation} $\sigma(A)$
$$
\sigma(A) = \sqrt{\textsf{Var}(A)} = \sqrt{\langle A^2 \rangle - \langle A \rangle ^2}.
$$
The standard deviation $\sigma(A)$ ($\sigma$: Greek letter sigma) is what we mean when we say that the average age of the physics class is 16.8 $\pm$ 0.8 years.

When we talk about \textbf{uncertainty} in physics, we always mean the standard deviation. The uncertainty is often denoted by the capital Greek delta; $\Delta x$ then
means the standard deviation of the position $x$: $\Delta x= \sigma(x)$, and similarly for the momentum $p$ and other quantities.

\end{mdframed} 

\subsubsection*{How uncertain are waves?}

After we have determined what exactly uncertainty is, we go back to the classical
waves in Section \ref{welle}, where we had already noticed a relationship between the
uncertainty of the wave length and that of the position: The wave in Figure \ref{wellenzug_kurz} had a
fairly well-defined position, but a wave length that could not be measured precisely;
the wave in Figure \ref{wellenzug_lang}, on the other hand, had a very precisely defined wave length,
but was spatially extended and thus had no precisely determinable position. These were
not specially constructed cases, but examples of a general mathematical relationship.

We will dispense with the details of the necessary higher mathematics here, but we
will hint at what it is about and why it has something to do with quantum mechanics.
For the sake of simplicity, we will again only talk about waves in one dimension.
Everything works the same way in more than one dimension, it just requires more
paperwork and offers more opportunities to miscalculate.

\begin{mdframed}[style=tpq]

\begin{center} 
\textsf{Amplitude}
\end{center}

The \textbf{amplitude} is the maximum displacement of a wave, e.g. in the case of a
water wave, the height of the wave crest above the calm water surface. In
addition to the amplitude, the \textbf{phase} of a wave is also important, as it
determines, for example, exactly where the wave crests are located.
When waves are superimposed, both the amplitudes and the phases of the
partial waves are important, keyword interference.

In the case of a superposition of several quantum mechanical wave
functions, one also speaks of the amplitudes of the different wave functions
and thus means the respective components (expressed as complex
numbers).

\end{mdframed}
\begin{mdframed}[style=tpq]

\begin{center} 
\textsf{Complex number}
\end{center}

A \textbf{complex number} $z$ is made up from two real numbers $x$ and $y$ according to the prescription $$ z= x+iy.$$ Here $i$  is the \textbf{imaginary unit} with the property $$i^2=-1.$$
$x$ is called the \textbf{real part} and $y$ the \textbf{imaginary part} of $z$. Complex numbers
can be represented in a plane with $x$ and $y$ as coordinates in the usual way (Gaussian number plane). The quantity $$|z|^2=x^2+y^2$$ is called \textbf{absolute value squared} of $z$. The absolute value $|z|$ of the number $z$ then indicates the distance between $z$ and the zero point of the Gaussian number plane.

To warm up, simply calculate the product of $3 + 4i$ and
$3 - 4i$. Hint: The result is real and a square.

\end{mdframed}

\textit{Any} waveform can be represented as a superposition of simple sinusoidal waves like
the one on the right in Figure \ref{brandung}. This requires several (or even many) waves with
different wave lengths, amplitudes and phases. For the mathematical
paperwork (and for the connection to quantum mechanics), it is more appropriate not
to use the wave length $\lambda$ itself, but its reciprocal value; that would be the number of
waves per meter. For mathematical and technical reasons, this quantity is multiplied
by $2\pi$, designated by the letter $k$, and named the \textbf{wave number}:
$$
k=\frac{2\pi}{\lambda}.
$$
Now you can specify for all different values of $k$ which amplitude and which phase
 the wave with the wave number $k$ must have in order to represent a desired
waveform as a superposition of all these different waves. Every physicist learns
exactly how to do this in the 3rd semester (at least in Dortmund) under the name
\textit{Fourier analysis}. For every waveform in space $x$, there is a clearly defined pattern of
amplitudes (and phases) of the different waves with wavenumber $k$. It can
then be seen from many examples that a narrow waveform in the spatial coordinate $x$
corresponds to a broad distribution of amplitudes in the wavenumber $k$, and vice
versa. 

\begin{figure}[h] 
\includegraphics[width=\textwidth,keepaspectratio=true]{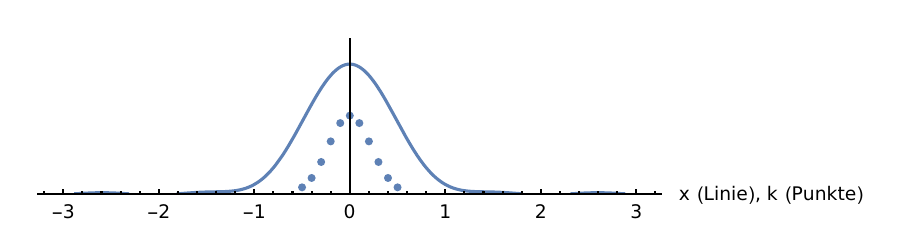}
\includegraphics[width=\textwidth,keepaspectratio=true]{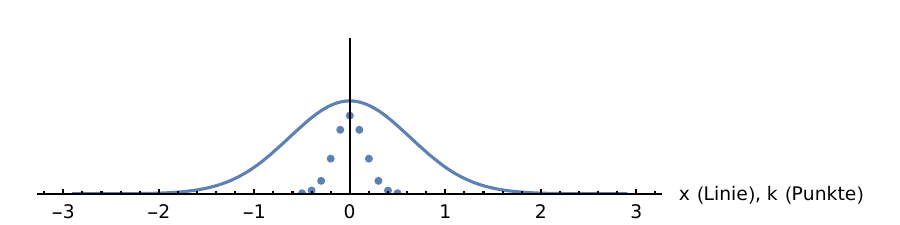}
\caption{Two waveforms (solid lines) in one dimension $x$. Both are created by
superimposing 11 waves with certain wave numbers $k$. Both the coordinate $x$ and the wave
number $k$ are plotted along the horizontal axis. The wavenumbers used are the same for both
waveforms, but the amplitudes associated with these wavenumbers (height of the points above
the horizontal axis) are different. A broad distribution of $k$ amplitudes (large standard
deviation of $k$, top) leads to a narrow $x$ waveform (small standard deviation of position $x$)
and a narrower distribution in $k$ leads to a broader waveform in $x$ (bottom). There is therefore
a relationship between the uncertainties (standard deviations) in $x$ and in $k$ already for \zitateng{completely
standard} (classical) waves. (Linie=line, Punkte=dots)
\tiny{(Graphic: Stolze)} 
}
\label{fourierreihe}
\end{figure}

A simple example for two cases is shown in Figure \ref{fourierreihe}. In both cases, 11
waves with certain wavenumbers $k$ were superimposed. The values of $k$ used are the same in both cases, but the amplitudes of
the 11 waves are different. The amplitudes are represented by the dots in Figure \ref{fourierreihe}:
The value of the amplitude is plotted upwards, the corresponding $k$ value to the
right/left. The solid lines show\footnote{In the two parts of the figure, the horizontal axes for the coordinate $x$ and the wave number $k$ have different
scales for display reasons; however, the scales do not differ between the two parts of the figure. All waves are
cosine waves with the same phase. A cosine wave with $k = 0$ (i.e. with infinite wave length) is
simply a constant contribution, without which the waveform would not decay to zero towards the outside. For
experts: Since only a finite number of $k$ values at equal intervals are used here, the waveform repeats itself
on the $x$-axis at regular intervals.} the waveform in the spatial coordinate $x$. You can
see clearly that the pattern of the amplitudes in the
the lower image is significantly narrower (\zitateng{more sharp}) than the upper image, while at the
same time the waveform is much wider ( \zitateng{less sharp}).

\subsubsection*{The uncertainty principle is actually old hat...}

As already mentioned in Section \ref{welle}, the uncertainty relationship between the
position $x$ and the wavenumber $k$ illustrated in Figure \ref{fourierreihe} is by no means new. If the
wave is not considered in space (at a fixed point in time), but in time (at a fixed
position), then the position $x$ must be replaced by the time $t$ and the wavenumber $k$
by the frequency $f$ (more precisely: by the angular frequency $\omega=2\pi f$, but this is not
too important). Then you can see: To transmit a particularly short signal, you need
 a particularly broad \zitateng{band} of frequencies. This is one of the most important
laws of message transmission and is well known to every electrical engineer.

If you can transmit particularly short signals, you can transmit a particularly high
number of signals per second. That's why 5G mobile communications needs much
higher frequencies and a broader band of frequencies than 4G, and the high data transmission rate of 5G means that you can watch movies in the finest
resolution on your smartphone somewhere deep in the woods without any jolting - provided
that someone has placed enough 5G transmitters in the woods.

Figure \ref{fourierreihe} makes it clear that for the waveforms shown, a
spatial uncertainty $\Delta x$, and for the amplitude distributions a \zitateng{wave number
uncertainty} $\Delta k$ can be defined, and for each example calculated one will find that the
product $\Delta x \Delta k$ is a constant that is of the order magnitude of one. As I said,
this is well known and doesn't upset anyone...

\subsubsection*{...but it gets exciting when the momentum enters the scene}

...and now (1924) Monsieur le Prince Louis Victor de Broglie comes along and
suggests that a particle with momentum $p$ can be assigned a wavelength $\lambda= \frac hp$
in order to describe quantum physics. Then the momentum is
$$
p = \frac h{\lambda} = \frac h{2\pi} \frac{2\pi}{\lambda} = \hbar k,
$$
i.e. proportional to the wave number! (With $\hbar$, the Planck constant  divided by 2$\pi$, as the
proportionality constant). This turns the well-known position-wavenumber uncertainty
relation into the position-momentum uncertainty relation that some people have been
getting worked up about for around 100 years:
$$
\Delta x \Delta p \sim \hbar.
$$
(The $\sim$ sign here means \zitateng{roughly in the ballpark of}).

When you arrive at the uncertainty principle as described above, you don't really see
what is supposed to be exciting about it. However, one should not underestimate how
groundbreaking and exciting the idea is that momentum and wavelengths should have
something to do with each other. Because the uncertainty principle is so important, we
now want to write it down in the precise form  which can be proven
mathematically\footnote{For details see physics study course, semester 4.}:
$$
\Delta x \Delta p \ge \frac{\hbar}2.
$$
The constant $\hbar=1,055 \cdot 10^{-34}$Js is so tiny that it is not possible to directly check
the uncertainty relationship between the position and momentum of a particle in a comparatively
simple laboratory such as the \textit{Treffpunkt Quantenmechanik}.

\subsubsection*{Uncertainty in the wave trough}

A simple experiment with the wave trough shows that uncertainty relationships are
already at work in classical physics, only they are not usually referred to as such
there. If you generate plane waves in the wave tank and let them pass through a \textit{wide}
slit, then a straight wave bundle spreads out behind the slit. If the slit is made narrower,
the straight wave bundle becomes a kind of fan that opens up all the more the narrower
the slit is.

In Section \ref{interferenz_quanten}, we analyzed how the diffraction pattern of light behind a single slit
(Figure \ref{einzelspalt}) comes about. The result was simple: the width of the diffraction pattern
is inversely proportional to the width of the slit.

If we think of Huygens' principle (Section \ref{interferenz}) that spherical (circular) elementary
waves emanate from each point of a wave front and build up a new wave front, then the slit width indicates the uncertainty in the position (in the \zitateng{transverse direction}) from which the elementary waves emanate. The
width of the \zitateng{wave fan} behind the slit in the wave trough or of the diffraction pattern
on the observation
screen in an experiment with light is a measure of how inaccurate the direction of light is
in which the waves propagate behind the slit. For light with a specific wavelength, the
\textit{magnitude} of the momentum is precisely defined, but the direction is not.
The momentum component in the \zitateng{transverse direction}\footnote{The component in the \zitateng{longitudinal direction} is of course also blurred, but the longitudinal uncertainty is
much smaller for the usual situation in diffraction experiments, with very small diffraction angles, than the
transverse uncertainty.} is therefore fuzzy.
Once again, we see that there is a connection between the wave nature of a process
and the uncertainty relation.
This makes it clear that the diffraction effects observed with electrons
or other particles are also closely logically related to the uncertainty relation.

There is not only an uncertainty relationship between position and momentum, but in general
between each pair of physical quantities that \zitateng{disturb each other} during
measurements. We will come back to this in Section \ref{was_bei_messung}. The position-momentum 
uncertainty relation is particularly easy to discuss graphically because of the
relationship between momentum and wavelength and because of the classical
references to superpositions of waves. For the more general version of the uncertainty
principle we have, as is often the case: the more general, the more abstract.

\subsubsection*{Meaning of the uncertainty principle}

If we think about how tiny the constant $\hbar$ relevant for the uncertainty relation is, we
ask ourselves why this relation should have any practical meaning at all. In fact, the
goalkeeper does not have to worry about the uncertainty relation when he wonders
where the ball will fly at a penalty kick. In that situation, there are completely different important factors that have nothing to do
with quantum mechanics. For (much) smaller objects than footballs, however, the
uncertainty principle is \zitateng{vital}.

Around the year 1900, it gradually became clear that atoms consist of a very small 
positively charged nucleus, which contains almost the entire
mass of the atom, and some very light negatively charged electrons. The force of
attraction between opposing charges holds the whole structure together. Since this
attractive force obeys the same law of distance (proportional to 1/square of the
distance) as the gravitational force between celestial bodies, one could imagine that the
electrons orbit the nucleus like the planets orbit the sun.

However, this idea ignores the crucial difference that the electrons, unlike the planets,
are electrically charged, and electrically charged particles in such an orbit
permanently radiate electromagnetic waves and thus energy, like small antennas. This
would cause the electrons to lose their kinetic energy within fractions of a
microsecond and to plunge into the nucleus on a spiraling path. Atoms as we know them, and with
them our entire world, would collapse. 

Of course, physicists around 1900 were also aware of this and wondered why this
form of the end of the world had not already taken place. Niels Bohr then made the
assumption that the electrons could - for whatever reason - move stably on very
specific orbits and thus developed Bohr's atomic model, which could indeed correctly
explain many observations.

\subsubsection*{Uncertainty prevents the end of the world}

This idea persisted until the development of quantum mechanics in the 1920s, when it
became clear that it is the uncertainty relation that saves us from the end of the world.
For the sake of simplicity, let's look at how this works using a hydrogen atom
consisting of a proton as a nucleus and an electron. If the electron were pushed onto
the nucleus, then we would know exactly where it is (namely at the nucleus, whose
diameter is 100,000 times smaller than that of the atom) and how fast it is moving
(namely not at all, if we assume that the nucleus is not moving). However, this would
be a contradiction to the uncertainty principle, because we would then have determined
both the position and the momentum of the electron.

If we now assume that the position of the electron is only known with an uncertainty $\Delta x$, then $\Delta x$  is also the order of magnitude of the distance between the nucleus and
the electron. The potential energy of the electron in the attractive field of the nucleus
is then proportional $-\frac 1{\Delta x} <0$. The negative value means that we would have to spend
energy to draw the electron away from the nucleus.

Because of the uncertainty relation $\Delta x \Delta p \ge \frac{\hbar}2$ we then have $\Delta p \ge \frac{\hbar}{2 \Delta x}$ and for the kinetic
energy we expect something in the range of$\frac{\Delta p^2}{2m} {\ge} \frac{\hbar^2}{8m (\Delta x)^2} > 0$ (As has often been the case, our \zitateng{calculation} here is only very crude.)

The total energy, i.e. the sum of kinetic and potential energy, should be negative, as
we expect a bound state of the atom in which the electron
does not leave its nucleus \zitateng{voluntarily} (i.e. without additional energy input). The
\zitateng{most tightly bound} state is achieved by varying $\Delta x$ in such a way that the total
energy is minimized. If you do this, you get $\Delta x$ in the order of magnitude
of 10$^{-10}$m and a total energy in the order of magnitude of -10eV\footnote{One electron Volt (eV) is the energy that an electron absorbs when it is accelerated with a voltage of
one Volt. 1eV=1.6 $\cdot 10^{-19}$J is the typical energy that is converted in atomic or molecular processes.}. As a matter of principle one should not try to get more precise numbers from such a crude estimate; however, these
figures correspond well with reality.

We keep for the record that without the uncertainty principle there is no stability of matter. The Nobel
laureate Richard Feynman writes about this in his \zitateng{Lectures on Physics}: \textit{So we now understand why we do not fall through the floor. As we walk, our shoes with their masses of atoms push against the floor with its mass of atoms. In order to squash the atoms closer together, the electrons would be confined to a smaller space and, by the uncertainty principle, their momenta would have to be higher on the average, and that means high energy; the resistance to atomic compression is a quantum-mechanical effect and not a classical effect.}

Things become even more interesting if we consider not just a single electron, as we have
just done, but several or many. Quantum mechanical uncertainty then explains the
structure of atoms, the periodic table of chemical elements and the fact that there are
insulators, metals and semiconductors among the many existing substances. But one
thing at a time!

\subsubsection{Two electrons cannot be distinguished because of the uncertainty}

Because of the uncertainty, the concept of the \textit{trajectory} of a particle is meaningless in
quantum mechanics, whereas in classical mechanics the determination of the trajectory
using Newton's laws of motion is precisely the goal. Let us now compare an
interaction process between two classical particles with that between two quantum
mechanical particles (Figure \ref{keinebahn}). In both cases, the particles move from left to
right. At the beginning they are far apart, then they come so close that they exert
forces on each other and in the end they are far apart again.

\begin{figure}[h] 
\includegraphics[width=\textwidth,keepaspectratio=true]{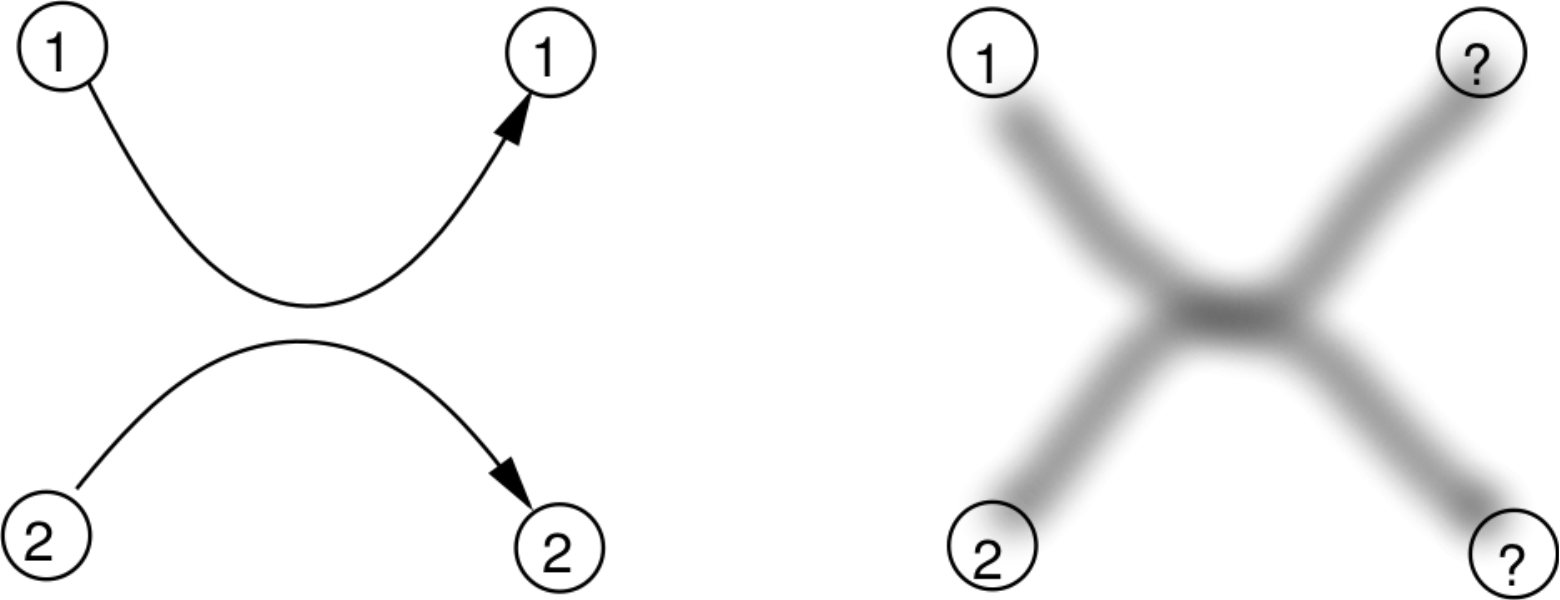}
\caption{An interaction process between two particles. At the beginning and end, the
particles are far apart, in between they come close to each other. Left: classical particles, right:
quantum mechanical particles.
\tiny{(Graphic: Stolze)} 
}
\label{keinebahn}
\end{figure}

In the classical case, it is always clear which particle is which, as the trajectories can be
distinguished from one another: The particle that was above the other at the
beginning is also the upper one at the end. In the quantum mechanical case, on the
other hand
the \zitateng{probability clouds} of the two particles get mixed up, and if they are similar
particles (e.g. two electrons), then it is completely unclear at the end,
whether the particles have swapped positions or not. Quantum mechanical particles
of the same type are \textbf{indistinguishable} in principle; they are also referred to as
\textbf{identical} particles. The question of whether two identical particles have swapped
positions is meaningless, and the quantum mechanical description of systems
consisting of two particles must therefore not offer any possibility of distinguishing
between two particles of the same kind.

In Section \ref{wellenfunktion} we saw already  that the state of a particle at time $t$ is described by
the wave function $\Psi(x,t)$. (As always, we assume for the sake of simplicity that the
particle can only move in one dimension, i.e. along the $x$-axis). To describe two
particles, we need two coordinates, which we will call $x_1$ and $x_2$. If one particle is
now in a state $\Psi_a$ and the other is in a state $\Psi_b$, then the overall state for both particles can be written as a simple product\footnote{The mathematical structure of the theory shows that the product form makes sense.} of the two wave
functions:
$$
\Psi(x_1,x_2,t)=\Psi_a(x_1,t) \Psi_b(x_2,t).
$$
However, if the two states $\Psi_a$ and $\Psi_b$ show physically measurable differences, the two
particles can be distinguished from each other in such a state,
but we wanted to rule this out. So the combination \zitateng{with reversed roles} 
$$
\Psi_a(x_2,t) \Psi_b(x_1,t)
$$
must also be admissible. The superposition principle discussed in Section \ref{wellenfunktion} states
that any superposition of these two combinations, i.e.
$$
\alpha \Psi_a(x_1,t)\Psi_b(x_2,t)  + \beta  \Psi_a(x_2,t) \Psi_b(x_1,t)
$$
is also an admissible state for the two particles. Here, $\alpha$ and $\beta$ are (complex) numbers
whose absolute values squared $|\alpha|^2$ and $|\beta|^2$ indicate the probabilities of finding the first or second combination of $\Psi_a$ and $\Psi_b$ in a measurement.

However, these two probabilities must be the same, otherwise it would still be possible to
distinguish between the two particles, for example by repeatedly measuring which
particle is more likely to be in the state $\Psi_a$ . It turns out that you only get no
mathematical contradictions if either $\alpha=\beta$ or  $\alpha=-\beta$. This results in two admissible
forms of the wave function for two identical particles, which we will call $\Psi_+$ and $\Psi_-$:
$$
\Psi_{\pm}(x_1,x_2,t)=\alpha \Psi_a(x_1,t) \Psi_b(x_2,t) \pm \alpha \Psi_a(x_2,t) \Psi_b(x_1,t).
$$
The number $\alpha$ is determined from the normalization of the wave function.

\begin{mdframed}[style=tpq]

\begin{center} 
\textsf{Complex number}
\end{center}

A \textbf{complex number} $z$ is made up from two real numbers $x$ and $y$ according to the prescription $$ z= x+iy.$$ Here $i$  is the \textbf{imaginary unit} with the property $$i^2=-1.$$
$x$ is called the \textbf{real part} and $y$ the \textbf{imaginary part} of $z$. Complex numbers
can be represented in a plane with $x$ and $y$ as coordinates in the usual way (Gaussian number plane). The quantity $$|z|^2=x^2+y^2$$ is called \textbf{absolute value squared} of $z$. The absolute value $|z|$ of the number $z$ then indicates the distance between $z$ and the zero point of the Gaussian number plane.

To warm up, simply calculate the product of $3 + 4i$ and
$3 - 4i$. Hint: The result is real and a square.

\end{mdframed}

\subsubsection*{Bosons, Fermions and the Pauli principle}

There are therefore two types of identical particles: Those with the plus sign are
called Bose particles or \textbf{bosons} after their discoverer and those with the minus sign
are called Fermi particles or \textbf{fermions}\footnote{Historically particularly correct people speak of Bose-Einstein particles and Fermi-Dirac particles.
However, most physicists are not particularly historically correct, but are lazy when it comes to speaking
and writing.}. Both types occur in nature and differ in another
physical property, the \textit{spin}, which we will discuss in Chapter \ref{VI}. Most of the particles
that make up matter (electrons, protons, neutrons...) are fermions; photons and some
other types of particles are bosons.

An interesting situation arises for fermions if the two \zitateng{one-particle states} are equal,
i.e. $\Psi_a=\Psi_b$. This then results in
$$
\Psi_-(x_1,x_2,t)=\alpha \Psi_a(x_1,t) \Psi_a(x_2,t) - \alpha \Psi_a(x_2,t) \Psi_a(x_1,t) = 0,
$$
This wave function is therefore \textbf{impossible}; two fermions cannot be in the same state.
This is the famous \textbf{Pauli principle}, also known as the exclusion principle.
Two bosons, on the other hand, have no problem finding each other in the same state;
this also applies to three or more bosons and is related to some interesting physical
phenomena (superconductivity, superfluidity, Bose condensation), which unfortunately
we cannot discuss here.

\subsubsection*{The Pauli principle and the periodic table}

The Pauli principle is fundamental to the understanding of atoms, molecules and
solids. In order to understand an atom, the Schrödinger equation must first be
solved. The eigenvalues $E_i$ determined in this process are the possible values of the
energy that an electron can have when it is in the state $\phi_i$ connected to that
energy. The electrons belonging to an atom then occupy these states by filling them up
from bottom to top (in energy); there can only be one electron\footnote{In truth, there can be not just one but two electrons in each state, which differ in the two possible values of the spin, which we will ignore here for the time being.} in each state since the electrons are Fermions.  Depending on how the states and their energies are organized, the atoms
differ in their chemical properties, e.g. react easily or less easily; the details are
researched in chemistry.

\begin{mdframed}[style=tpq]

\begin{center} 
\textsf{Schrödinger equation}
\small
Caution: Contains higher mathematics.
\normalsize
\end{center}

The \textbf{time-dependent Schrödinger equation} for the wave function $\Psi(x,t)$ of a quantum mechanical particle moving along the $x$ axis is a differential equation in the variables $x$ (position) and $t$ (time), by which the time evolution
 of the wave function is determined:
$$
i \hbar \frac{\partial}{\partial t} \Psi(x,t) = - \frac{\hbar^2}{2m}
\frac{\partial^2}{\partial x^2}  \Psi(x,t) + V(x) \Psi(x,t).
$$
Most important here are the derivatives $\frac{\partial}{\partial t} \Psi(x,t)$ and $\frac{\partial^2}{\partial x^2}  \Psi(x,t)$,
 i.e.
the first derivative with respect to the time $t$ and the second derivative with respect  to the position $x$. Whenever several variables are involved, the derivatives are not denoted
by the usual $d$, but by $\partial$, in order to keep in mind that there are several variables involved, but only one of them is considered, while the others are regarded as constant; this is referred to as \textbf{partial} derivatives and the Schrödinger equation is a \textbf{partial differential equation}.

The \textbf{imaginary unit} $i$ has the property $i^2=-1$. It is a \textbf{complex number} with real part zero and imaginary part one.

$\hbar$ (read: \zitateng{h-bar}) is the Planck constant $h$ divided by $2\pi$, $m$ is the mass of the particle.

$V (x)$ is the potential energy of a \textit{classical} particle at location $x$, which is related to the forces acting on the particle; the term with the second derivative with respect to the position $x$ has to do with the kinetic energy of the particle.

Usually, a shorthand notation is introduced for all operations that appear on the right-hand side of the Schrödinger equation, i.e. that are performed on the wave function. One writes
$$
\hat H =- \frac{\hbar^2}{2m}
\frac{\partial^2}{\partial x^2}   + V(x)
$$
and calls $\hat H$ the \textbf{Hamiltonian operator} or Hamilton operator. 

The Hamiltonian operator contains the forces acting on the particle via the potential energy and thus defines the physical situation. It determines the time-dependent Schrödinger equation as the \textbf{equation of motion} for the wave function. In addition, it also determines the possible energy values
of the particle via the \textbf{time-INdependent Schrödinger equation}
$$
\hat H \phi_i(x) = E_i \phi_i(x).
$$
The energy $E_i$ can only assume certain values $E_0 , E_1 , E_2 , ...$ with  different corresponding wave functions $\phi_0(x), \phi_1(x),\phi_2(x),...$ (read: \zitateng{Phi-zero of x, ...}). Such an equation is called an \textbf{eigenvalue equation}, the numbers
$E_i$ are called \textbf{eigenvalues} and the functions $\phi_i(x)$ are called \textbf{eigenfunctions}.
The eigenfunctions $\phi_i(x)$ of the Hamilton operator describe the \textbf{stationary states} of the particle in which the probability of finding the particle at a point $x$ does not change
over time.

Incidentally, the mathematical situation occurring here is not at all
extraordinary. The vibrations of a guitar string are also described by a partial differential equation, and the guitar string can also only vibrate in very specific forms that correspond to very specific tones (frequencies), the fundamental and the higher harmonics. If, instead, any frequencies were possible,
the guitar would no longer sound like a guitar, but very weird.

\end{mdframed}

For a  \textbf{solid} consisting of very many atoms, this results in a huge number of
possible states for the electrons, with a correspondingly large number of possible energies.
Again, the different states are filled up by the available electrons  from the lowest energy
upwards, and the properties of the solid depend on the possibilities open to the electrons
with the highest energy (the \textbf{Fermi energy}). If there is a gap in the energy spectrum
above this energy, an electron has to absorb a lot of energy to reach an \textbf{excited state};
we are then dealing with a non-conductor (\textbf{insulator}). If, for example, the energy gap is
larger than the energy of photons of visible light, then these photons cannot transfer
their energy to the electrons and the solid is therefore transparent; think of a salt
crystal or a diamond. If the energy gap is small, the electrons can be excited with some
energy from the thermal motion, or additional electrons can be introduced through
deliberately introduced \zitateng{impurities}, thus facilitating excitation; this opens up 
the broad field of \textbf{semiconductor} technology. If there is no energy gap at all, the
electrons can react very easily to external forces and flow. This is the reason why
\textbf{metals} conduct electricity well and are not transparent, but reflect light and shine
more or less beautifully.

\subsubsection*{The Pauli principle and the uncertainty in space}

The Pauli principle also applies in the \zitateng{infinite expanses} of space. When a star the
size of our sun, for example, has spent its fuel (hydrogen, later helium)
in nuclear fusion reactions, it essentially consists of oxygen, nitrogen and carbon.
However, these atoms are completely ionized by the high temperature, i.e. the
electrons have become detached from the atomic nuclei,
move freely and form a \zitateng{Fermi gas} in which the atomic nuclei also float. The force of
gravity holds the spherical object together and compresses its
diameter from about $10^6$km to about $10^4$km, i.e. about as small as the Earth; this is
referred to as a \textit{white dwarf}. Because of the Pauli principle, the many electrons flying
around more or less freely all have to assume different states, they have great energy
and exert great pressure, which counteracts gravity. If the mass of the burning-out
star is even greater (about one and a half solar masses), the pressure of the electrons
can no longer withstand gravity. The flying electrons and protons from the atomic
nuclei are then squeezed together to form neutrons, leaving only neutrons. These then
form a $\sim$10km sphere, a \textit{neutron star}. The neutrons are again fermions and, due to the
Pauli principle, exert a counter-pressure to gravity. If the mass becomes too large and
thus the gravity too strong, this pressure is no longer sufficient and the enormous gravity creates
a \textit{black hole}.

Quantum mechanical uncertainty therefore has spectacular consequences for the
structure of matter, from the atom to the black hole shortly before collapse. We will
shortly see what role uncertainty plays in the \textit{observation} of black holes.

The uncertainty relation is an inequality,
$$
\Delta x \Delta p \ge \frac{\hbar}2,
$$
it therefore indicates the \textit{minimum} size of the product of the two uncertainties.
How large it really is depends on the state the quantum mechanical system is in; $\Delta x \Delta p$
 can also be much larger than $\hbar$. For high-precision measurements, in which
quantum mechanical methods are often used, the uncertainty must be taken into
account and made as small as possible. Such states of minimal uncertainty\footnote{The uncertain quantities are no longer the position and momentum of a particle, but more complicated
parameters that describe the wave field of the laser.} can be generated
in modern laser systems, with which very small changes in optical path lengths or
path differences can be measured by measuring interferences (Section \ref{interferenz}).

Einstein's theory of general relativity predicted that there would have to be
\textit{gravitational waves}, i.e. changes in the force of gravity that propagate periodically in
space and time. Such gravitational waves should cause tiny changes in distances, which
were long sought in vain. In 2015, a signal less than a second long was then
observed using several kilometer-long laser arrays,
which can be found on the Internet under \zitateng{GW150914}. By comparison with
theoretical calculations, it was possible to prove that the cause was the fusion
of two black holes with 29 and 36 solar masses, respectively. Fortunately, this cosmic catastrophe
took place at a great distance from us. The discovery was rewarded with the 2017
Nobel Prize in Physics.

After it became clear that gravitational waves are observable in principle, further
improvements were made to the observation instruments. One of these is based
is based on the \zitateng{actually simple} idea that the uncertainty relation only concerns the
\textit{product} of two uncertainties, but not the two uncertainties individually. One should
consequently be able to reduce one uncertainty at the expense of increasing the
other uncertainty. Thus, one could make the measurement of an interesting quantity
more accurate at the expense of the accuracy of a less interesting quantity. To do this,
the quantum mechanical states involved must be suitably deformed; one speaks of \textit{squeezed}
states. The use of such states has been investigated for some time in smaller
experimental setups (e.g. in Hanover), and since 2019 the
largest gravitational wave interferometers have also been squeezing the accuracy to be
able to measure even better what is happening in space.

\chapter{Is everything just random?}
\label{III}
\section{How does a particle move?}
\label{wie_bewegt}

In Chapter \ref{II}, we familiarized ourselves with the fact that the terms \zitateng{particle} and \zitateng{wave} in quantum physics are no longer as unambiguous as we are used to in our everyday world. We have discussed in detail how the wave phenomenon of \textit{interference} affects quantum mechanics. We have also seen that \textit{uncertainty relations} already occur with classical waves, i.e. that there are pairs of quantities that \zitateng{disturb} each other during measurements: the more accurately you can measure one quantity, the less accurate the other becomes.

So far, however, our description is completely static and we do not yet know how
waves/particles \textit{move} in quantum physics and how this can be reconciled with the
laws known from classical physics. Our introduction to quantum \textit{mechanics} proper begins
with these questions.

We will first get to know \textbf{wave packets}; these are objects that offer an acceptable
compromise between the descriptions as waves and as particles and with the help of
which it becomes clear that the laws of motion of classical physics also apply in
quantum mechanics if only the right quantities are considered. That is somewhat
reassuring!

Less reassuring is the fact that particles can \zitateng{melt} without any external influence. We
will see that very different time scales can be involved here.
It also takes some getting used to the fact that quantum-mechanical 
particles can escape from \zitateng{classical prisons} by \textit{tunneling} under walls.

However, if the prison walls are made thick enough, this no longer happens; instead,
another phenomenon occurs that has no classical explanation, namely that a particle
trapped between such walls can only \zitateng{live peacefully} in very specific states.
 These \textbf{stationary states} have very specific energies that provide the explanation for the optically observed line spectra of the chemical elements. Particles with very high energy can, metaphorically speaking, fly past the prison. Physics has learned a lot about how particles interact with \zitateng{walls} or other
particles from the way they do this.

\begin{mdframed}[style=tpq]

\begin{center} 
\textsf{Spectrum (plural: spectra), spectroscopy
}
\end{center}
A \textbf{spectrum} is the set of the wavelengths (or frequencies) of
electromagnetic radiation that can be observed in a range, together with
their respective intensities. 
If all wavelengths can occur within some range, the spectrum is called
\textbf{continuous}, e.g. in the case of thermal radiation from a hot object. If only
very specific wavelengths can occur, the spectrum is called \textbf{discrete}. This is
the case for pure chemical elements, for example. Discrete spectra are also
called \textbf{line spectra}, named after the image of colored lines on a dark
background that appears in an optical spectrometer. By examining such line
spectra in \textbf{spectral analysis}, it is possible to determine which chemical
elements the substance under investigation contains. The study of
spectra is called \textbf{spectroscopy}.

In more general terms, the possible values of the energy of a quantum system
are also referred to as the \textbf{energy spectrum}.

\end{mdframed}

\subsection*{Wave packet: wave and particle reconciled}

In Chapter \ref{II} we have already seen that there are waves that are restricted to a small
region of space (Figures \ref{wellenzug_kurz}, \ref{fourierreihe}). If we think of the unavoidable uncertainty in the measurement of position, we can accept these objects as
somewhat \zitateng{fuzzy} particles; such an entity is usually referred to as a \textbf{wave packet}.

In Section \ref{unschaerfe} we have argued that wave packets can be represented as  superpositions of a certain number of simple sine waves like the one on the
right in Figure \ref{brandung}. These simple waves differ in their respective wavelengths and
amplitudes, i.e. the proportions with which they contribute to the superposition. In
Figure \ref{fourierreihe} we have seen examples for the construction of  wave packets.

\begin{mdframed}[style=tpq]

\begin{center} 
\textsf{Amplitude}
\end{center}

The \textbf{amplitude} is the maximum displacement of a wave, e.g. in the case of a
water wave, the height of the wave crest above the calm water surface. In
addition to the amplitude, the \textbf{phase} of a wave is also important, as it
determines, for example, exactly where the wave crests are located.
When waves are superimposed, both the amplitudes and the phases of the
partial waves are important, keyword interference.

In the case of a superposition of several quantum mechanical wave
functions, one also speaks of the amplitudes of the different wave functions
and thus means the respective components (expressed as complex
numbers).

\end{mdframed}

Instead of the wavelength $\lambda$, we have used the wavenumber $k= \frac{2\pi}{\lambda}$ because $k$ is
proportional to the momentum, $p=\hbar k$, with $\hbar$, the Planck constant  divided by $2 \pi$.
Which wavenumbers with which amplitude must be packed into the packet in order to
achieve a certain shape is the subject of Fourier analysis. That is a field of higher
mathematics that is not only important for physics, but also for all engineering sciences
and computer science (e.g. image processing). However, we do not have to deal with
the technical details here.

How does such a wave packet move? This is governed by the appropriate equation of motion. For quantum mechanics, this is the Schrödinger equation, and for light 
or other electromagnetic waves,  the wave equation. Both equations are \textit{linear}, i.e. if
two solutions are known, a combination (sum) of both solutions with arbitrary
proportions is also a solution of the equation. This property makes the construction of
wave packets meaningful in the first place and immediately provides the answer to
the question from above: You can simply follow how the individual waves move and
how the moving waves combine again in the end.

\begin{mdframed}[style=tpq]

\begin{center} 
\textsf{Equations of motion}
\end{center}

In physics, equations of motion are always differential equations,
i.e. they indicate at a certain point in time $t$ how the state changes over
time (mathematically: derivative with respect to time). From this, the state
at a later point in time can be calculated.

For a \textit{classical} particle, the equation of motion is given by Newton's laws, e.g. in the form \zitateng{force = mass times acceleration} If the force is known, the acceleration can be used to
calculate how the velocity changes, and from this, how the position
changes. Then you can follow the path of the particle in space and time, e.g. for a
satellite orbiting the earth.

The equation of motion for a \textit{quantum mechanical} particle is the Schrödinger
equation, a differential equation for the wave function. Due to Heisenberg's
uncertainty principle, the concept of orbit does not exist for a quantum
mechanical particle (e.g. an electron in an atom).

\end{mdframed}

Each individual wave moves at a very specific speed, which depends on the wave number. This speed is measured, for example, by following a specific wave crest. The situation is particularly simple for light or other electromagnetic waves: All
these waves move in a vacuum at the speed of light $c$, and therefore the entire light
wave packet also moves at this speed. This is different for a quantum mechanical wave
packet. If no external forces are acting, the Schrödinger equation shows that the speed
of a wave increases proportionally to the wave number:
$$
v_{\mathsf{wave}}= \frac{\hbar}{2m} k,
$$
Shorter waves have a larger wavenumber $k$ and are therefore faster. For a wave
packet, this means that the parts with shorter wavelengths overtake the other waves
and accumulate at the front end of the wave packet. The wave packet
corresponding to a free particle will therefore change its shape. This will be
examined in more detail in the next section.

The fact that different wavelengths lead to different speeds is also known for light waves traveling in a material (glass, water, etc.). There, however, the longer waves usually travel faster and are therefore less strongly refracted when they enter the
material, which leads to the formation of rainbows, for example.

The construction of wave packets offers the possibility of describing objects within the framework of quantum mechanics that \zitateng{feel like particles}, i.e. are localized to some
extent in space. This is only possible because superpositions of possible wave functions are also possible wave functions, i.e. through the superposition principle. The waves involved in the wave packet interfere destructively everywhere except in a
small spatial region. In this way, a wave packet, i.e. an object that is \textit{localized} in space, can be created by superimposing many waves that \textit{extend infinitely far} in space.

The question is whether this object also moves in the way we are used to from classical physics. Let us recall Section \ref{wellenfunktion}, where we saw that the wave function $\Psi(x,t)$
 of the wave packet, or more precisely, its magnitude squared $|\Psi(x,t)|^2$, describes the probability  of finding
 the particle at position $x$ at time $t$; Figure \ref{aufenthalt}
shows an example. If this probability is known, the mean value $\langle x \rangle$ of the position for a wave packet can be calculated. Since a wave packet is composed of parts with different wavenumbers $k$, the mean
  value of $k$ can also be calculated, and since the
wavenumber is linked to the momentum $p$ by $p=\hbar k$, the mean value $\langle p \rangle$ of the
momentum can also be obtained in this way.

 If one solves the Schrödinger equation for the wave function $\Psi(x,t)$ of a particle, i.e. if one
follows the temporal development of $\Psi(x,t)$, then it becomes apparent that the \textit{mean values} of the fundamental physical quantities behave in the same way as described by
the classical equations of motion. The momentum $p$ is the product of mass $m$ and
velocity, where the velocity is the temporal change in (mean) position; the
following therefore also applies in quantum mechanics
$$
 m \cdot \mbox{  temporal change of  } \langle x \rangle = \langle p \rangle.
$$
Newton's time-honored law \zitateng{temporal change of momentum $=$ acting force} also
applies in quantum mechanics, in the form
$$
 \mbox{  temporal change of  } \langle p \rangle = \langle F(x) \rangle,
$$
where $F (x)$ is the force acting at location $x$\footnote{The force does not appear in the Schrödinger equation, but the potential energy does. The force is linked to it by minus the spatial change (derivative).} These two equations show that
quantum mechanics is not quite as crazy as is sometimes believed, but has reassuring
similarities with classical mechanics. These laws of motion for mean values apply to
\textit{arbitrary} wave functions, but the close connection to the usual behavior of particles
only applies if the occurring mean values are not subject to too large fluctuations.
Sufficiently narrow wave packets therefore behave like classical particles. If, on the
other hand, the uncertainties, i.e. the standard deviations of position or momentum,
become too large, it no longer makes much sense to talk about the mean values.

Particularly attentive readers may have noticed a problem with the concept of velocity,
which will be discussed here in a moment. A little further
above, we gave the velocity of a single wave from the wave packet; quite precisely, it was the velocity at which a wave maximum (or a zero crossing, in any case a point with a certain fixed phase of the wave) moves. If we equate the product $\hbar k$ with the momentum $p$ in the formula for this velocity $v_{\mathsf{wave}}$, the result is
$$
v_{\mathsf{wave}}= \frac p{2m}.
$$
If, on the other hand, we think of a whole wave packet and define its speed by the
temporal change of $\langle x \rangle$, then we have
$$
v_{\mathsf{packet}}= \frac {\langle p \rangle}{m},
$$
as we are used to classically. So there are obviously two different speeds here!
In addition, both speeds also depend on the wavenumber $k$: for $ v_{\mathsf{wave}}$  this is the wavenumber linked to
the sharply defined momentum $p$, for $v_{\mathsf{packet}} $ it is the mean wavenumber defined by the
\textit{mean} momentum $\langle p \rangle$. What we called $v_{\mathsf{wave}}$ is generally called \textbf{phase velocity} in
physics, and $v_{\mathsf{packet}}$ is called  \textbf{group velocity}. The fact that these two velocities are not
the same often occurs in physics, for example for water waves. Electromagnetic waves
in empty space are almost the only waves that propagate at a uniform speed for any
given wavelength, namely at the speed of light $c$.

The fact that for a quantum mechanical wave packet the phase velocity is drastically smaller than the group velocity can be observed by solving the Schrödinger equation for a force-free wave packet such as in Figure \ref{wellenzug_kurz} and making a movie out of it. While the whole packet is in motion, new waves are constantly created at the front end, which are then \zitateng{passed through to the rear} and disappear again at the rear end.

A well-designed simulation for a force-free wave packet can be found at
\url{https://demonstrations.wolfram.com/WavepacketForAFreeParticle/} \phantom{x} 
\begin{wrapfigure}{r}{3cm}
  \includegraphics[width=3cm]{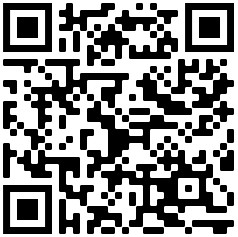}
\end{wrapfigure}
Either the real part or the imaginary part of $\Psi(x,t)$, or also
$|\Psi(x,t)|^2$ are shown there; the latter quantity (probability density) is called $P (t)$ in the simulation. By clicking on one of the gray fields with the
plus sign, you can change the time and other parameters. If
you choose \zitateng{momentum uncertainty, $\Delta p$} larger than the default setting, you can see very clearly that the short waves
are faster than the long ones, but that the overall packet moves
even faster. In the online version, you can only 
\zitateng{click on} the time in leaps; if 
you want a continuous simulation, you must use the
Download and install the demonstration and the associated free WolframPlayer display software.

\begin{mdframed}[style=tpq]

\begin{center} 
\textsf{Wave function}
\small
Caution: Contains higher mathematics.

\normalsize
\end{center}
The \textbf{wave function} of a quantum mechanical particle moving on the $x$ axis depends on the position $x$ and the time $t$; it is usually denoted by $\Psi(x,t)$
(read: \zitateng{Psi of $x$ and $t$}).

$\Psi(x,t)$ ) is a \textit{complex} function, i.e. it consists of two parts, $ \textsf{Re}\Psi(x,t)$ and $\textsf{Im}\Psi(x,t)$
(read: \zitateng{real part of Psi and imaginary part of Psi}). 
The parts play a similar role to the coordinates $x$ and $y$ of a point in a plane, where, for example, the distance $r$ of the point $(x, y)$ from the zero point can be calculated using Pythagoras' theorem:
$$
r^2=x^2+y^2 .
$$
In a similar way, the wave function yields the important quantity 
$$
| \Psi(x,t)|^2 = ( \textsf{Re}\Psi(x,t))^2 + (\textsf{Im}\Psi(x,t))^2
$$
(read: \zitateng{Psi absolute value squared}), which indicates how likely it is to find the particle at position $x$ at  time $t$. (More 
 precisely, $| \Psi(x,t)|^2$ is the
probability density of finding the particle).

\end{mdframed}

\subsection*{Free particles melt}

We know by now that wave packets describing a force-free particle are composed of
components with different wave numbers $k$ or momenta $p=\hbar k$, and that these
components move at different speeds. Therefore, the shape of such a wave packet
changes over time. Let us now see what this means in detail.

For this purpose, we consider a so-called Gaussian wave packet that is supposed to
describe a force-free particle with mass $m$. This is a wave packet for which
 $|\Psi(x,t)|^2$ has the shape of a Gaussian bell curve at the initial time $t = 0$, very similar to the
curves in Figure \ref{fourierreihe}. The wave packet should move with a certain average
momentum $\langle p \rangle$ and at time $t = 0$ it should start at  position $\langle x \rangle=0$. This is an
  example for which the temporal development of  the wave function $\Psi(x,t)$ can be completely calculated using the
Schrödinger equation\footnote{Many physics students have spent several instructive hours with this homework problem.} .

The wave packet can be set up in such a way that at time $t = 0$ the spatial uncertainty is
$\Delta x=\frac a2$ and the momentum uncertainty $\Delta p= \frac{\hbar}a$. The constant $a$ is a length which specifies the initial position uncertainty. It can be chosen arbitrarily; we
will see examples shortly. The uncertainty product is then
$$
\Delta x \Delta p = \frac a2  \frac{\hbar}a = \frac{\hbar}2,
$$
the smallest value permitted by the uncertainty relation. What
development do we now expect over time?

Since the wave packet is composed of waves with certain wave numbers (momenta), and since their distribution does not change as the waves travel on, we do \textit{not} expect
any change in the momentum fluctuations. However, because the waves with a short
wavelength travel faster than those with a long wavelength, we expect the wave
packet to become \textit{wider} over time. The uncertainty product $\Delta x \Delta p$ consequently becomes larger. This is exactly what the solution to the Schrödinger equation
provides: $\Delta p = \frac{\hbar}a$  remains
 constant and $\Delta x$ becomes time-dependent:
 $$
\Delta x = \frac a2 \sqrt{1+\left(\frac{2 \hbar t}{m a^2} \right)^2 }
= \frac a2 \sqrt{1+\left(\frac{ t}{T} \right)^2 }.
$$
Here 
$$
T= \frac{ma^2}{2\hbar} 
$$
 is an abbreviation for the time scale on which $\Delta x$ changes. A small table shows what
this means in concrete terms:

\begin{tabular}{c|c}
$t$ & $\Delta x$ \\
\hline
0 & $1,00 \; a/2$ \\
$T$ & $1,41 \; a/2$ \\
$2T$ & $2,24 \; a/2$ \\
$3T$ & $3,16 \; a/2$ \\
$4T$ & $4,12 \; a/2$ \\
...& ...\\
$9T$ & $9,06 \; a/2$ \\
$10T$ & $10,05 \; a/2$ \\
\end{tabular}

After the time $10T$, $\Delta x$ has therefore increased approximately tenfold. The time $T$ ,
which indicates how quickly $\Delta x$ increases, is the longer the greater the mass $m$ of the
particle. Classical
(large and heavy) objects will therefore \zitateng{stay in focus} longer than, for example,
electrons. Since the denominator of the formula for $T$ is the tiny quantity $\hbar$, $T$ could well be large. We now calculate the time $T$ for three objects from different size and
weight classes.

We start small, with an electron ($m=9,1 \cdot 10^{-31}$kg),  with an initial uncertainty that
corresponds approximately to the size of an atom ($a=10^{-10}$m). This then yields\footnote{Note: $1\mathrm{J}=1\mathrm{Nm}=1\frac{\mathrm{kg \; m}^2}{\mathrm{s}^2}$.}
for the \zitateng{meltdown time}
$$
T= \frac{9,1 \cdot 10^{-31}\mathrm{kg} \cdot (10^{-10}\mathrm{m})^2}{2 \cdot 1,055 \cdot 10^{-34}\mathrm{Js}}
= 4,3 \cdot 10^{-17}\mathrm{s},
$$
The electron thus delocalizes extremely quickly. The oscillation period of visible
light, a very short time by all human standards, around 10$^{-15}$s, is still much longer in comparison.

The next object is still very small, but undoubtedly already classical: a dust particle
with a diameter of 10$^{-6}$m, i.e. one thousandth of a millimeter or one micrometer (this 
is the very finest class of fine dust, as classified in environmental legislation). The volume of such a particle is of the order of
$10^{-18}\mathrm{m}^3$ and the mass thus m = 10$^{-15}$kg, if we assume the density of the dust
material to be approximately the same as that of water. We then assume that we
could localize the dust particle on a scale that corresponds to the resolution of an
optical microscope, i.e. approximately the wavelength of light, say $500 \cdot 10^{-9}$m.
We then obtain
$$
T= \frac{10^{-15}\mathrm{kg} \cdot (500 \cdot 10^{-9}\mathrm{m})^2}{2 \cdot 1,055 \cdot 10^{-34}\mathrm{Js}}
= 1,18 \cdot 10^{6}\mathrm{s},
$$
that is 13.7 days. So once we have seen the dust particle under the microscope, we
can take a few days' vacation and the dust particle will still be there.

An object that is still not particularly large, but which you can already touch with
your hand, is a metal ball with a diameter of about 2 mm and a mass of 50 mg. We
localize this to the measuring accuracy of good mechanical measuring devices
(micrometer screw), i.e. about one thousandth of a millimetre, and calculate
$$
T=\frac{50 \cdot 10^{-6}\mathrm{kg} \cdot (10^{-6}\mathrm{m})^2}{2 \cdot 1,055 \cdot 10^{-34}\mathrm{Js}}
= 2,37 \cdot 10^{17}\mathrm{s},
$$
that is $7,5 \cdot 10^9$ years. The age of the universe is currently estimated to be around $13,8 \cdot 10^9$ years.

Free particles of reasonably macroscopic size will therefore behave as usual and be at
a certain location at all times without melting into a kind of cloud. Then it is clear that
we can use the usual laws of classical mechanics to describe the motions of such
particles; we have already seen above that the mean values of position and
momentum satisfy the classical
equations of motion. However, we will investigate how \zitateng{genuine quantum mechanical particles} behave  under the influence of external forces  in the
following sections, with some surprising results.

\subsection*{Obstacles and tunnels}

As usual, we will make life easy for ourselves and only examine the motion of a
particle in one dimension $x$. In Section \ref{teilchen}, we saw that for a classical particle, we
only need to consider the kinetic and potential energies. If a force acts at position $x$, the
potential energy depends on $x$; to characterize this, we write it as $V (x)$. The force
always acts in the direction in which $V (x)$ decreases and is the stronger the faster this
decrease is. The sum of kinetic and potential energies remains constant during the entire motion.

\begin{mdframed}[style=tpq]

\begin{center} 
\textsf{Kinetic and potential energy, quantum mechanical}

\end{center}

In quantum physics, too, kinetic and potential energy can be defined, and just as in classical physics, the total energy as the sum of kinetic and potential energies is constant; the law of conservation of energy also applies in quantum physics.

As in the classical case, the potential energy is defined by a function $V (x)$ that depends on a spatial coordinate. (For the sake of simplicity, we only consider motion in one dimension.) However, the value of the potential energy of a quantum mechanical particle in a certain state also depends on the wave function $\Psi(x,t)$, which describes this state, as the classical potential energy function $V (x)$ must be  averaged (over the coordinate $x$) using the probability density $|\Psi(x,t)|^2$ of finding the particle as a weighting function.

The kinetic energy of a particle in the state $\Psi(x,t)$ depends on how much $\Psi(x,t)$ changes spatially (i.e. on the derivative of $\Psi(x,t)$ with respect to $x$).
This quantity naturally depends on $x$ and must also be averaged over the entire spatial region in which the particle is located.

Specifically for a \textit{free} particle with momentum $p$ and mass $m$, the wave function is a wave with a certain (De Broglie) wavelength $\lambda=\frac hp$. ($h$ is the Planck constant.) The shorter the wavelength, the greater the momentum $p$ and thus the kinetic energy $E_{\textrm{kin}} = \frac{p^2}{2m}$.
The shorter the wave length, the greater the spatial change in the wave function and the connection to the more general definition of kinetic energy can be seen.

To determine the wave function $\Psi(x,t)$, you have to solve the Schrödinger equation. However, in order to get a rough picture without a lot of arithmetic, you can also consider what the wave function must look like so that the kinetic and potential energies add up to the correct value of the total energy everywhere in space.

\end{mdframed}

\begin{figure}[h] 
\includegraphics[width=\textwidth,keepaspectratio=true]{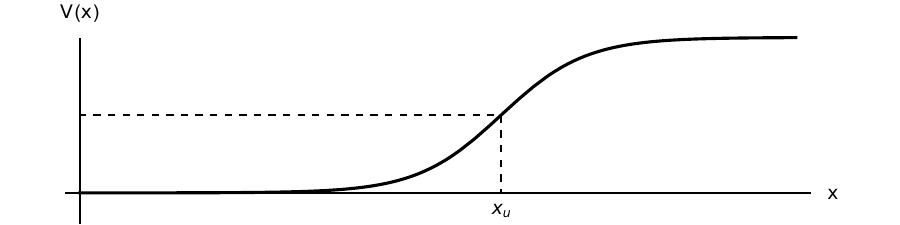}
\caption{ A classical particle with a given energy (horizontal dashed line) reaches the 
classical turning point $x_u$. The range $x > x_u$ is classically forbidden.  
\tiny{(Graphic: Stolze}) 
}
\label{umkehrpunkt}
\end{figure}

Figure \ref{umkehrpunkt} shows a simple situation for a one-dimensional motion. The motion is
determined by the shape of the potential energy, which has a constant value of zero
on the left-hand side of the image and a certain larger value on the right-hand side. In
the transition region, it increases from zero to this larger value. A classical particle
approaching from the left with a certain (total) energy (horizontal dashed line) is
decelerated in the range in which the potential energy $V (x)$ increases until its kinetic
energy and thus its velocity is zero. This happens at the point $x_u$ and there the particle
reverses its motion and flies back; this is why $x_u$ is called the classical turning point.
The particle cannot reach the region to the right of $x_u$: This is the classically forbidden
area. How large this range is  depends of course on how much energy the particle has at the beginning.

\begin{figure}[h] 
\includegraphics[width=\textwidth,keepaspectratio=true]{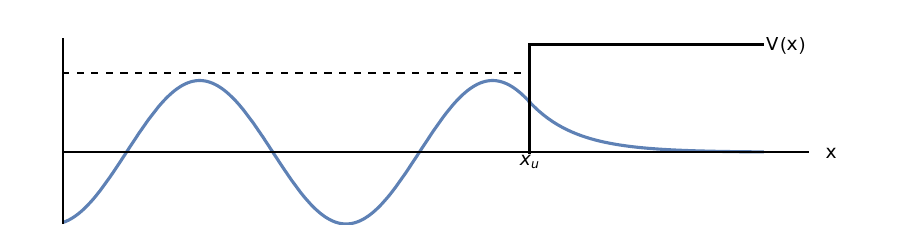}
\caption{A quantum mechanical particle with a given total energy (horizontal dashed line)
is represented by a wave to the left of the classical turning point $x_u$. In the range $x > x_u$, the
wave function decays exponentially.
\tiny{(Graphic: Stolze}) 
}
\label{stufe}
\end{figure}

To investigate the corresponding quantum mechanical situation, we simplify greatly
and make the shape of the potential energy much steeper, as in Figure \ref{stufe}. A
classical particle would again only get as far as $x = x_u$ . To find out what a quantum
mechanical particle does, we would have to solve the Schrödinger equation. We replace
this mathematical procedure with a simple energy argument. A particle that is
located in the area to the left of $x_u$ and has a certain kinetic energy (= total energy,
since the potential energy $V (x)$ is zero there) also has a certain momentum and thus also a certain (de Broglie) wavelength. However, if it has a precisely determined momentum, its location must be
completely indeterminate according to the uncertainty relation. We therefore expect a
wave with a certain wavelength that extends over the entire region to the left of $x_u$.
This is also what the Schrödinger equation delivers if you do the exact calculation.

\begin{mdframed}[style=tpq]

\begin{center} 
\textsf{Exponential function}
\end{center}

The exponential function is given by
$$f(x)=e^x,$$
where $e=2.71828...$ is Euler's number. You can see that this function 
increases very strongly when $x$ becomes larger: 
$e^0=1$, $e^1=e$, $e^2=e\cdot e=7.389$, ... $e^{10}=22026.5$, $e^{20}=485,165,195$.
 Each time $x$ grows by 1, $e^x$ is multiplied by $e$.
 This is called an exponential \textit{growth}.

The function $e^{-x}=\frac 1{e^x}$ shows an exponential \textit{decay}. 
Each time $x$ increases by 1, it decreases by a factor $\frac 1e=0.3679$, so that $e^{-10}=0.000045$.

Radioactive decay provides an example of exponential decay (in time $t$).
The number $N (t)$ of radioactive particles present at time $t$ is
$$
N(t)=N_0 e^{-t/\tau},
$$
where $N_0$ is the number at the initial time $t = 0$. $\tau$ (Greek letter tau) is the average \zitateng{lifetime} of a radioactive particle until its decay. $t_{1/2}=0.6931 \tau$ is the half-life after which half of the particles
present at the beginning are still there. After ten half-lives, only 1/1024 of
the radioactive particles remain.

\end{mdframed}

Since the Schrödinger equation is a differential equation for the wave function $\Psi(x,t)$,
the latter must fulfill certain mathematical conditions\footnote{ To be precise: The wave function and its first derivative with respect to $x$ must be continuous, otherwise
the second derivative with respect to $x$ occurring in the Schrödinger equation is not defined.
}. This is why $\Psi(x,t)$ cannot suddenly become zero at the classical turning
point $x_u$. In the region to the right of $x_u$, the total energy (dashed line) is smaller than
the potential energy $V (x)$. It turns out that the Schrödinger equation also provides solutions for the wave
function for such a {classically forbidden} situation. These solutions are 
increasing or decreasing exponential functions. An increasing exponential function
would grow beyond all limits to the right of $x_u$ and could not be meaningfully related to
the probability density, as explained in Section \ref{wellenfunktion}. This leaves only an
exponential function decreasing towards the right, which must be connected
\zitateng{smoothly} to the wave to the left of $x_u$ at the point $x_u$.

In contrast to a classical particle, a quantum mechanical particle can therefore penetrate a little way into the classically forbidden range
$x > x_u$. The penetration depth depends on how much energy the particle is \zitateng{missing}; the greater 
the deficit of
energy, the faster the wave function decays into the classically forbidden range.

However, if the energy of the particle is greater than the height of the \zitateng{threshold} in
the potential energy, then the range $x > x_u$ is no longer forbidden and the particle coming from the left
can continue its journey, but of course with a lower kinetic
energy and thus a lower momentum and longer wavelength.

At first glance, this looks like classical physics again: at $x_u$ , the kinetic energy of the
particle decreases, but it continues to fly in any case. In quantum mechanics, however,
the previously mentioned mathematical
\zitateng{connection conditions} for the Schrödinger equation at $x_u$ must be obeyed  and it turns
out that this is only possible if the particle also  with a certain
probability flies back to the left, i.e., as we say, is \textbf{reflected}.

A classical particle will therefore be reflected with certainty at a potential threshold
as in Figure \ref{stufe} if the energy is too small and \textbf{transmitted} with certainty if the
energy is sufficient. A quantum mechanical particle can penetrate a little way into the
classically forbidden region, even if the energy is too small; if it is large enough, the
particle can be transmitted, but with a certain (calculable) probability it can also be
reflected.

A particularly interesting situation arises if the step in the potential energy $V (x)$
in Figure \ref{stufe} is not infinitely wide, but jumps back to zero at $x_o$, a certain distance to the right
of $x_u$, as in Figure \ref{wall}. The shape of the potential energy is then reminiscent of a
wall. A classical particle that is located to the left of the wall (i.e. at $x < x_u$ ) and
whose total energy corresponds to the dashed line never reaches the right side ($x >x_o$) of the wall. For a quantum mechanical particle, the  wave function on the left is a simple  wave whose wavelength is related to the total energy, just as in the case of
Fig. \ref{stufe}.
Just as there, the wave function at $x_u$ cannot suddenly drop to zero, but the
exponential solution in the classically forbidden range \zitateng{within the wall} must be connected \zitateng{smoothly} to the wave solution of the Schrödinger equation. In the region to the right of $x_o$ , a quantum mechanical particle may be present
with the energy specified before (dashed line), i.e. the Schrödinger
equation has a solution there. This solution is again a wave, because the total energy is
equal to the kinetic energy (the potential energy is zero)
and the wavelengths to the right and left of the \zitateng{wall} are the same, as the kinetic
energy remains the same, and so does the momentum. (The \zitateng{sacred} 
conservation of energy!) However, the amplitude of the wave is significantly smaller
on the right than on the 
left, since the wave function has decreased exponentially \zitateng{under the wall} and since
the wave functions at $x_o$ must be smoothly connected to each other again. The smaller the amplitude of the wave function to the right of the wall, the greater the probability that the
particle will not penetrate the wall, but will bounce off and fly back to the left.

\begin{figure}[h] 
\includegraphics[width=\textwidth,keepaspectratio=true]{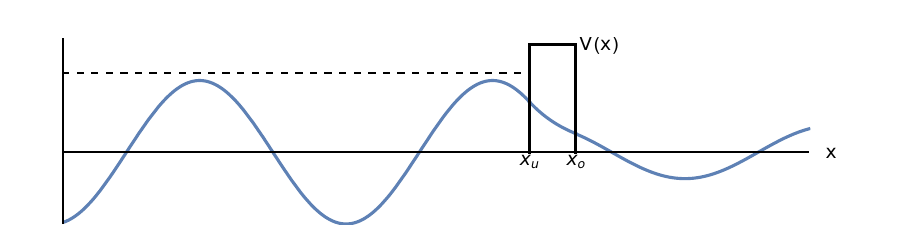}
\caption{
A quantum mechanical particle cannot jump over a \zitateng{wall} in potential energy if it does not have enough energy, but it can \zitateng{tunnel} through the wall.
 \tiny{(Graphic: Stolze}) 
}
\label{wall}
\end{figure}

A particle can therefore penetrate a wall or a mountain in potential energy, even if its
energy is not sufficient to climb over the obstacle in the classical way. Instead,
according to the illustrative picture, the particle bores a tunnel at the height
corresponding to its energy. This phenomenon is known as the \textbf{tunnel effect}. The tunnel
probability with which the particle makes it through the wall becomes exponentially
smaller when the wall becomes higher (because the wave function in the wall then falls
off more strongly) and when it becomes thicker (because the wave function then falls
off a longer distance).

We have illustrated the tunnel effect using a particularly simple case. The potential
energy $V (x)$ was constant in a whole range, and then the Schrödinger equation for the
wave function has particularly simple solutions. If the total energy is greater than the
potential energy, the solution is a wave; if it is smaller, the solution is an exponentially
increasing or decreasing function. At the points where the potential energy changes,
these simple solutions must be adapted to each other. Of course, the tunnel effect also
exists for more complicated shapes of the potential energy, but then the calculation
is not as simple as in the case shown in Figure \ref{wall}.

Shortly after its discovery, the tunnel effect provided a theoretical solution to a
problem that had long preoccupied physicists. It concerns the so-called \textit{$\alpha$ decay}. In
this process, an atomic nucleus decays by emitting an $\alpha$ particle (composed of two
protons and two neutrons). For a certain isotope (a nucleus with a certain number of
protons and neutrons), the $\alpha$ particle has a certain energy and the decay takes place
with a certain half-life. The energies are roughly between 2 and 8 MeV (mega-
electron volts, i.e. millions of electron volts), but the half-lives vary greatly, from
10$^{-12}$s to billions of years.

The decisive idea for explaining $\alpha$ decay was the following: The $\alpha$ particles already
exist within the atomic nucleus, move freely there and already possess the (kinetic)
energy that can be measured after decay. As a result, they fly back and forth within the
nucleus, but are repeatedly stopped by the very strong nuclear forces and bounce back.
The potential energy $V (x)$ of the nuclear forces is practically constant within the
nucleus, so there is no force acting, but it rises very steeply at the edge of the nucleus
and holds the $\alpha$ particle in this way in place. The $\alpha$ particle therefore repeatedly runs up against a high wall of potential
energy, which it can penetrate with a certain, very small probability; then an $\alpha$ decay
has taken place. The probability of decay depends very strongly (exponentially, in
fact) on the kinetic energy of the $\alpha$ particle and the dimensions of the potential wall;
this is how the vastly different half-lives come about.

If the tunnel effect allows a particle to leave an atomic nucleus, then it should also be
possible to enter an atomic nucleus. This is precisely the process by which energy is
released in the \textit{sun}: Nuclear fusion. The first step is the fusion of two hydrogen nuclei
(protons). The electrical repulsion between the positively charged protons creates a
high barrier in the potential energy. Due to the high temperature inside the sun, the
protons move very quickly; nevertheless, this thermal kinetic energy is not nearly
sufficient to overcome the barrier. Only the tunnel effect makes the process possible
and thus ensures that the sun can shine. However, the probability of this tunneling
process (and the further steps of the nuclear reactions in the sun) is so low that each
proton \zitateng{lives} about one billion years on average  before it participates in a fusion reaction. Fortunately! If the probability were greater,
the sun would radiate much more energy and the earth would be much too
hot. In addition, the sun would burn out much faster and there would not be enough
time for higher life forms to develop through evolution on Earth or another planet.
Without the tunnel effect, we wouldn't be here.

Today, many people rely on the tunnel effect every day without even knowing it: The
fact that information can be written to an electronic flash memory (as found in every
USB memory stick) and then be stored for a long time is also thanks to the tunnel
effect. This is just one of many examples of the use of the tunnel effect in electronics,
which we will not go into in any detail here.

A less common device that also uses the tunnel effect is the \textit{scanning tunneling microscope} (STM), which can be used to \zitateng{see} individual atoms, or actually rather feel them. A very thin needle tip is located very closely above a surface
that is to be examined. Between the tip and the surface an electrical voltage is present, but no current is expected to flow, as
there is no electrical contact between the tip and the surface. Only
the tunnel effect leads to a very small current flowing, which depends very strongly
(exponentially) on the distance between the tip and the surface and on the type of
atoms on the surface. By scanning the surface, very fine details of the surface
can be examined and even individual atoms can be identified or moved back and forth
on the surface, thus producing nanometer-sized structures. The investigation of
surfaces and of small particles on surfaces is also a field of research at TU Dortmund
University; Figure \ref{STM} was taken in Prof. Hövel's research group.

\begin{figure}[h] 
\includegraphics[width=\textwidth,keepaspectratio=true]{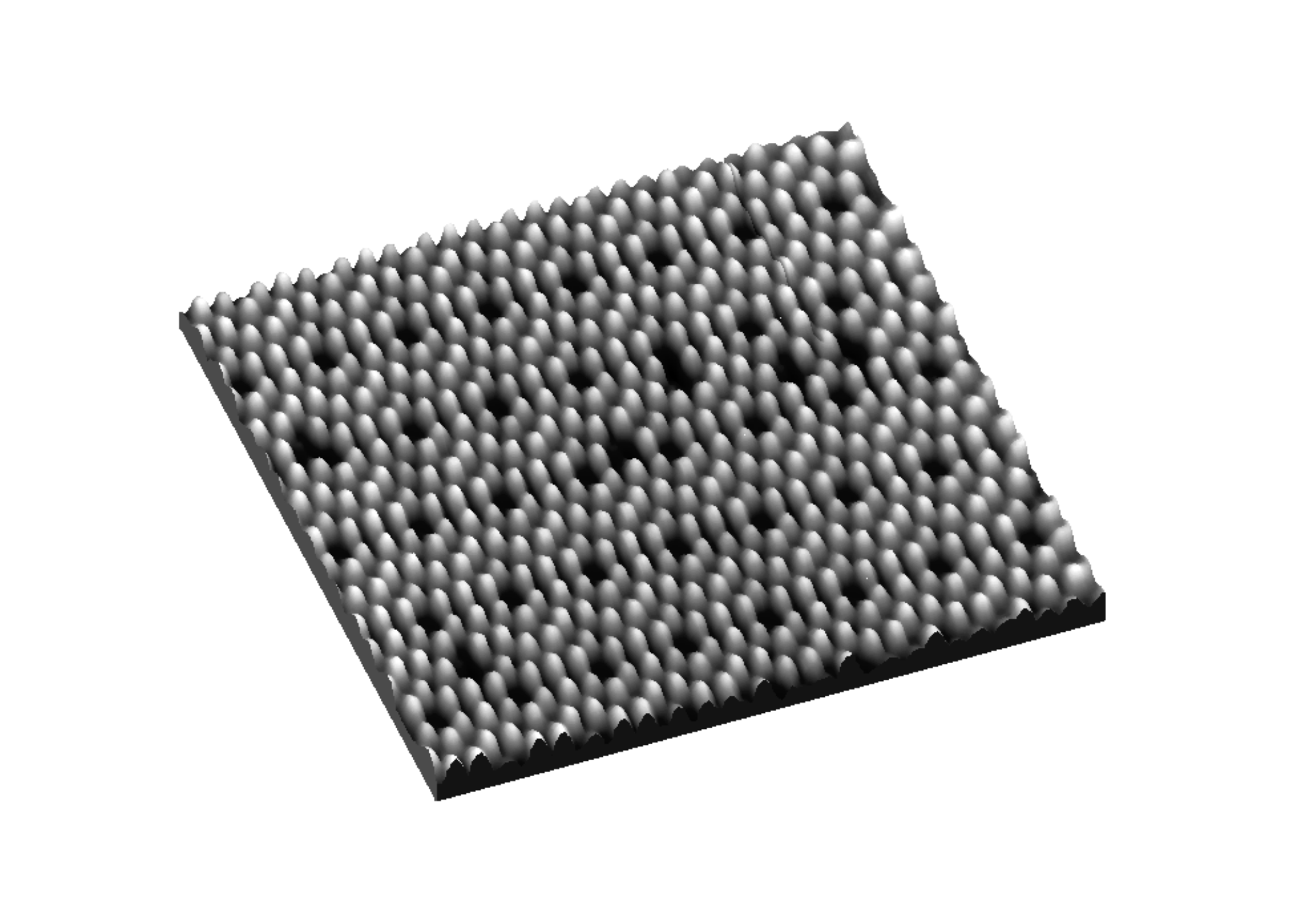}
\caption{Scanning tunneling microscope image of the surface of a silicon crystal, taken at 100 Kelvin or -173$^{\circ}$C. The image size is 21 nm$\times$13 nm. The silicon atoms
 on the surface are
arranged differently to those inside the crystal, as they have less
neighbors to form a chemical bond with. In this special regular arrangement, irregularly arranged
individual \zitateng{missing} atoms are detected. These defects in the structure show that the
microscope actually images each individual atom.
 \tiny{(Image: Prof. Dr. Heinz Hövel}) 
}
\label{STM}
\end{figure}

\subsection*{Trapped particles and stationary states}

So far, we have looked at particles that could move more or less freely, except for
obstacles that could either be overcome or tunneled under, depending on the energy
of the particle. Now we will look at particles that are bound, i.e.
for example, electrons in an atom; after all, the spectra of atoms were one of the  reasons
for the development of quantum mechanics.

We simplify again as far as possible and consider a particle in one dimension. We
imagine the potential energy $V (x)$ to be very simple again; it should be zero
everywhere except between the points $x = 0$ and $x = a$, where it should have the 
negative value $-U$.
  (Figure \ref{grube}) The potential energy now forms a rectangular pit, just
as it formed a wall in our simple model for the tunnel effect.

\begin{figure}[h] 
\includegraphics[width=\textwidth,keepaspectratio=true]{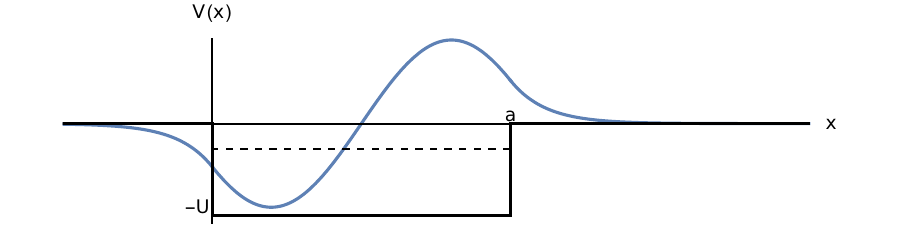}
\caption{A simple shape of the potential energy that makes bound states possible. The energy of one of the bound states is shown as a dashed line, the wave function of the state as a solid blue line. 
\tiny{(Graphic: Stolze}) 
}
\label{grube}
\end{figure}

We already know what the solutions of the Schrödinger equation for constant potential
energy look like: Waves, if the total energy is greater than the potential energy
and exponential functions if it is smaller. If a particle has an energy $E > 0$,
then its wave function everywhere in Figure \ref{grube} is a wave and therefore the particle can
be present in the entire space and is not bound. A bound particle must therefore have
a negative energy. The corresponding
wave function \zitateng{in the pit} has a certain wavelength, which is the smaller the higher
the energy above the bottom $-U$ of the pit,
 because the distance to the bottom of the pit corresponds to the kinetic energy, which in turn depends on the momentum,
and this determines the De Broglie wavelength.

We have already discussed what happens at the edges of the pit in Figure \ref{stufe}: In the
classically forbidden areas $x < 0$ and $x > a$, the wave function behaves exponentially increasing or decreasing, in such a way that it does not grow indefinitely far away from the pit, but on the contrary tends towards zero,
because otherwise there are problems with the relationship between the wave function
and the probability of finding the particle (keyword: normalization). The strength of this drop is also
fixed by the energy.

We thus obtain
\begin{itemize}
\item to the left of the pit  ($x<0$) an exponential function increasing from left to right
\item in the pit ($0<x<a$) a wave whose wavelength is fixed by the value of the energy, and
\item to the right of the pit ($x>a$)  a decreasing exponential function.
\end{itemize}
These three functions must be smoothly connected to each other at the edges of the pit
so that certain mathematical conditions are met, just as with the step in Figure \ref{stufe} and
the wall in Figure \ref{wall}. In this calculation, which we do not want to carry out here\footnote{Every physics student will spend a few entertaining hours with this little example.
} it
turns out that this is only possible for very specific values of energy. Only then is
the wavelength in the pit such that the connection conditions can be met
simultaneously at both ends of the pit.

We have thus understood the \textbf{quantization of  energy} for bound states of a
particle using this simple example. Of course, this is quite abstract and mathematical,
but it couldn't be cheaper. If you are looking for an illustrative analogy, you can think
of a guitar or another stringed instrument: there, too, the \zitateng{wave function} of the
vibrating string is fixed to very specific wavelengths, simply because the string is
clamped at the ends. (Incidentally, the Schrödinger equation is mathematically not
so far removed from the equation that describes a vibrating string).

What we have seen here in a very simple example also works for more complicated
cases. In his first work on quantum mechanics, Schrödinger himself took on the
interaction between an electron and a proton, which holds the hydrogen atom together.
The associated potential energy looks more complicated than the one in Figure \ref{grube}, and
above all the problem is not
one-dimensional, but three-dimensional. Nevertheless, Schrödinger was able to explain
the previously mysterious energy spectrum of hydrogen \zitateng{at the first attempt}.

A hydrogen atom can therefore assume different states with very specific different
energies. During the transition from one state to another, the energy changes by
$$
\Delta E = E_{\mathrm{after}} - E_{\mathrm{before}}
$$
and the energy difference is emitted in the form of electromagnetic waves of frequency
$f$  (if $\Delta E$ is negative, i.e. the atom emits energy) or absorbed (if $\Delta E$ is positive). The relationship already introduced by Planck (Section \ref{wellenfunktion}) applies here:
$$
|\Delta E| = hf.
$$
($h$ is the Planck constant, and the absolute value bars are necessary because $\Delta E$ can be
both positive and negative, whereas frequencies are usually positive.)

In the time between two such transitions, the atom should \zitateng{behave calmly}; we
therefore expect no dynamics. This is also confirmed: The solutions $\Psi(x,t)$  of the
Schrö\-dinger equation for fixed energies have the property that the probability $|\Psi(x,t)|^2$ is \textit{independent of the time $t$}. Therefore, the states with  well-defined
energies are also called \textbf{stationary states}.

In the early days of quantum mechanics, the transition between two stationary states of
an atom, during which a (tiny amount of) energy is exchanged, was often referred to as
a quantum leap. Later, this word has taken on a life of its own. 
So if you hear today that some development is a \zitateng{quantum leap in the field of XY},
i.e. a huge step forward, then you know that the originator of this
formulation definitely has \textit{no idea} about quantum physics.

How many stationary bound states with well-defined energies are possible depends, of course, on
the potential energy in each individual case. In the case shown in Figure \ref{grube}, the
deeper and wider the pit, the larger the values of $U$ and $a$, the more such states exist.
The lowest possible energy must be greater than $-U$ because of the uncertainty relation. If the total energy of the particle had the value $-U$ , its kinetic energy
would  have to have the value zero. However, this would mean that the momentum
would be exactly zero and the location completely undetermined, so that the particle
could not be bound.

On the other hand, the depth $U$ and the width $a$ of the pit must also fulfill certain
minimum conditions so that a bound state is possible at all. This can be understood
as follows: For a bound particle, it is expected that the spatial uncertainty $\Delta x$ is
roughly as large as $a$, because the particle is located in or near the pit. Using the
uncertainty relation, this means a minimum size for $\Delta p$ (inversely proportional to $\Delta x$
or $a$). Then remember that, by definition, the square of $\Delta p$ is
$$
(\Delta p)^2 = \langle p^2 \rangle - \langle p \rangle^2
$$
where the angle brackets $\langle ... \rangle$ are mean values. However, the mean value  $\langle p \rangle$ is zero because the
particle is bound and therefore on average
at rest. The particle of mass $m$ therefore has the
kinetic energy $\frac{(\Delta p)^2}{2m}$.
 Due to the uncertainty principle, this is proportional to $1/a^2$ and
therefore increases as $a$ decreases. If it is then greater than $U$ (because $U$ or $a$ are
small), the particle has a positive total energy and therefore cannot be bound.

We have already used the uncertainty principle in a very similar way in Section \ref{unschaerfe} to
show why not all electrons fall into the atomic nuclei and why matter is stable in the
state we know. In quantum mechanics, the state of lowest energy is called the \textbf{ground
state}. We have seen that, due to the uncertainty principle, the ground state energy is
always a little greater than the lowest value of the potential energy $V (x)$; this
difference is also called the \textbf{zero-point energy}.

The quantization of energy, i.e. the fact that energy can only have certain values, is
one of the most astonishing and important phenomena in the microscopic world. As
early as the 19th century, spectroscopic investigations showed that atoms can only emit
or absorb light with very specific wavelengths. These wavelengths are related to each
other in very specific ways that could not initially be explained.

Niels Bohr provided a first explanation with his atomic model, in which the
electrons orbit the nucleus like the planets orbit the sun. In order to explain that an
atom constructed in this way is stable at all and can emit or absorb light with the
known wavelengths, Bohr had to assume that the electrons can only move along very
specific orbits. This \textbf{Bohr model of the atom} actually provided the observed wavelengths and is therefore still used today as an \zitateng{explanation} of the atomic spectra. Actually, however, it is \textit{wrong}; we already know
that the concept of the \textit{orbit} of an electron  is meaningless because of the uncertainty principle.
The assumptions about the \zitateng{permitted orbits} are not justified by
anything except the fact that the correct numbers come out in the end. For more
complex systems, such as molecules,
Bohr's atomic model fails completely.
However, the correct description by the Schrödinger equation is unfortunately not
quite as descriptive and also somewhat more mathematically demanding than Bohr's model.

\begin{mdframed}[style=tpq]

\begin{center} 
\textsf{Spectrum (plural: spectra), spectroscopy
}
\end{center}
A \textbf{spectrum} is the set of the wavelengths (or frequencies) of
electromagnetic radiation that can be observed in a range, together with
their respective intensities. 
If all wavelengths can occur within some range, the spectrum is called
\textbf{continuous}, e.g. in the case of thermal radiation from a hot object. If only
very specific wavelengths can occur, the spectrum is called \textbf{discrete}. This is
the case for pure chemical elements, for example. Discrete spectra are also
called \textbf{line spectra}, named after the image of colored lines on a dark
background that appears in an optical spectrometer. By examining such line
spectra in \textbf{spectral analysis}, it is possible to determine which chemical
elements the substance under investigation contains. The study of
spectra is called \textbf{spectroscopy}.

In more general terms, the possible values of the energy of a quantum system
are also referred to as the \textbf{energy spectrum}.

\end{mdframed}

In the \textit{Treffpunkt Quantenmechanik}, there are two experiments that demonstrate
the quantization of the energy of atoms, the \textit{Franck-Hertz experiment} and the \textit{Optical
spectroscopy} experiment.

In the \textit{Franck-Hertz experiment}, free electrons are accelerated by an electrical voltage
(i.e. provided with kinetic energy), sent through a tube of diluted neon gas and, at the end, the
electrons passed through the gas are measured as an electric current. If the accelerating
voltage is slowly increased, the current of the electrons passed through slowly
increases until it abruptly decreases at a certain voltage $U_0$ . At this voltage, the
electrons have just enough energy to promote an electron in a neon atom from the
ground state of the atom (the lowest energy state) to an \textit{excited state} with higher
energy. As the free electrons have given up their kinetic energy in this process, they
can no longer pass through the gas.

However, the current of the electrons allowed to pass does not drop completely to
zero, as not every electron releases its energy. If the accelerating voltage is increased
further, more electrons are allowed to pass through until the current decreases again
abruptly at a voltage of $2U_0$. An accelerated electron can now release its energy to an
atom, then be accelerated again and release the same amaount of energy to another atom. If the
acceleration voltage is increased further, the process continues in a similar way. The
atoms excited by the collisions with the free electrons can return to the ground state,
emitting light whose frequency depends on the voltage $U_0$ . (Can you state what this
relationship looks like as a formula?)

In the \textit{Spectroscopy} experiment, you use interference on a grating (Section \ref{interferenz}) to split
light into its components with different wavelengths. You also use a completely
different grating in the electron diffraction experiment (Section \ref{interferenz_quanten}) to demonstrate
the wave properties of electrons. If you are dealing with both  of these experiments, it is
interesting to compare the wavelengths involved and also the dimensions of the
gratings used, i.e. the distances between the \zitateng{grating bars}.

In the \textit{Spectroscopy} experiment, you will learn how to set up the device used correctly
and how to use it to determine the wavelengths of different spectral lines of the elements
mercury (Hg) and cadmium (Cd) and to calculate the energies of the associated
transitions between different stationary states of the atoms.

If you have mastered this technique, you could, for example, determine that the light
source used contains the elements Hg and Cd (if we have not already told you).
Spectroscopic techniques can be used to determine the chemical composition of
samples that contain too little material for conventional analysis, or that are not
directly accessible at all because they are far away in the Earth's atmosphere or even
in space.

However, the methods for spectroscopic analysis, for example of air pollutants or
industrial production processes, are largely automated nowadays and are carried out
with commercial devices; nobody has to adjust a spectrometer anymore. However, the
basis is still quantum mechanics, and physical precision is still required for special
situations or if extreme accuracy is desired.

\subsection*{Discrete and continuous energy spectra}

So far, starting from the simple \zitateng{pit} potential in Figure \ref{grube}, we have first considered
the bound states, which can have only very specific energies. Such well-separated energy values are called \textbf{discrete} energies in physics; in our case, the energies of the bound states were negative. In contrast, a particle with
positive energy is not bound in the potential shown in Figure \ref{grube}. Since the (total)
energy along the entire $x$-axis is greater than the potential energy, the Schrödinger
equation has a wave as a solution in all regions and thus the particle can stay
everywhere. Its wavelength has the same value to the left and right of the pit. In the
region of the pit, the kinetic energy of the particle is greater because the potential
energy is smaller, and therefore the wavelength is shorter.
As before with the bound states, the waves with the different wavelengths  must be joined to each other at the points $x= 0$ and $x = a$ in such a way that the Schrödinger equation is fulfilled. It then turns out that a wave coming from the left is not completely transmitted
into the region on the right, but is partially reflected. This behavior is quite different
from that of a classical particle: if that comes from the left with sufficient kinetic
energy, it will definitely continue flying to the right. We had discussed this difference
between classical
 and quantum mechanical behavior  already when treating the \zitateng{step} in the potential energy (Figure \ref{stufe}).

The Schrödinger equation can be solved here for arbitrary (positive) values of the energy;
the reflected and transmitted parts of the wave depend on the energy. The
corresponding states are not bound, as the wave functions extend over the entire $x$ axis, and the associated energy spectrum is \textbf{continuous}: all positive energies are
possible. A continuous energy spectrum always exists if the potential energy does not
increase indefinitely in both directions. Below the continuous spectrum, there \textit{can} exist bound states with a discrete energy spectrum. However, it is also possible that the
potential 
energy is \zitateng{not strong enough} to have a bound state; in our example, if the pit is too
shallow or too narrow.

The fact that an incoming wave does not simply continue to run undisturbed, but is
partly reflected and partly transmitted, is a simple example of a \textbf{scattering process}.
In physics, a scattering process is any process in which a wave (whether a classical
light wave or a quantum mechanical matter wave) is modified by
the interaction with a \zitateng{target object}. Many experiments in physics are
scattering experiments, for example in particle physics, but also in
other areas of physics, for example whenever the structure of particles, atoms,
molecules, biological systems or solids needs to be elucidated.

In real, three-dimensional life, there are many more possibilities than just reflection
or transmission. The incident wave can be deflected in different directions to different
degrees and this pattern can change with the wavelength; it gets even
more complicated when the wave exchanges energy with the object under
investigation and thus causes changes. Large parts of physical research consist of
predicting such scattering processes theoretically (with more or less complicated
calculations) or measuring them precisely (with more or less complicated
measurements). Information about the structure of the scattering object and about the
type of interaction between the incident wave (the incident particle) and the object under investigation is obtained from the measured scattering data.

\section{What happens in a measurement?}
\label{was_bei_messung}

What happens during a measurement in quantum mechanics is so unfamiliar to our
minds trained in classical phenomena that it should be explained in more
detail here, even though the basic facts have already been described in Sections \ref{besonders}
and
\ref{wellenfunktion}.

In the classical world, we never think about whether a measurement influences the
state of the system under investigation. We also always assume that the measured
quantity has a certain value, regardless of whether we measure it or not: For example,
a battery supplies a voltage of 1.5V, regardless of whether I measure it or not. My
mass, for example, is 77kg, regardless of whether I step on the scales or not; it does
not change when I step off the scales again after I have read the value.

A simple system in classical mechanics consists of particles that move in space under
the influence of certain forces. The state of such a system is completely known if the
position and momentum of each individual particle is known. The classical equations
of motion then provide the further evolution of the state, i.e. the future values of
the positions and momenta of all particles. All classical measurable quantities
depend on the values of these positions and momenta (and nothing else). Thus, if the
state of a classical mechanical system is known, the values of all measurands are also
fixed: A classical measurable quantity already has a certain value before the measurement and
that value is not changed by the measurement.

\begin{mdframed}[style=tpq]

\begin{center} 
\textsf{Equations of motion}
\end{center}

In physics, equations of motion are always differential equations,
i.e. they indicate at a certain point in time $t$ how the state changes over
time (mathematically: derivative with respect to time). From this, the state
at a later point in time can be calculated.

For a \textit{classical} particle, the equation of motion is given by Newton's laws, e.g. in the form \zitateng{force = mass times acceleration} If the force is known, the acceleration can be used to
calculate how the velocity changes, and from this, how the position
changes. Then you can follow the path of the particle in space and time, e.g. for a
satellite orbiting the earth.

The equation of motion for a \textit{quantum mechanical} particle is the Schrödinger
equation, a differential equation for the wave function. Due to Heisenberg's
uncertainty principle, the concept of orbit does not exist for a quantum
mechanical particle (e.g. an electron in an atom).

\end{mdframed}

In quantum mechanics, the relationship between the state of a system, its evolution in time,  and measurements is different, as already explained in the Sections \ref{besonders} and \ref{wellenfunktion}.
This can be seen from the example of a single particle moving in one
dimension. The state of the particle is completely determined if the wave function $\Psi(x,t)$
at a certain time $t$ is known. 
The Schrödinger equation determines the further evolution of the state, i.e. the
wave function $\Psi(x,t)$ for later times, but only \textit{as long as no measurement takes place}.
A measurement changes the state $\Psi(x,t)$ of the particle, depending on the result of
the measurement. A quantum mechanical measurable quantity therefore  \textit{does not already} have
a specific value before the measurement. The state $\Psi(x,t)$ immediately after the
measurement is determined by the measured value and then continues to develop
according to the Schrödinger equation -- until the next measurement.

\begin{mdframed}[style=tpq]

\begin{center} 
\textsf{Measurements and probabilities}
\end{center}

The state of a particle in one dimension at time $t$ is \textit{completely} determined by the wave function $\Psi(x,t)$. Nevertheless, the result of a \textbf{single measurement} of the position $x$ (or another measurable quantity) is not predictable. However, if the following operations are carried out very often one after the other:
\begin{enumerate}
\item Prepare the state $\Psi(x,t)$,
\item Measure the position $x$ and
\item Register the result of the position measurement,
\end{enumerate}
then the distribution of the measurement results of the position, which is given by the probability density $|\Psi(x,t)|^2$, is gradually obtained.

In contrast to a single measurement, this is a statistical measurement; experts often speak of \textbf{ensemble measurements}. It is important that the system is always prepared anew, i.e. the initial state $\Psi(x,t)$ is created.
Alternatively, one could also imagine to prepare a  large number of particles (an ensemble of particles) all in the desired state $\Psi(x,t)$ and measure each particle once.

After a single measurement, the state of the particle is usually \textit{no longer} $\Psi(x,t)$, but a different one, depending on which physical quantity was measured and which value this measurement produced.

\textit{A measurement changes the state of a system.}

\end{mdframed}

Apart from \zitateng{technical details}, this actually says everything about the role of
measurements in quantum mechanics. However, it is worth looking at the details in
the following
to take a closer look in order to understand (or get used to) what  distinguishes the 
quantum world from the classical everyday world. The differences are so important
and so surprising that the measurement process has been discussed in physics and
philosophy for as long as quantum mechanics has existed in its current form, i.e. for
about a century.

The comparison between a classical-mechanical and a quantum-mechanical system
shows an important difference: In the classical system, the state always develops
\textit{deterministically} (i.e. predictably), and measurements have no influence on it. In the
quantum mechanical system, there are \textit{two} types of state changes: The Schrödinger
equation also provides a \textit{deterministic}
change in the state $\Psi(x,t)$, but the measurement causes a \textit{random} change,
with the state after the measurement depending on the measured value. Chance
therefore comes into play during the measurement. \footnote{ The distinction between two types of state change alone gives rise to some of the aforementioned
discussions, which we will not describe in detail here. Our point of view corresponds to the interpretation which is
used by most practicing scientists and which reflects the results of experiments well.}

To understand what is going on here, we need to take up the terms used in the
Info box \zitateng{Measurements change states}. For each measurable quantity, there are special quantum
mechanical states (i.e. wave functions) in which that quantity  has a very specific, clearly defined value. Such states are called
\textbf{eigenstates} of the measured variable and the corresponding measurement values are called
\textbf{eigenvalues}. If the eigenvalues of all eigenstates are different from each other, then
\textit{every} possible state of the system under consideration can be represented as a superposition of the eigenstates of our measured variable. \footnote{Only the simplest
possible case is presented here; more complicated cases require more formal
and mathematical effort, but do not provide any new insights.} Each eigenstate enters this superposition with a certain amplitude, and
that amplitude determines \textit{the probability of measuring the corresponding eigenvalue.} 
Each measurement can only provide one of the eigenvalues; other measurement
results are not possible.

\begin{mdframed}[style=tpq]

\begin{center} 
\textsf{Amplitude}
\end{center}

The \textbf{amplitude} is the maximum displacement of a wave, e.g. in the case of a
water wave, the height of the wave crest above the calm water surface. In
addition to the amplitude, the \textbf{phase} of a wave is also important, as it
determines, for example, exactly where the wave crests are located.
When waves are superimposed, both the amplitudes and the phases of the
partial waves are important, keyword interference.

In the case of a superposition of several quantum mechanical wave
functions, one also speaks of the amplitudes of the different wave functions
and thus means the respective components (expressed as complex
numbers).

\end{mdframed}

A single measurement therefore provides one of the possible eigenvalues, and the
state immediately after the measurement is the corresponding eigenstate. All other
eigenstates that were previously present in the superposition are eliminated,
i.e. their amplitudes are zero after the measurement. After this \textit{sudden and random}
change, the state of the system continues to develop \textit{continuously and
deterministically}, as described by the Schrödinger equation. In the course of that evolution
the state will change and the system will not remain in the previously measured initial
state. For example, a particle will change its position or direction of movement under
the influence of forces. The amplitudes of the states eliminated by the measurement
will therefore not remain zero, and a second measurement after some time will again
provide any of the possible eigenvalues, whereby the
state of the system is \zitateng{reduced} again to the corresponding eigenstate. (Sometimes
this is also referred to somewhat dramatically as the \zitateng{collapse} of the state).

The fact that measurements provide random results can therefore be attributed to the
fact that states of the system can be represented as superpositions of eigenstates of the
measured quantities. This sounds quite abstract, but we have just seen in detail what
this means. Richard Feynman (Nobel Prize 1965), who thought intensively about
quantum mechanics, says at the beginning of his first lecture on quantum physics that
the superposition principle (any superposition
of possible states is also a possible state) is \zitateng{the only mystery of quantum mechanics}.

Because measurements change states, they also influence each other. We consider two
different measurement variables, which we will call $A$ and $B$. Each of the two
quantities has eigenvalues and eigenstates; there are therefore $A$ eigenstates and $B$ eigenstates. We can represent any state of the system under consideration both as a superposition of $A$ eigenstates and as a superposition of $B$ eigenstates. \footnote{This is generally unproblematic; we will not discuss the exact mathematical conditions for this here.} We now start from some arbitrary
state and measure $A$. As a measurement result, we obtain, as described
above, one of the
eigenvalues of $A$, and after the measurement the system is in an $A$ eigenstate.
This state, like all states, can be represented as a superposition of $B$ eigenstates, and if
we measure the quantity $B$ immediately after the $A$ measurement, we obtain one of the
eigenvalues of $B$ and the system is in a $B$ eigenstate. So far, so good. However, if
we measure $B$ first and then $A$, we end up with an $A$ eigenstate, unlike when performing thw two
measurements in reverse order!

The measurements of $A$ and $B$ are therefore not interchangeable, unless every $A$
eigenstate is also a $B$ eigenstate and vice versa. In that special case the measurements are
interchangeable and if the measurement of $A$ has a \zitateng{sharp} value
(because the system is in an $A$ eigenstate), then the measurement of $B$ also returns
a sharp value because the $A$ eigenstate is also a $B$ eigenstate. The two
quantities are therefore, as we say, \textit{simultaneously precisely measurable}.

If, however, the measurements of $A$ and $B$ are \textit{not} interchangeable, then the $A$
eigenstate that results after the measurement of $A$ is composed of \textit{several} $B$
eigenstates, so that the result of the $B$ measurement is random; the two quantities
are therefore \textit{not} simultaneously precisely measurable. We have thus learned the
cause of the \textbf{general uncertainty relation}, which applies to any pair of quantities that
do not have a system of common eigenstates. This is now much less descriptive than
the considerations on the uncertainty relation between location and momentum in
Section \ref{unschaerfe}, but it has a much more general validity.

If you take a closer look at the mathematical formalism of quantum mechanics, you
can understand these statements using an example. The states of a quantum
mechanical system can be represented by wave functions, as we have already seen.
Measured quantities are represented by \textbf{operators} that perform certain mathematical
operations on the states - as the name suggests. In the representation using wave
functions, however, this is not technically easy. Therefore, we will discuss operators
later, namely in Section \ref{vektoren}, where we will use a version of quantum mechanics that
represents the states by \textbf{vectors} and the operators by \textbf{matrices} and which is
mathematically so simple that the above statements about measurements and
uncertainties can be easily recalculated.

To confirm these abstract considerations in a concrete experiment, the experiment
\textit{Polarization of light} in the \textit{Treffpunkt Quantenmechanik} is perfectly suitable.

\begin{mdframed}[style=tpq]

\begin{center} 
\textsf{Polarization of light}
\end{center}
Electromagnetic waves consist of coupled electric and magnetic fields that change in space and time. In the case of a monochromatic (only one frequency or wavelength) plane wave in free space, the electric and magnetic fields are perpendicular to each other and both are perpendicular to the direction in which the light propagates. If the electric field only oscillates back and forth in one direction, it is referred to as a \textbf{linearly polarized} wave. If you look in the direction of wave propagation, you can
distinguish between two fundamentally different polarizations (directions of oscillation): \zitateng{right and left} ($x$ polarization) and \zitateng{up and down} ($y$ polarization). All
more complicated waveforms can be composed from these basic forms.

\end{mdframed}

In that experiment, we consider a light source, a laser, whose light we can classically
regard as an electromagnetic wave, or quantum mechanically as a collection of
photons. Both approaches can explain the experimental results well, as we will see. The
quantum mechanical consideration also shows us an example of how a
measurement influences the state of a system.

The laser  delivers an electromagnetic wave that we can observe on a screen
as a bright spot. From the perspective of quantum mechanics, the laser excites
photons that produce a bright spot on the screen. (Warning: 
As already mentioned at the end of Section \ref{interferenz_quanten}, we should not think of photons as
being too \zitateng{particle-like}. They are energy quanta of a certain oscillation form of the
electromagnetic field and are only localized at a specific position when 
registered by the eye or another detector.)

Polarization filters are used to measure the polarization of light. A polarization filter
consists of a substance (usually a special plastic film) that only allows light waves
with a certain polarization to pass through. In most cases, polarization filters are
rotatably mounted so that the direction of the transmitted polarization can be
adjusted. This is also the case in the experiment at \textit{Treffpunkt Quantenmechanik}. For
the discussion here, we want to represent the settings with directional arrows; a $\uparrow$
 filter then only lets pass electromagnetic waves or photons whose electric
field oscillates \zitateng{up and down}. Accordingly, a $\rightarrow$ filter only allows waves through
whose electric field oscillates \zitateng{back and forth}.

Then we can start experimenting and understand the results in a \textit{classical} way. First, a $\uparrow$
filter is placed in the beam.  After the light has passed through the filter, it is completely $\uparrow$-polarized. A second 
 filter
 in the $\uparrow$-direction  of course allows this
  light to pass, as can be seen on the observation screen.

However, if the second filter  is turned to the $\rightarrow$-direction, no more light passes through
and the observation screen becomes dark. If, on the other hand, the  second filter is
oriented in the $\nearrow$-direction, light still reaches the screen because the electric field of the $\uparrow$-light transmitted by the  first filter can be divided into a component in the 
 $\nearrow$-direction that is
transmitted and an equal component in the $\nwarrow$-direction that is
  not transmitted.

 None of this is surprising once you have understood what polarization actually means,
namely the direction of the electric field. The following experiment is, however,
initially baffling: a $\rightarrow$ filter is placed behind
  the $\uparrow$ filter; the screen is then of course
dark. However, if a third filter is placed \textit{between} the two filters, which is set in the $\nearrow$-direction, \textit{the screen becomes bright again}.

To understand this, you simply have to follow what happens to the electric field. After
the first filter, it points in the $\uparrow$-direction. The
  second filter splits this $\uparrow$ field into a part
in the $\nearrow$-direction (transmitted)
 and a part in the $\nwarrow$-direction (blocked). The field then
points in 
 the $\nearrow$-direction; it can be split into a $\uparrow$-component  (blocked by the third filter) and a $\rightarrow$-component that \zitateng{survives}.
 
 The \textit{quantum mechanical} analysis of the experiments looks at individual photons. A $\uparrow$ filter performs a measurement of the \zitateng{$\uparrow\rightarrow$-polarization}
 on each photon. The possible measurement results are then either $\uparrow$, and the photon is transmitted, or $\rightarrow$
 and the photon is absorbed. A $\rightarrow$ filter measures the same quantity, but the
other polarization is transmitted. Behind the $\uparrow$ filter, all photons are therefore in  the $\uparrow$
state and are therefore certainly transmitted by another $\uparrow$ filter, but certainly not
by a $\rightarrow$ filter.

If the second filter is oriented in the $\nearrow$ direction, a measurement of the \zitateng{$\nwarrow \nearrow$-polarization}
is carried out. The incoming $\uparrow$ photon is in a superposition state of $\nwarrow$ and $\nearrow$.
 If the measurement result is $\nwarrow$, the photon is in the $\nwarrow$ state after the
measurement, but is absorbed by the filter. However, if the measurement result is $\nearrow$,
then the photon is in the $\nearrow$ state and is transmitted.

The experiment with three filters in the order $\uparrow$, $\nearrow$, $\rightarrow$, has a simple explanation within
quantum mechanics, too. 
As we have just seen, the photon is in the $\nearrow$ state after the first two filters (or it has been absorbed). Then the last
filter ($\rightarrow$) again performs a measurement of the \zitateng{$\uparrow\rightarrow$ polarization}. The $\nearrow$ state is a superposition of equal parts of the $\uparrow$ state and the $\rightarrow$ state. If the
measurement result of the last filter is $\uparrow$, the photon is in the $\uparrow$ state after the
measurement, which is not transmitted by the filter. If the measurement result is $\rightarrow$,
the photon is in the $\rightarrow$ state after the measurement and reaches the screen.

The quantum mechanical analysis of the polarization experiments also makes it clear that
\textit{measurements are generally not interchangeable}. Two filters in the
 sequence \zitateng{first $\uparrow$, then $\nearrow$} results in a photon that is either in state $\nearrow$ or in the
state $\nwarrow$ (and unfortunately
 gets absorbed). The sequence \zitateng{first $\nearrow$, then $\uparrow$} results  in either the state $\rightarrow$ (absorbed) or the state $\uparrow$, i.e. definitely a
different state than in the previous sequence of measurements.

One may find it annoying that when measuring polarization on individual photos with
polarization filters, one of the possible measurement results is always immediately \zitateng{destroyed} by absorption in the filter, but this is precisely how the perfect 
agreement with the classical analysis is achieved:
Behind the filter there remain only
the \zitateng{correctly} polarized photons, and the \zitateng{correctly} polarized electric field of the
classical wave is composed of very many of these.

In the end, we note for the record that there are \textit{two types of state changes} in quantum mechanics. The
Schrödinger equation describes the time evolution of the wave function
deterministically, i.e. predictably; measurements, on the other hand, change the state
of a system randomly. The possibilities and the extent of the randomness are
determined by the state before the measurement. Only if the state before the
measurement is an eigenstate of the measurable quantity is the measurement result
completely fixed, and with it the state after the measurement. An example of this are
the stationary states (Section \ref{wie_bewegt}), e.g. the bound states of an atom. Stationary states have a
certain energy, i.e. they are eigenstates of the energy and therefore you can
measure the energy as often as you want, you will always obtain in the same value and
the state will not change. If, on the other hand, the state before the measurement is not
an eigenstate of the measured quantity, the measurement result is random and the
state after the measurement depends on this random measurement result.

As already mentioned, the measurement process is still the subject of debate, and there are various approaches to making it \zitateng{more beautiful} in one respect or another, or to describe it \zitateng{more precisely}. The description given here is accepted by most
practicing physicists, is comparatively simple, 
and describes the experiments well.

\chapter{How it all started: Discovering quanta}
\label{V}

Electromagnetic radiation consists of quanta, matter consists of atoms, and these in
turn consist of electrons, protons and neutrons. Today, these statements are part of the
self-evident foundations of a modern world view. Until around 1900, however, nobody
could have made sense of these terms.
A series of discoveries and \zitateng{inexplicable} observations then led to a
complete reorganization of the foundations of physics.

You can reproduce some of the fundamental experiments in the \textit{Treffpunkt Quantenmechanik}. They will be briefly discussed in this chapter. The technical details for carrying
out the experiments can then be found in the individual experiment instructions.

The contradictions in the theories of the spectrum of \textit{Blackbody radiation} forced Max
Planck to assume that the electromagnetic radiation field
can only  absorb or emit energy in very specific \zitateng{portions}, i.e. the quanta. The
explanation that Albert Einstein gave for the measurement results of the \textit{Photoelectric effect}
was also based on the existence of quanta. The \textit{Millikan experiment} finally showed
that the electric charge can only occur in multiples of a smallest unit, i.e. it is also
quantized.

In order to make discoveries, certain observation techniques and devices are required.
\textit{Spectroscopy} is a technique that had been known for a long time, but its findings were
only later explained by quantum mechanics. The \textit{Detection of single photons}, on the
other hand, has only become possible through the use of modern components whose
function is itself based on quantum mechanical effects.

Photons not only have energy, but also momentum. This is demonstrated by the
\textit{Compton effect}, the \zitateng{collision} of a single photon with a single electron.
Unfortunately, we cannot show this effect at \textit{Treffpunkt Quantenmechanik} because it
requires X-ray radiation, which may only be used under special safety precautions
by trained personnel. For the sake of completeness, however, the Compton effect will
be explained at the end of this chapter.

\section{Blackbody radiation}

\label{schwarzkoerper}

When a piece of iron is heated, it first appears red to the eye and then yellow to white
at an even higher temperature. The composition of the radiation emitted by the
glowing metal therefore changes, and in the steel industry, for example, the
temperature of the material is measured using the radiation. Since each piece of iron
has a different shape and surface texture, it is more practical to relate measurements
to a standard system, the ideal black body. As the name suggests, such a body absorbs
all incident radiation. An example
of a \zitateng{pretty good} black body is a small hole in the wall of a cavity whose wall (and
thus also its interior) has a certain temperature.
 Radiation that falls in from the outside disappears in the
hole and only the radiation that is present inside due to the temperature comes out.
This is blackbody radiation, formerly also known as cavity radiation.

If one measures the spectrum, i.e. the distribution of this radiation over the different
wavelengths, one finds little radiation intensity at particularly short and particularly
long wavelengths, and a maximum intensity in between. This maximum shifts with
increasing temperature towards shorter wavelengths, for example from red to yellow, as described for the piece of iron above.

With the means of the classical  theories of heat and of electromagnetic
phenomena, the behavior at large wavelengths could be explained well. For small
wavelengths, however, there was a problem. Since, roughly speaking, the smaller the
wavelength, the more waves fit into a radiating cavity, the maximum intensity should
lie at very short wavelengths, in contradiction to observation. Although there was
another theory that correctly described the behavior at short wavelengths, it failed for long wavelengths.

Max Planck developed the idea that cavity radiation is generated by oscillating
systems (oscillators), each of which oscillates at a certain frequency $f$ and can only emit or absorb
energy in multiples of the energy quantum
$$
E=hf.
$$
($h$ is the Planck constant.)

In the \textit{Treffpunkt Quantenmechanik}, you do not use a cavity in which the
radiation is generated, but the (simple but sufficient) substitute for the ideal black body is the filament of a halogen lamp, which has a very high temperature. An optical grating,
as used in the \textit{Spectroscopy} experiment, splits the radiation in different directions
depending on the wavelength. With a small photoelectric element\footnote{A photoelectric element is a light sensor in which an incident photon transfers energy to an electron in
a semiconductor material, thereby generating an electric current. Photovoltaic systems also use this principle
(\textit{internal photoelectric effect}) to convert light into electrical energy. The \textit{external photoelectric effect} is described in Section \ref{photoeffekt} , the internal photoelectric effect in Section \ref{einzelphotonen}.} you can then measure the
intensity as a function of the wavelength.

\section{Photoelectric effect}
\label{photoeffekt}

If a metal surface is irradiated with light, charges can be released from it; this was first
observed in the 1880s. More detailed investigations
showed that it is electrons that leave the metal. This photoelectric (light-electric)
effect was investigated in more detail in
the following years to determine how the measurement data depended on the
frequency and intensity of the light.

These investigations showed the following results:
\begin{itemize}
\item
 The number of electrons observed increases with the intensity of the light, but
only if the frequency of the light exceeds a certain value, which depends on the
metal used. If the frequency is lower than this limit frequency, no electrons are
observed even if the intensity of the light or the irradiation time is increased so
that the total energy irradiating the metal becomes very high.
\item
 The kinetic energy of each escaping electron depends only on the frequency of
the light, not on the intensity or the duration of irradiation. It is proportional to
the difference between the irradiated frequency and the cut-off frequency
specific to the metal.
\end{itemize}

The explanation for these observations was provided by Albert Einstein in 1905,
using the quantization of energy proposed by Planck: Light with frequency $f$ consists of energy quanta of the magnitude $E = hf$ and each energy quantum is
completely transferred to an electron. To leave the metal, the electron must
but apply the \zitateng{work function} $W_A$ , which is different for different metals. The
kinetic energy of the emitted electrons is then $hf- W_A$ , but
of course only if $hf$ is greater than $W_A$ . If this is not the case, the electrons cannot
leave the metal.

In the \textit{Treffpunkt Quantenmechanik}, the actual experiment takes place in a vacuum so that the escaping electrons cannot collide with air molecules. The electrons
emerge from the metal surface (the cathode) and then fly in all possible directions.
They are slowed down on their way to the anode by a counter voltage, so that as the
counter voltage increases, fewer and fewer electrons reach the anode and are
registered there as current. If the countervoltage is so high that even the electrons flying straight
towards the anode have  lost their kinetic energy (and converted it into potential
energy), the current at the anode becomes zero, and the kinetic energy of the
electrons can then be calculated from the value of the countervoltage.

A mercury spectral lamp is used to generate light that is split into different parts of
different frequencies using a prism. This makes it possible to investigate the
dependence of the photoelectric effect on frequency in experiments. The
countervoltage method described here was used from 1912 by Robert Andrews
Millikan to confirm Einstein's theory of 1905, which was rewarded with the Nobel Prize for Physics in 1921. We will encounter Millikan again the next section.

\section{Millikan experiment}
\label{millikan}

The last two sections dealt with the quantization of energy; now it is about the
quantization of charge: In the whole of nature one finds only\footnote{You may have heard of quarks, which make up protons and neutrons and which do not have an integer charge (in units of $e$)
but charges which are multiples of one  third. However, due to the nature of their interactions, quarks can never
occur individually, but only in combinations which then again have an integer total charge.} charges that are multiples of the elementary charge $e=1,602 \cdot 10^{-19}$C. An electron has the charge $-e$, a proton has the charge $+e$.

Since the elementary charge is so small, sensitive measurement methods are required.
Harvey Fletcher, who experimented under the guidance of Robert Andrews
Millikan for his doctoral thesis, was able to improve the methods already used by other researchers to such an extent that a fairly accurate determination of $e$ was
possible in 1910, for which Millikan received the Nobel Prize for Physics in 1923. \footnote{On the decisive publication in the journal \textit{Science}, only Millikan was listed as the author.
He and Fletcher had contractually agreed that Fletcher would be allowed to publish another part of his research as the sole author and thus obtain a doctorate.}

The basic structure of the experiment at \textit{Treffpunkt Quantenmechanik} is very similar
to the apparatus used over 100 years ago. It all begins with the creation of small
droplets of oil using an atomizing nozzle. Many of these droplets are charged as a result of frictional electricity. The amounts of charge on the droplets are different, as are their sizes. The droplets are so small that they only sink slowly in the air, similar
to a steel ball in honey or another viscous liquid; they thus form an aerosol. The
sinking speed of the droplets is determined by the balance between the force of gravity
and the force of air friction, which depends on the radius of the droplet, its falling speed
and also on the temperature of the air.

The falling velocity of a single droplet is determined by observation with a microscope. As the air temperature can change due to the necessary illumination, it must always be measured at the same time. Of course, there must not be any air flow
in the measuring chamber that could move the droplet and thus falsify the
measurement. One of the ideas that led to the success of the original experiment was
the use of oil instead of other liquids, whose droplets evaporate far too quickly, so that
the droplets change from one moment to the next. The force of gravity is proportional
to the volume of the droplet and the force of air friction is proportional to the radius
of the droplet and its falling velocity. The size of the droplet, and thus its mass, can
then be determined from the speed of fall.

In the second step, an electric field is applied to balance the gravitational force so that
the droplet remains exactly at rest. Then the frictional force is absent and the
charge of the droplet can be determined from the balance between the gravitational
force and the electric force, because its size and thus its mass and the strength of the
electric field are known. Repetition with many droplets provides many values for the
charge of the droplets, which should all be multiples of $-e$, within the limits of measurement precision.

\section{Spectroscopy}
\label{spektroskopie}

\begin{mdframed}[style=tpq]

\begin{center} 
\textsf{Spectrum (plural: spectra), spectroscopy
}
\end{center}
A \textbf{spectrum} is the set of the wavelengths (or frequencies) of
electromagnetic radiation that can be observed in a range, together with
their respective intensities. 
If all wavelengths can occur within some range, the spectrum is called
\textbf{continuous}, e.g. in the case of thermal radiation from a hot object. If only
very specific wavelengths can occur, the spectrum is called \textbf{discrete}. This is
the case for pure chemical elements, for example. Discrete spectra are also
called \textbf{line spectra}, named after the image of colored lines on a dark
background that appears in an optical spectrometer. By examining such line
spectra in \textbf{spectral analysis}, it is possible to determine which chemical
elements the substance under investigation contains. The study of
spectra is called \textbf{spectroscopy}.

In more general terms, the possible values of the energy of a quantum system
are also referred to as the \textbf{energy spectrum}.

\end{mdframed}

Spectra and spectroscopy have already been mentioned several times in this text, for
example in Sections \ref{interferenz}  and \ref{wie_bewegt}. The \textit{Spectroscopy} experiment in \textit{Treffpunkt
Quantenmechanik} shows how an optical spectrum can be measured; this results in a line spectrum that demonstrates the quantization of energy. The technique used has
been tried and tested for over two centuries and is still used today, now of course
supplemented by automatic control, data recording and evaluation by computer.

In the experiment, the light of a lamp is analyzed in which atoms of mercury (Hg)
and cadmium (Cd) are put into excited states. A short time later, they fall back to the
ground state (the lowest energy state) and emit a photon with a certain energy. This
energy corresponds to a certain wavelength. Since there are several excited states with
different energies, the light emitted by the lamp contains several, but only very
specific wavelengths, unlike the light emitted by a glowing piece of iron, which
contains completely arbitrary wavelengths. The fact that only a few different
wavelengths are observed is an indication that the atoms in the lamp can only have
very specific different energies, i.e. that their energy is quantized. (The quantization
of energy was explained in Section \ref{wie_bewegt} using a simple example).

In order to be able to observe and measure the different wavelengths of light, they must
first be separated from each other. There are two standard techniques for this. You
can send the light through a prism, as in the \textit{Photoelectric effect} experiment, or
through a grating, as in this experiment. Prisms and gratings utilize different physical
effects. In the glass of the prism, the speed of the light depends on its wavelength. This
\textit{refraction} of the light ensures that the light leaves the prism at  different angles
depending on the wavelength. In contrast, \textit{diffraction} occurs in the grating, which is based on interference discussed in Section \ref{interferenz} .

Here, diffraction at the grating is briefly discussed once again; more details can be found in Section \ref{interferenz}, from Figure \ref{zweilambdas}. A grating is an arrangement of very many parallel slits
in an opaque screen. The light waves are allowed to hit the screen perpendicularly
and the light is observed after passing through the screen at a certain angle to the
direction of incidence. Then the waves emanating from two neighboring slits have
traveled different paths, as sketched in Figure \ref{zweiquellen}; this is called a \textit{path difference}. If the path difference is exactly one wavelength (or two wavelengths, or three...), the two waves  amplify each other. The many slits of a grating are arranged in equal distances,
 so that at the \zitateng{correct} angle, the waves from all the
slits add up. For other angles, the waves from the many slits cancel each other. However, as the \zitateng{correct} angle has a different value for each wavelength, the
different wavelengths are separated from each other by the grating. 
The relationship between angle, wavelength and slit distance is given by simple
geometry, so that the wavelength can be calculated from the angle. (If, on the other
hand, a prism is used, the exact shape of the prism and the properties of the type of glass used must be known).

In the experiment, the light to be examined is first formed into a narrow bundle
through a slit (the so-called entrance slit), then it falls onto the grating. Behind the
grating is a telescope, which is rotatably mounted so that it can be swiveled around
the grating and thus see the light that has passed through the grating at different
angles. The telescope is mounted on a ring with a scale on which you can read the
angle. Before you can do this, however, the apparatus must be precisely adjusted: The
telescope must be adjusted so that you can see the entrance slit (without the grating)
sharply and the grating must be aligned exactly so that the light really comes in at right angles.
The width of the entrance slit can be adjusted. If the slit is wide, you have a lot of light and can see it well, but the angle measurement is less accurate than with a narrow slit. It is important to find a good compromise.

Today, there are methods for producing optical gratings at very low cost (unlike
prisms); a grating with 1000 slits per mm is so cheap on the Internet that the shipping
costs are higher than the price of the grating. If you are handy and can find the
instructions on the Internet, you can then use the camera of your smartphone as a spectrometer with a cardboard housing and admire the spectrum on the screen (cell
phone spectrometer, Max Planck Institute for Plasma Physics). Of course, the device is not a quantitative precision instrument, but it can distinguish the spectra of different
light sources quite well.

\section{Detection of single photons}
\label{einzelphotonen}

This experiment in \textit{Treffpunkt Quantenmechanik} demonstrates how a simple detector
for single photons works. The detector itself is based on quantum mechanical effects,
namely the quantization of energy on the one hand and the photoelectric effect on the
other.

Both the detector and the light source for individual photons are light-emitting diodes
(LEDs). A diode is a semiconductor component that only allows current to pass in one direction. This is because it consists of two different materials. The possible energies for the electrons are quantized and differ between the two
materials. There are energy ranges (\zitateng{bands}) containing very 
many possible energies and areas that are energetically \zitateng{forbidden}, which are known as
band gaps. When a voltage is applied to the diode in the forward direction,
electrons flow and a current is generated. In a light-emitting diode, the electrons move
from a higher energy level in one material to a lower energy level in the other material
and each electron emits the energy difference in the form of a photon.

A voltage in the reverse direction, on the other hand, does not cause a current because
the electrons are then displaced from the transition region between the two materials,
and where there are no moving charges, no current can flow. However, this only
applies up to the so-called breakdown voltage. A voltage shifts the energy levels of
the two materials towards each other. (Remember that voltage is nothing else than
energy per charge.) At the breakdown voltage, the energy levels of the two materials
are shifted in such a way that a current can flow in the reverse direction.

A single photon can promote an electron from a bound state to a state of higher
energy in the transition region of the diode, so that it can move freely in the material.
As the electron does not leave the material, this is referred to as the \textit{internal
photoelectric effect}, in contrast to the \textit{external photoelectric effect} from Section \ref{photoeffekt} .
When a high voltage is applied in the reverse direction, the corresponding
electric field accelerates the electron so strongly that it   \zitateng{liberates} further electrons (by collison processes), which are then also accelerated, and so on. The
result is an avalanche-like
increasing current, which indicates that a photon has been registered. Suitable
measures can be taken to quickly slow down this avalanche so that a single photon
only results in a short current pulse. The light-emitting diode is then ready to register
the next photon.

The experiment is carried out in two steps. In the first step, the detector is exposed to
daylight or room lighting so that a large number of photons are constantly incident.
The breakdown voltage can be determined in this arrangement. In the second part, the
detector is then shielded from the external illumination so that only the light from
another LED can be incident. The intensity of the light emitted by this LED can be
controlled and calculated via the applied voltage. In a series of measurements with
different operating voltages, it is then possible to determine how the number of
photons registered per second depends on the light intensity.

\section{Compton effect}
\label{comptoneffekt}

The photoelectric effect (see Section \ref{photoeffekt}) shows that light only delivers its energy to
electrons in very specific \zitateng{portions}. Einstein explained the effect by assuming that light is made up of photons which
behave like particles when interacting with electrons. Since a moving particle not only has (kinetic) energy, but also
momentum, photons should also be able to transfer momentum. The proof of this is the
Compton effect, as already mentioned in Section \ref{wellenfunktion}. Here we will briefly
explain what the effect is and why it unfortunately \textit{cannot} be shown in the
\textit{Treffpunkt Quantenmechanik}.

The discovery of X-rays (1895) was a scientific sensation and many physicists began
to study this new phenomenon. Gradually, evidence accumulated that X-rays are a
wave phenomenon, similar to light, but with a much shorter wavelength. Many
materials, e.g. body tissue, are penetrated by X-rays. Other materials, e.g. bone or
lead, absorb them to a greater or lesser extent. However, \textit{scattering} of the rays was also
 observed, i.e. a change in direction due to the interaction with the material. This
is not unusual, as light is also scattered, e.g. sunlight on the atmosphere, and because
the short-wave light is scattered more strongly, the sky appears blue during the day.
However, there was evidence from the scattering of X-rays that the scattered
radiation had a greater wavelength than the incident radiation. This had never been
observed with visible light and was therefore unusual.

The situation was initially very unclear, as both the sources for X-rays and the
analytical equipment did not work precisely enough. It was not until 1922 that Arthur
H. Compton achieved experiments with clear results. After failing to explain the
phenomenon within the framework of classical physics, he tried Einstein's hypothesis
of light quanta or photons and was able to explain the experiments as collisions
between an electron and a photon. He received the Nobel Prize for this in 1927.

In his experiment, Compton directed X-ray beams with a specific wavelength onto
graphite (carbon) and measured the angle and wavelength of the deflected beams.
The electrons of carbon are only relatively weakly bound to the atoms, and the energy
of the incident X-ray photons is very high.
As a result, the electrons are \zitateng{effortlessly} released from the material and the scattering process is
effectively a collision between a quasi-free electron and the X-ray photon.

\begin{mdframed}[style=tpq]

\begin{center} 
\textsf{Kinetic and potential energy, classical}
\end{center}

Kinetic energy is the energy of motion. If a particle of mass $m$ moves at velocity $v$, it has the energy
$$
E_{\textrm{kin}} = \frac 12 m v^2.
$$
If the particle transfers its energy to a spring, for example, which is tensioned by the impact, then the energy is stored in the spring as potential energy $E_\textrm{pot}$. The total energy, i.e. the sum of kinetic and potential energies, is constant. This law of conservation of energy  is one of the most important principles of physics.

$E_\textrm{pot}$ often depends on a spatial coordinate $x$, in our example on the position at
which the particle is stopped by the spring. The faster it was at the
beginning, the further the spring is compressed and the greater the potential
energy. In physics, such a potential energy dependent on the location $x$ is
usually referred to as $V (x)$ for short and is also briefly called potential.
(A somewhat misleading term, as in electricity theory the potential energy
\textit{per charge} is referred to as electric potential).

The kinetic energy is often expressed by the momentum $p = mv$; then one has the relationship
$$
E_{\textrm{kin}} = \frac{p^2}{2m}.
$$
However, this relationship only applies if the particle is moving at speeds that are very small compared to the speed of light $c$. If this is not the case, the theory of
relativity must be employed, with the result
$$
E_{\textrm{kin}} = \sqrt{m^2 c^4 + p^2 c^2} - m c^2.
$$
In this text, however, the theory of relativity only becomes important at one
point, namely in the Compton effect.

\end{mdframed}

To analyze this collision process, one must consider the laws of conservation of energy
and momentum: The total energy and the total momentum must be the same before
and after the collision. What the photon releases in energy and momentum must be
absorbed by the electron. This applies to this collision in the same way as to the
collision between two billiard balls. The difference to the billiard balls is that the
photon can only move at the speed of light, whereas the electron can only move at
less than the speed of light. The energy $E$ of a photon with momentum $p$ is $E = pc$ ($c$
= speed of light). However, the kinetic energy of a particle with mass $m$ and
momentum $p$ is always less than the energy of a photon with momentum $p$. This
means that the electron cannot absorb the entire momentum and the entire energy of
the photon at the same time, so the photon cannot disappear completely\footnote{ This is different from the photoelectric effect; there, a single electron from the metal irradiated
with light absorbs the entire energy of the photon, and the momentum balance is equalized by the many atoms of the metal piece whose total mass $M$ is so big that the kinetic energy $p^2/2M$ transmitted by the photon (with momentum $p$) is neglibible.} but must
leave the place of collision in a changed form.

The photon can transfer the maximum momentum if, after the collision, it flies back in
exactly the same direction from which it came. This situation can be analyzed
relatively easily, as the electron after the collision then moves exactly in the photon's original direction
of flight. During this process, the photon loses a maximum amount of its energy, so
that its frequency  decreases (maximally) and thus its wavelength
increases (maximally) by exactly $2 \lambda_C$, where $\lambda_C = \frac h{mc} = 2.42 \cdot 10^{-12}$m is the so-called
Compton wavelength. $h$ is, as always, the Planck constant, $m$ the electron mass and $c$ the speed of light. If the photon flies in a direction other than \zitateng{backwards} after the
collision, there is also a component of the momentum perpendicular to 
 the original direction of flight, which, however, is less than the photon momentum of the backward scattering. This smaller component is compensated for by a corresponding momentum component of the electron, so that the law of conservation
of momentum remains fulfilled. Overall, the frequency of the scattered photon then
decreases less strongly and the wavelength increases less strongly; however, we will
spare ourselves the exact calculation.

Why is the Compton effect observed with X-rays but not with ordinary light? To find
out, you have to understand the wavelength ratios: visible light has wavelengths of (roughly) 400 nm to 800 nm, i.e. $4\cdot10^{-7}$m to $8\cdot10^{-7}$m \quad (1nm = $10^{-9}$m). However, the Compton wavelength is over a hundred
thousand times smaller! This means that although the effect is present, it is so tiny that
it is practically impossible to observe. In his experiment, Compton used X-rays with a wavelength of $70,8 \cdot 10^{-12}$m, so that the Compton wavelength $\lambda_C$ accounted for about 3.4\% of it and thus the change in wave length (up to $2 \lambda_C$) was easily
measurable.

This also explains why the Compton effect is not set up as an experiment in \textit{Treffpunkt Quantenmechanik}: to observe it, you need X-rays whose wavelength is comparable to the
Compton wavelength, and these should only be used by specially trained personnel
under strict safety precautions. After all, we only want to instruct you, not irradiate
you.

\chapter{Superposition explains (almost) everything.}
\label{IV}
In this section, the unclear and confusing aspects of quantum physics will be revisited
without referring to individual experiments of the \textit{Treffpunkt Quantenmechanik}  in detail. So if you are \zitateng{only} looking for information on the
experiments, you can skip this section. 
However, if you still want to shed more light on the quantum mechanical \zitateng{mysteries} and their
interconnections, you should read on. To really understand these
connections,
however, you need a bit of math, which is quite feasible and at the
same time quite exciting, as mathematics has some surprising insights to offer.

In section \ref{was_bei_messung} we have already quoted Richard Feynman's remark that the
superposition principle (a superposition of allowed states is also an allowed state) is the
\zitateng{only mystery of quantum mechanics}. If, for example
\zitateng{flying to the right} and \zitateng{flying to the left} are two possible states of a particle,
 then a superposition of these two states is also
possible. In the superposition state, the particle no longer has a definite direction of
flight. If the
direction is measured, the result is randomly either \zitateng{right} or \zitateng{left}, but after the measurement, the direction of flight is suddenly uniquely defined again. That was
a very simple example; the double-slit experiment in Section \ref{interferenz_quanten} offers even more
food for thought, and the more you think about it, the more you find that Feynman is
right. In this chapter, we want to analyze in more detail how the superposition
principle gives rise to the random character and many other properties of quantum
mechanics. To do this, we will need some mathematical tools and will get to know the
most important properties of \textbf{vectors} and \textbf{complex numbers}.

The easiest way to imagine vectors is as \zitateng{directed lines}, with a start, an end and a
specific direction in space. You can add them to each other, 
by appending the beginning of one vector to the end of the other. The result is again a
vector. You can also add a part of one vector to a multiple of the other; then you have
formed a \textit{superposition}. We will illustrate this using two-dimensional vectors.
Vectors
thus fulfill the superposition principle, just as we have seen before for wave functions,
in Sections \ref{wellenfunktion} and \ref{interferenz_quanten}. So there are similarities between vectors and wave functions;
in fact, wave functions can be completely replaced by vectors as a description of quantum mechanical states, and this has many advantages, which is why we want to get to know the very simplest vectors in more detail here.

Vectors also help to clarify which types of state changes are compatible with the
superposition principle: It must not matter whether two states are first changed in the
same way and then superimposed, or whether they are first superimposed and then
the superposition is changed. The permissible changes of vectors according to the
superposition principle are described by \textbf{matrices}, which we will also get to know
using the simplest example. We will see that \textit{the order} of several consecutive changes
(matrices)  \textit{is important}. This corresponds to the fact that measurements of different
quantities can interfere with each other, so that the final state after several measurements
depends on the order.

To see exactly how this works, we need the terms \textit{eigenvalue} and \textit{eigenvector} (or eigenstate), which were already mentioned in Chapter \ref{III}, but which we will get to know in concrete terms here using the simplest example. In the end, this leads to the
\zitateng{mysterious} uncertainty relation, which we will then not only have talked about, but which we really will have
understood insofar as we are able to calculate it and see how
 it results from the superposition principle. These simple considerations with
matrices and vectors are part of the basic framework of quantum mechanical theory;
even the thick academic textbooks on quantum mechanics have nothing more to offer,
apart from mathematical subtleties and many, many applications.

The second new mathematical concept in this chapter is the \textit{complex number}. For the
mathematical description of quantum mechanics, complex numbers are not only
practical, but unavoidable. Working with them does not pose a particular challenge
once you have become accustomed to vectors.

Just as a two-dimensional vector can be defined by its length and its direction, a
complex number can be defined by its magnitude and its \textit{phase} (which is an angle,
i.e. also a direction). In a superposition of two states, the share of each state is
defined by a complex number. The difference in the phases between the shares
of the two states is very important here, as can be seen from the simple example at the
beginning of Section \ref{interferenz}. There, the interference of two waves with the same
wavelength (and the same amplitude) was shown. The decisive factor here was how
the two waves lie relative to each other, i.e. how far the crests of one wave are shifted
in relation to the crests of the other. This is precisely expressed by the phase difference:
phase difference zero means constructive interference, phase difference $180^{\circ}$ means
destructive interference.

The wave functions for a particle in physically interesting situations look more
complicated than the simple waves at the beginning of Section \ref{interferenz}. When two such
wave functions are superimposed, the phase difference can determine whether the
particle is preferably found in a certain area of space or in another. If this phase
difference changes over time, the particle will shift its probability of location, i.e. it will \textit{move}. Thus quantum mechanics explains the motion of particles in space, which we have
already discussed in Section \ref{wie_bewegt} using traveling waves. We will then see how the two descriptions, with traveling waves and with time-dependent phase differences, relate to each other.

\section{Vectors and the superposition principle}
\label{vektoren}

Up to now, we have described the state of a quantum mechanical particle using a wave function, which provided simple information about the probability of the particle's position in space. The quantum mechanical interference phenomena were
also easy to visualize due to their similarity to the behavior of water waves. Other physical quantities
can only be extracted from the wave function with somewhat greater mathematical
effort, and some questions cannot be answered by the wave function at all.

\begin{mdframed}[style=tpq]

\begin{center} 
\textsf{Wave function}
\small
Caution: Contains higher mathematics.

\normalsize
\end{center}
The \textbf{wave function} of a quantum mechanical particle moving on the $x$ axis depends on the position $x$ and the time $t$; it is usually denoted by $\Psi(x,t)$
(read: \zitateng{Psi of $x$ and $t$}).

$\Psi(x,t)$ ) is a \textit{complex} function, i.e. it consists of two parts, $ \textsf{Re}\Psi(x,t)$ and $\textsf{Im}\Psi(x,t)$
(read: \zitateng{real part of Psi and imaginary part of Psi}). 
The parts play a similar role to the coordinates $x$ and $y$ of a point in a plane, where, for example, the distance $r$ of the point $(x, y)$ from the zero point can be calculated using Pythagoras' theorem:
$$
r^2=x^2+y^2 .
$$
In a similar way, the wave function yields the important quantity 
$$
| \Psi(x,t)|^2 = ( \textsf{Re}\Psi(x,t))^2 + (\textsf{Im}\Psi(x,t))^2
$$
(read: \zitateng{Psi absolute value squared}), which indicates how likely it is to find the particle at position $x$ at  time $t$. (More 
 precisely, $| \Psi(x,t)|^2$ is the
probability density of finding the particle).

\end{mdframed}

However, there is an alternative description of quantum mechanics that is better suited
for other aspects, such as determining the probabilities for the various possible
measurement results when other quantities than the position of the particle are
measured. This formalism uses \textbf{vectors} in an abstract space instead of wave
functions. This is much simpler than it sounds, as we will see in a moment. The two
descriptions were developed in parallel in the early days of quantum mechanics:
Schrödinger's \zitateng{wave mechanics} and Heisenberg's \zitateng{matrix mechanics}. \footnote{What matrices are and what they have to do with vectors will be explained in a moment.} However, it soon became clear that the two theories were completely
equivalent. Nevertheless, each has its own advantages for certain purposes.

 \subsection*{Ordinary vectors...}
 
 \zitateng{Ordinary} vectors are mathematical tools for describing geometry in two dimensions (e.g.
on the surface of the earth) or in three dimensions.
For example, you can describe a route by saying: \zitateng{Walk 1000m to the east and 1000m to
the north.} The destination of this route can also be reached after the instruction: \zitateng{Go 1414m to the northeast.} These two instructions are already examples of
the two common ways of describing a vector in two dimensions:
You can either specify the \textbf{components} of the vector with respect to the axes of
a coordinate system (e.g. the cardinal points), or you can specify the \textbf{magnitude}
(length) and the \textbf{direction} of the vector. You can also add vectors. 
In our example, this would be done with an additional instruction: \zitateng{Go another 500m north from the destination point.} Then you are a total of 1500m to
north and 1000m to the east by adding the second vector to the end of the first. In
component notation, the components of the sum vector are simply obtained by adding
the components of the individual summands.

\begin{figure}[h] 
\includegraphics[width=\textwidth,keepaspectratio=true]{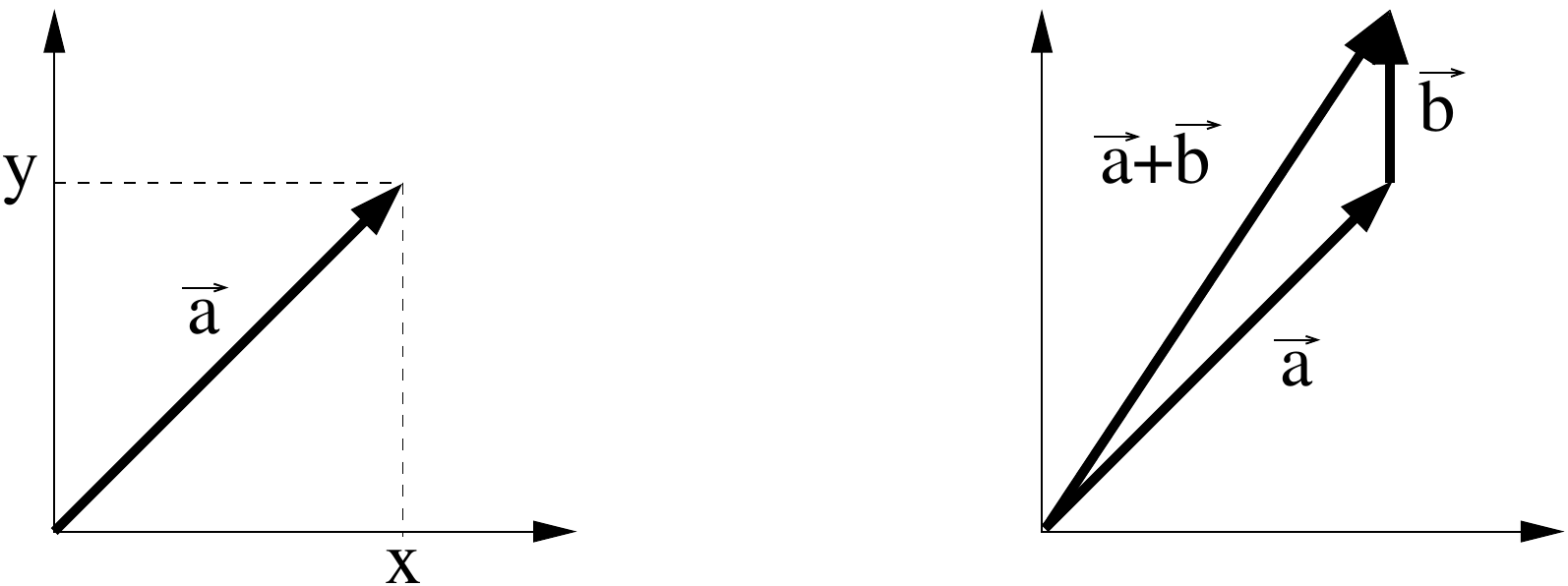}
\caption{Vectors in two dimensions can be represented as arrows and therefore a letter with an
arrow is also used as a mathematical symbol.
\newline
Left: A two-dimensional vector $\vec a$ can be specified by its components $x$ and $y$ with respect to
two axes.
\newline
Right: Two vectors $\vec a$ and $\vec b$ can be added graphically by joining the arrows  to each
other. The individual components of the two vectors can be added mathematically.
\tiny{(Graphic: Stolze}) 
}
\label{vektoren_1}
\end{figure}

There are many different ways of writing vectors; we will use the arrow notation
here: If a small arrow appears above a symbol, a vector is meant (in two or three
dimensions, depending on the case). This is how it is done in Figure \ref{vektoren_1}. For \textit{calculating}
with vectors, the notation with components is best suited. You write the components
one above the other and enclose them in
brackets; usually the $x$ component (\zitateng{east}) comes first, and below it the $y$ component
(\zitateng{north}). For our example, this yields (if we agree on meters as the unit):
$$
\vec a = \vecz{1000}{1000} \quad,\quad 
\vec b = \vecz{0}{500} \quad,\quad
\vec a + \vec b = \vecz{1000}{1500}.
$$
The notation with components immediately makes two things clear: The order of
addition is unimportant.\footnote{If you understand the vectors as walking instructions, the wrong order can lead you into a swamp, but we do not want to deal with such practical questions here.} You can also multiply vectors by numbers, which leads to an elongated or shortened arrow, depending on whether the number is greater or less than 1. In this way, \textbf{linear combinations} (\zitateng{weighted
sums}) of vectors can be constructed:
$$
1,3 \vec a = \vecz{1300}{1300} \quad,\quad
-0,4 \vec b = \vecz{0}{-200} \quad,\quad
1,3 \vec a -0,4 \vec b = \vecz{1300}{1100}.
$$
A linear combination of two vectors is a vector in the same space\footnote{The totality of all vectors is called a space or, more precisely, a vector space. The dimension of the space is
the number of components of the vectors.
}; here that is the two-dimensional plane. This is exactly the \textbf{superposition principle}: A superposition
(linear combination) of two vectors is also a vector, just as a superposition of two wave functions is again a wave function. The superposition principle (together with a few other properties) is \textit{the decisive reason} for mathematicians to call an object a vector. The notation with components, by the way, basically also describes a superposition, namely a superposition of two vectors that point exactly in the two axis directions.

We have illustrated the properties of vectors in two dimensions here. In three
dimensions, however, everything works in the same way, but involves a little more
writing and the drawings are a little more difficult. However, we are about to move
on to more abstract spaces with \textit{many} more dimensions anyway. But as always: \textit{Don't
panic!}

\subsection*{...and abstract vectors}

You can go directly from two-dimensional ordinary vectors to abstract vectors. A two-dimensional ordinary vector is a superposition of two vectors that point exactly in the
two coordinate directions $x$ and $y$ or east and north. A vector in a two-dimensional
abstract space of quantum mechanics is a superposition of two \zitateng{genuinely different}
quantum 
mechanical states, for example the ground state and the first excited 
state of an atom, and any superposition of these two states is then a two-dimensional
abstract vector that can be treated computationally in the same way as a two-
dimensional ordinary vector. Since atoms and other quantum mechanical systems often
have many more than just two different states, physicists have to get used to vectors
with many more than two components. However, all important properties of vectors
can be explained very well in two dimensions.

\subsection*{A short course in vector algebra}

We rewrite a vector with the components $x$ and $y$:
$$
\vecz xy = x \vecz 10 + y \vecz 01 = \vecz x0 + \vecz 0y.
$$
That is the decomposition into two vectors along the coordinate axes which we mentioned already. The two vectors $\vecz 10$ and $\vecz 01$ both have length one and are perpendicular to each other. \textbf{Unit vectors} with these properties are called  \textbf{basis vectors}. Such a basis can be found in any vector space, regardless of dimension. The ground state and the excited states are then the basis vectors in the
abstract space of the states of an atom\footnote{We will not discuss in detail the dimension of this space and what \zitateng{perpendicular to each other} means in this
space because it is not relevant here. But it is not really difficult.
}.

In Section \ref{wellenfunktion} , we learned about the relationship between the wave function and the
probability of a particle being present. It became clear that physically meaningful wave
functions for a particle must be \textit{normalized},
i.e. the probability of finding the particle \textit{somewhere} in space should be one. This
property should still hold if we no longer regard the states of a quantum mechanical
system as wave functions, but as vectors. The role of normalization is assumed by the
\textit{length} of the vector. The basis vectors already have a length of one, and we now demand the same from
all other (state) vectors.
 The length of a vector $\vecz xy$ in two dimensions is $\sqrt{x^2+y^2}$ and we
require that physically meaningful states fulfill
$$
x^2+y^2=1 .
$$
An obvious (and, as it turns out, correct) idea is, to interpret the number one again as the sum of probabilities for different possibilities. Since our system only has two basis states available, $x^2$ is the probability of finding the system in the basis state $\vecz 10$ on measurement, and $y^2$ is the probability of finding it in the other basis state $\vecz 01$. \footnote{For know-it-alls: $x$ and $y$ can also be complex numbers; then you have to use $|x|^2$ and $|y|^2$ as probabilities. We will deal
with complex numbers later, but for now we are making life as easy as possible.}

Physically meaningful state vectors therefore have a length of one. According to the
superposition principle, two (or more) such admissible state vectors can be
superimposed, i.e. a weighted sum can be formed. Such a superposition again
represents an admissible state if the weighted sum has a length of one.

Now quantum mechanical states, i.e. abstract vectors, can change over time, for
example through the effects of external forces or through interactions between
particles. These changes cannot be completely arbitrary, because the length of a state
vector must always remain equal to one. In the simplest case, in two dimensions, only
rotations around the zero point are possible as changes; only then is it ensured that
the components $x$ and $y$ of the vector always fulfill the condition $x^2+y^2=1$ (the
equation of a circle with radius one around the origin). Because, according to the
superposition principle, superpositions (of length one) of admissible states are
again admissible states, it must not matter whether I first superimpose two vectors
and then carry out a change to the superposition, or whether I carry out this change to
the two individual vectors and then superimpose them. \footnote{Emphatically, this does \textit{not} apply to the changes that a state undergoes as a result of a \textit{measurement}, since each measurement changes the state in a \textit{random} way, as explained in Section \ref{was_bei_messung}.} In two dimensions, this
means that it makes no difference whether I rotate two vectors by a certain angle
and then superimpose them or first form the superposition and then rotate (by the same
angle, of course).

This idea can be translated into the language of mathematics: If a
vector $\vecz xy$ develops into another vector $\vecz rs$ over time:
$$
\vecz xy \longrightarrow \vecz rs ,
$$
then the superposition principle can only be fulfilled if the \zitateng{new} components $r$ and $s$ depend \textbf{linearly} on the \zitateng{old} components $x$ and $y$, and that means
$$
r=ax+by \quad \mbox{and} \quad s=cx+dy ,
$$

with certain numbers $a, b, c$, and $d$. Mathematicians have invented a compact notation
for this type of linear transformation of vectors:
$$
\vecz rs =      \matz abcd \vecz xy   .
$$ 
The number scheme with the two rows $(a \quad b)$ and $(c \quad d)$
is called a \textbf{matrix} (plural: matrices). Since this matrix has two rows and two columns $\vecz ac$ und $\vecz bd$
it is known as a \zitateng{2 $\times$ 2 matrix} (read: \zitateng{two by two matrix}). This suggests that matrices also exist in other sizes; however, we
do not need those, as we are only interested in vectors with two components. We see that the first row of the matrix \zitateng{takes care} of the first component of the resulting vector $\vecz rs$ and the second row generates the second component. \footnote{Werner Heisenberg reinvented matrix algebra for himself when he was taking a cure on the remote North Sea island Heligoland 
for his hay fever. When he returned to Göttingen, his boss Max Born informed him that mathematicians
already knew something like this. Erwin Schrödinger is also said to have developed important ideas about wave functions during a stay at a resort.}
What we have just done is called \textbf{multiplying} a matrix by a vector.

The operation \zitateng{matrix times vector} results in a new vector. The fact that the superposition principle
 is actually fulfilled in this operation can be checked as follows:
If you multiply the matrix $\matz abcd$ with the two vectors $\vecz xy$ and $\vecz tu$ individually and add the two resulting vectors, you obtain the same result as when you multiply the matrix with the sum vector $\vecz {x+t}{y+u}$. \footnote{You are welcome to do the math to become more familiar with matrices and vectors.} As each matrix performs a mathematical operation with the vector, which represents a quantum
mechanical state, it can also be referred to as a quantum mechanical \textbf{operator}. In the
vector representation of quantum mechanics, all physical measurement quantities are
given by certain matrices.

To get used to matrices, let's do a few simple calculations; first:
$$
\matz 0110 \vecz xy = \vecz yx .
$$
This matrix effects an exchange of the two components of the vector; since we shall use this matrix again in more calculations, we introduce an abbreviation and call it $\hat{\mathbf{X}}$:
$$
\hat{\mathbf{X}} = \matz 0110 .
$$
So our calculation from above is now
$$
\hat{\mathbf X} \vecz xy = \vecz yx .
$$
We call another simple matrix $\hat{\mathbf{Z}}$:
$$
\hat{\mathbf{Z}} = \matz 100{-1} , 
$$
and this one ensures that the second component of a vector is given a minus sign:
$$
\hat{\mathbf{Z}} \vecz xy = \vecz x{-y} .
$$
These two matrices\footnote{ They belong to the Pauli matrices, named after Wolfgang Pauli, who invented them in connection
with the spin of the electron.
} are sufficient to demonstrate some central properties of quantum
mechanics.

We multiply the matrix $\hat{\mathbf{X}}$ by the vector $ \hat{\mathbf{Z}} \vecz xy $ and obtain
$$
\hat{\mathbf{X}} \left(\hat{\mathbf{Z}} \vecz xy \right) = \hat{\mathbf{X}} \vecz x{-y} = \vecz {-y}x.
$$
So far, so simple! Now we do \zitateng{the same, but the other way around}, so
$$
\hat{\mathbf{Z}} \left(\hat{\mathbf{X}} \vecz xy \right) = \hat{\mathbf{Z}} \vecz yx = \vecz {y}{-x} .
$$
So: \zitateng{First $\hat{\mathbf{Z}}$, then $\hat{\mathbf{X}}$} results in something different than \zitateng{first $\hat{\mathbf{X}}$, then $\hat{\mathbf{Z}}$}!
If
you apply several quantum mechanical operators one after the other, the sequence of
the operators is important. Quantum mechanical \textbf{operators are not interchangeable!} In more scholarly
language, we also say that they do not \textit{commute} with each other, or that they are not
\textit{commutative}.

If we go back a few steps in our argument, we can see that the properties of operators, in particular
their non-interchangeability, ultimately arise from the superposition principle.
Operators are also used in quantum mechanics to describe measurable quantities.
Each measurable physical quantity corresponds to an operator. The mathematical fact
that two different operators are as a rule not interchangeable corresponds to the
physical fact that two measurements of different quantities can disturb each other,
which leads to measurement results that are not precisely determined. This was the
subject of Section \ref{was_bei_messung} , and what was only described in words there will be explained
in the following sections with the help of matrices and vectors.

\subsection*{Eigenvectors, eigenvalues, and uncertain measurements} 

In Section \ref{was_bei_messung} , we became familiar with the terms \zitateng{eigenvalue} and \zitateng{eigenstate}: For
each operator, there are special states, which only change by a numerical
factor when the operator acts on them. Such states are called eigenstates, and
the factors are called eigenvalues. \textit{The eigenvalues of an operator are the possible
measured values of the associated physical quantity. Other values cannot show up
in a measurement.} After the measurement, the system is in the eigenstate of the
operator that belongs to the measured eigenvalue. Since we now represent states by
vectors and operators by matrices, we try to find 
\textbf{eigenvectors} to our simple operators $\hat{\mathbf{Z}}$ and $\hat{\mathbf{X}}$ and to determine
the corresponding eigenvalues.

We are doing \zitateng{small-scale kindergarten quantum mechanics}
 here: We are only
interested in what happens in a two-dimensional space that is formed, for example, by
the superpositions of the ground state and the lowest excited state of some atom.
That is already completely sufficient to understand the \zitateng{strange} quantum mechanical fact of the uncertainty relation mathematically. You see immediately that the vectors $\vecz 10$ and $\vecz 01$ are normalized eigenvectors of $\hat{\mathbf Z}$:
$$
\hat{\mathbf Z} \vecz 10 = 1 \vecz 10 \quad , \quad 
\hat{\mathbf Z} \vecz 01 = -1 \vecz 01.
$$
The eigenvector $\vecz 10$ is reproduced by the operator $\hat{\mathbf Z}$ and is multiplied by the eigenvalue 1; the other eigenvector $\vecz 01$ is also left untouched, but is multiplied by the eigenvalue -1.  There are no more normalized eigenvectors. This is
due to the fact that every two-dimensional vector is a superposition of these two
eigenvectors, and that the two eigenvalues are different from each other.

The eigenvectors of $\hat{\mathbf X}$ are not much more difficult: If you remember that $\hat{\mathbf X}$ swaps
the two components of a vector, you can see that for an eigenvector these two
components should be either equal or opposite. If you then take into account that
vectors should be normalized for quantum mechanical states, the result is
$$
\hat{\mathbf X} \vecz{\scriptstyle{1/{\sqrt 2}}}{\scriptstyle 1/{\sqrt 2}} = 1 \vecz{\scriptstyle 1/{\sqrt 2}}{\scriptstyle 1/{\sqrt 2}} \quad , \quad
\hat{\mathbf X} \vecz{\scriptstyle 1/{\sqrt 2}}{\scriptstyle -1/{\sqrt 2}} = -1\vecz{\scriptstyle 1/{\sqrt 2}}{\scriptstyle -1/{\sqrt 2}}.
$$
The two matrices (operators) $\hat{\mathbf Z}$ und $\hat{\mathbf X}$ therefore have the same eigenvalues, namely
+1 and -1, but different eigenvectors. The eigenvectors of $\hat{\mathbf X}$ are superpositions of
the eigenvectors of $\hat{\mathbf Z}$ in equal parts:
$$
\vecz{\scriptstyle 1/{\sqrt 2}}{\scriptstyle 1/{\sqrt 2}} = \frac 1{\sqrt 2} \left( \vecz 10 + \vecz 01\right) 
\; , \;
\vecz{\scriptstyle 1/{\sqrt 2}}{\scriptstyle -1/{\sqrt 2}} = \frac 1{\sqrt 2} \left( \vecz 10 - \vecz 01\right),
$$
and vice versa, the eigenvectors of $\hat{\mathbf Z}$ are superpositions of the eigenvectors of $\hat{\mathbf X}$
 in equal parts, as you can calculate yourself if you wish.
 
 \subsection*{Uncertainty relation, quite simply}

These mathematical considerations seem somewhat abstract, but they enable us to
analyze the decisive consequences they have for measurements and their
uncertainty. Let's assume that we have carried out a measurement of $\hat{\mathbf X}$\footnote{At the moment, we are not interested in which concrete physical quantity corresponds to the abstract
operator $\hat{\mathbf X}$. The two dimensions of the state space could, for example, correspond to two states of an
atom or two polarizations of a photon, but we are not interested in that at the moment because we are
thinking about \textit{very general} properties of quantum mechanics .
} and
obtained the value +1. Then we know from Section \ref{was_bei_messung} that after the measurement our
atom is in the eigenstate of $\hat{\mathbf X}$ that belongs to the eigenvalue +1, i.e. in the state
$$
\vecz{\scriptstyle 1/{\sqrt 2}}{\scriptstyle 1/{\sqrt 2}} = \frac 1{\sqrt 2} \left( \vecz 10 + \vecz 01\right) ,
$$
which contains the two eigenstates of $\hat{\mathbf Z}$
 in equal parts. A measurement of $\hat{\mathbf Z}$ in this
state will yield one of the two possible values +1 and -1 with equal probability, i.e. it
will be \textit{maximally uncertain!} The same result (maximum uncertainty of the second
measurement) is obtained if the $\hat{\mathbf X}$ measurement gives the value -1, and also if you
first measure $\hat{\mathbf Z}$ and then $\hat{\mathbf X}$. If two operators are not interchangeable, they
influence each other during measurements. This is the root of the \textbf{uncertainty relation}, and we now have a better understanding of it, or can at least understand
mathematically how it comes about.

Two operators $\hat{\mathbf A}$ and $\hat{\mathbf B}$ which are not interchangeable \textit{always} fulfill 
an uncertainty relation, that is, the product of their uncertainties (standard deviations) in
a given state is greater than a
number which results from applying first $\hat{\mathbf A}$, then $\hat{\mathbf B}$ to the state (vector) and comparing the result to that which is obtained with the reverse order of the two operators. \footnote{For all those who want to know the fine details: Apply first $\hat{\mathbf A}$, then $\hat{\mathbf B}$ to a vector and calculate the scalar product of the result with the original vector. Subtract the result of the same procedure with the other order of operators and calculate the absolute value. (The scalar product of two vectors $\vecz{x_1}{y_1}$  and $\vecz{x_2}{y_2}$ is the number $x_1 x_2 + y_1 y_2.$)}
 The greater the difference between the two
results, the greater the uncertainty product must be (at least!). In some cases (such as
with position and momentum), it turns out that this minimum value is always the same,
\textit{regardless of the state used}. Then the
uncertainty relation is  \textit{particularly robust} and thus also particularly useful for
simple estimates and calculations that apply to \textit{arbitrary states}, as in Section \ref{unschaerfe} , when we had used the uncertainty principle to justify the stability of atoms (and
thus our own existence).

\section{A little more math for know-it-alls:
Complex numbers}
\label{komplex}

Complex numbers are by no means \zitateng{complex} in the everyday sense, i.e.
complicated. They are no more difficult to understand than the two-dimensional
vectors that were described in the
previous Section \ref{vektoren} , and they help to clarify various terms and relationships that may have remained
somewhat blurred in the previous chapters.

\subsection*{Why complex numbers at all?}

We have seen again and again what a central role the superposition principle plays in
quantum physics: Every superposition of two possible wave functions of quantum
mechanics is again a possible wave function of quantum mechanics; according to
Richard Feynman, this is \zitateng{the only mystery of quantum mechanics}.

In connection with the superposition principle, terms such as \zitateng{amplitude} and \zitateng{phase shift} appeared, and it was occasionally mentioned that an amplitude can  be
complex. The wave function $\Psi(x,t)$ itself is also complex,
and all this can lead to ambiguity and confusion. With a little more precise knowledge about complex numbers,
however, the confusion can be dissolved.

\subsection*{Complex numbers are not complicated}

The only new thing about complex numbers is actually the \textbf{imaginary unit} $i$. It has the
property
$$
i^2 = -1
$$
and thus allows something that most of you previously thought was forbidden, namely
taking the square root of a negative number:
$$
\sqrt{-1}=\pm i.
$$
An \textbf{imaginary number} is any multiple of $i$, such as $42i$ or $\pi i$. A
\textbf{complex number} $z$ is simply the sum of a real and an imaginary number
$$
z=x+iy,
$$
where $x$ and $y$ are real numbers. We call $x$ the \textbf{real part} of $z$ and $y$ the \textbf{imaginary
part}:
$$
x = \textsf{Re}z \quad ; \quad y=\textsf{Im}z.
$$
The easiest way to calculate with complex numbers is to add brackets and deal with
$(x + iy)$ in the same way as you learned to deal with bracket expressions at school.
This way you can \textbf{add} two complex numbers:
$$
(p+iq)+(r+is) = p+r + i(q+s),
$$
The real part of the sum is therefore the sum of the real parts of the two numbers, in
the same way for the imaginary part. Subtraction works in the same way.
\textbf{Multiplication} takes a little more work:
$$
(p+iq)\cdot(r+is) = p\cdot r + p \cdot is + iq \cdot r + iq \cdot is
= p\cdot r - q \cdot s + i(p \cdot s +q \cdot r),
$$
The real part of the product is therefore  $pr-qs$, the imaginary part $ps + qr$, which is
not really clear at first sight.

For our purposes, it is useful to have a descriptive representation of the complex
numbers. For this purpose, the real part and the imaginary part are treated similarly to
the components of a two-dimensional vector in Section \ref{vektoren} , thus defining the
\textbf{Gaussian number plane} in which $x = \textsf{Re}z$ is plotted along the $x$-axis and $y=\textsf{Im}z$ along
the $y$-axis (Figure \ref{zahlenebene}). The two axes are also referred to as the real
and imaginary axes.

\begin{figure}[h] 
\includegraphics[width=0.5\textwidth,keepaspectratio=true]{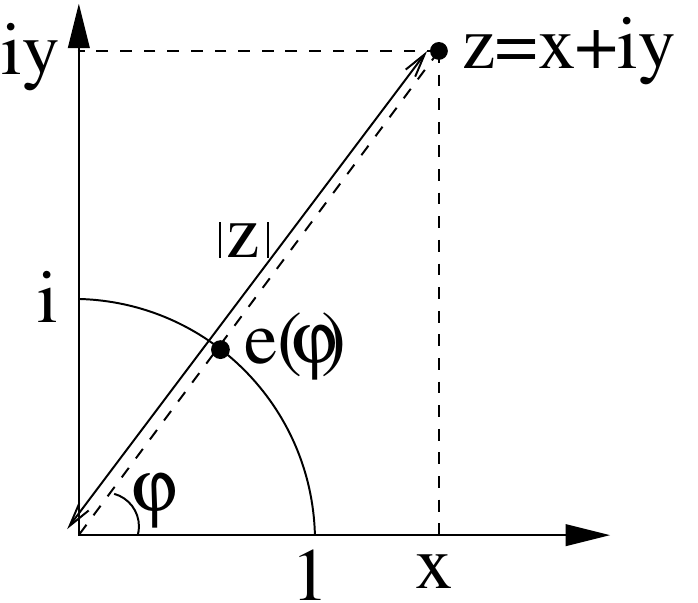}
\caption{The plane of complex numbers. Similarly to a two-dimensional vector, a complex
number $z = x + iy$ can either be represented by its \zitateng{components} $x$ (real part) and $y$ (imaginary part), or by its absolute value (length) $|z|$ and its direction (angle) $\varphi$. The complex
number $e(\varphi)$ has the absolute value one and the angle $\varphi$, i.e. it is located on the unit circle around
the zero point of the complex plane.
\tiny{(Graphic: Stolze}) 
}
\label{zahlenebene}
\end{figure}

The similarity to the representation of two-dimensional vectors in Section \ref{vektoren} is obvious. In the one case $x$ and $y$ are the components of the vector, in the other case
real and imaginary parts of the complex number, and just like a vector, a complex
number can be represented by a length and a direction. 
The length is simply the distance of the point $z = x+iy$ from the zero point of the
complex plane. This length is called $|z|$ (read \zitateng{absolute value of $z$}) and according to the theorem of Pythagoras $|z|=\sqrt{x^2+y^2}$. The direction is given by the angle $\varphi$ (Greek
letter Phi), and as will be shown shortly, it is practical for many applications to define a complex number $e(\varphi)$ that has the absolute value one and the direction $\varphi$. Using the angle
functions cosine and sine, you can write
$$
e(\varphi)= \cos \varphi + i \sin \varphi,
$$
and since the relation $(\cos \varphi)^2+(\sin \varphi)^2=1$ holds for any value of the angle $\varphi$,
it is clear that $|e(\varphi)|=1$, as desired. This means that the complex number $z$ can also
be expressed in the form
$$
z=|z| e(\varphi)
$$
write. Physicists like to call $\varphi$ the \textbf{phase} of the complex number $z$, and $e(\varphi)$ a \textbf{phase factor}. In this representation, the multiplication of two complex numbers
looks simpler, which is particularly useful when describing interference and other
superposition phenomena. If the complex number 
$z_1=|z_1| e(\varphi_1)$ 
is multiplied by
the complex number 
$z_2=|z_2| e(\varphi_2)$,
the result is\footnote{If you want to do the math, you should split
$e(\varphi_1)$ and $e(\varphi_2)$ into real and imaginary parts, multiply the bracket expressions
and use the addition theorems for sine and cosine.
The relationship $e(\varphi_1) e(\varphi_2) = e(\varphi_1 + \varphi_2)$ between the phase factors is reminiscent of a property of
the exponential function. That is no coincidence, but we do not need the connection to the exponential function here.} 
$$
z_1 z_2 = |z_1||z_2| \; e(\varphi_1) e(\varphi_2)=|z_1||z_2| \; e(\varphi_1+\varphi_2),
$$
in words: \textit{In the product of two complex numbers, the absolute values of the two factors are multiplied and the phases are added.}

However, the illustrative representation with absolute value and phase in the complex
plane is also useful for addition. To add two complex numbers, they can simply be \zitateng{joined together} in the number plane, just like two vectors in Figure \ref{vektoren_1}. If the phases of the two numbers then differ by 180$^\circ$, the numbers are exactly opposite to each other and the sum
will be smaller in absolute value than the larger (in absolute value) of the two summands. The exact
opposite is true for equal phases; then the absolute value of the sum is maximum.
This is the origin of  expressions such as \zitateng{in antiphase}, \zitateng{in phase} or \zitateng{phase difference}, which were repeatedly used in the description of interference phenomena
in Chapter \ref{II}.

Occasionally you need the number $z^*=x-iy$ corresponding to the  complex number $z = x + iy$.
One calls $z^*$ the number \textbf{complex conjugate} to $z$. Geometrically, $z^*$ arises from $z$ by
reflection on the real axis. If $z=|z|e(\varphi)$ then $z^*=|z|e(-\varphi)$, and you  can  see that
$$
z z^* = |z| |z| e(\varphi-\varphi) = |z|^2
$$
which can also be seen by just multiplying the bracket expressions $(x+iy)(x-iy)=x^2+y^2$.

 \subsection*{The superposition principle: Functions are vectors, too, after all}
 
 In Section \ref{vektoren}, we had gotten used to the idea that quantum mechanical states of a particle
can be understood not only as wave functions (wave mechanics), but also as vectors
(matrix mechanics). In both cases, the superposition principle applies, as we had
convinced ourselves: A superposition (\zitateng{weighted sum}) of two permitted states is again a permitted state. Allowed are,
as always, only normalized states; this must be ensured.

The two types of representation, with wave functions and with vectors, have different
advantages and disadvantages, depending on the question to be investigated. The wave
function provides information about the probability of the particle under
consideration being at a certain position. The representation by vectors is better suited
to determine the probabilities of measurement results for other physical quantities.
Depending on the type of physical quantity to be measured, different systems of
vectors can be useful for describing the physical processes; we have seen this in the
example of matrices $\hat{\mathbf{X}}$ and $\hat{\mathbf{Z}}$
and their eigenstates and eigenvectors in Section \ref{vektoren} (in the \zitateng{short course in vector algebra}). Changes or measurements of states are
described in different ways: Vectors can be modified by the use of matrices (hence the
name matrix mechanics), and wave functions by other mathematical operations from
differential and integral calculus. In general, we speak of \textit{operators} that act on states.

As already mentioned in Section \ref{vektoren}, in mathematics all objects are considered vectors which fulfill the superposition principle, be they \zitateng{ordinary} vectors in two or three
dimensions or wave functions.
The \zitateng{weights} in a superposition (a weighted sum) can also be complex numbers. For the example of two-dimensional vectors we have learned how to represent vectors by means of \textit{basis vectors} which are normalized, i.e. have length 1, and perpendicular to each other. That is all clear so far, but what are basis vectors in the space of functions and how in the world should \zitateng{length 1} and \zitateng{perpendicular to each other} be interpreted for functions?

We already learned about the normalization (\zitateng{length 1}) of a wave function $\Psi(x,t)$ in Section \ref{wellenfunktion} : $|\Psi(x,t)|^2$ specifies how the probability of finding 
 the particle is distributed over the $x$-axis, and since the particle \textit{mu}st be found
somewhere on the $x$-axis, the area between the curve of $|\Psi(x,t)|^2$  and the $x$ axis  must
be equal to one.

It remains to consider how functions can be perpendicular to each other. To
understand this, let's look, as a simple example, at functions of the form
$$
f(x)=a_0 \cdot 1 +a_1 \cdot x + a_2 \cdot x^2 +a_3 \cdot x^3
$$
with arbitrary numbers $a_0, a_1, a_2, a_3$. Such functions are called \textbf{polynomials of degree 3}, after the highest occurring power of $x$. A weighted sum of two polynomials
of degree 3 is again a polynomial of degree 3, so they obey the superposition
principle and form a vector space. Since four numbers $a_0, a_1, a_2, a_3$ are required to
define such a vector, the resulting vector space can be assigned the dimension 4. (To
define a two-dimensional ordinary vector, we needed two components $x$ and $y$).
Unfortunately, our vectors $f (x)$ are not normalized, because for large values of $x$, $f (x)$
grows beyond all bounds and the area enclosed by the graph of $|f(x)|^2$ and the $x$ axis becomes infinitely
large. However, this can be prevented by restricting our consideration to only a part
of the $x$ axis, e.g. from $x = -1$ to $x = +1$. Then normalization can be enforced on the four \zitateng{basic functions} $1, x, x^2$, and $x^3$ by  suitable pre-factors.
All four of them are obviously necessary to construct a polynomial of degree 3,
but they are not perpendicular to each other, if you want to use one
single notion of \zitateng{perpendicular} for all kinds
of vectors. To 
achieve this, certain combinations of these four functions must be formed,
namely the normalized versions of the functions $1, x, 3x^2-1$, and $5x^3-3x$. \footnote{The general definition is: Two vectors are perpendicular (orthogonal) to each other if their scalar
product is zero. For two ordinary vectors, the scalar product is $x_1 x_2 + y_1 y_2$; for two wave functions $\Psi_1(x,t)$ and $\Psi_2(x,t)$, it is the integral over all $x$ of $\Psi_1^*(x,t) \Psi_2(x,t)$, where the asterisk denotes the complex
conjugation. However, we do not need these details of the mathematical formalism.}

The polynomials of degree 3 on the interval between $x = - 1$ and $x = +1$ thus form a
four-dimensional vector space and the functions just mentioned are basis vectors in this
space. Of course, we can also look at polynomials of higher degree, for example of
degree 99, which then form a 100-dimensional vector space. You think that's a bit
wild? Well, even the simplest of all atoms, the hydrogen atom, has an \textit{infinite number}
of bound states that are all perpendicular to each other. The associated wave functions
therefore form a vector space with an infinite number of dimensions; that's a lot to get
used to, at least for non-mathematicians.

The superposition principle allows an infinite variety of weighted sums of any
number of wave functions, with complex numbers as coefficients (\zitateng{weights}), so that
arbitrary phase shifts between the \zitateng{components}
are possible. However, interesting effects can already be seen when considering
superpositions of only two wave functions, as we will now see in an example.

\subsection*{Phase shifts shift particles}

Figure \ref{oszillatorfunktionen} shows the wave functions of the ground state and the first excited state of a harmonic oscillator in the top image. \footnote{The harmonic oscillator is a model system used in both classical and quantum physics to describe
oscillation processes, for example the oscillations of a CO$_2$ molecule, which are responsible for the fact
that CO$_2$ has such a strong influence on the climate. A classical harmonic oscillator is simply a mass that
bounces up and down on a spiral spring or rubber band.
}. The ground state
 wave function is the one with the \zitateng{hump} near $x = 0$; we call it $\Psi_0$,
and the wave function of the first excited state we call $\Psi_1$. Both states are
stationary, so their probability densities are constant over time.

If you simply add the two wave functions together, i.e. form $\Psi_0+\Psi_1$ 
\footnote{Here and below we always refer to the \textit{normalized} sum, i.e. $(\Psi_0+\Psi_1)/\sqrt{2}$, if $\Psi_0$ and $\Psi_1$ are normalized.}
you will observe that on the left-hand side
(i.e. for negative $x$) the two functions will partially cancel each other and on the
right-hand side they will amplify each other. This can be seen particularly clearly if you
plot not the wave function itself, but its absolute value squared, i.e. the probability of finding the particle.
 This results in the black curve in the middle image. If you add the two
wave functions in antiphase (i.e. subtract them from each other), you get the red curve
for the  probability density, which shows large values on the left-hand side.

If you calculate the probability of finding the particle for the wave function $\Psi_0 + e(\varphi) \Psi_1$  and
vary the phase difference $\varphi$ between 0$^{\circ}$ und 180$^{\circ}$, you obtain images that show
how the probability density with increasing phase difference \zitateng{flows} from the right to the left side of the image. This is shown
 in the bottom image in
the figure, in which the phase difference increases in 45$^{\circ}$ steps from 0$^{\circ}$ (black) to
180$^{\circ}$
(red).

While in quantum mechanical terms one cautiously speaks of the oscillator's
probability density flowing, classically one can simply say that the oscillator is
moving. The movement is caused by the fact that the phase shift $\varphi$ in the
superposition wave function $\Psi_0 + e(\varphi) \Psi_1$  changes. In the next section, we will see
how the Schrödinger equation ensures that the phase shift changes regularly over
time.

\begin{figure}[h] 
\centerline{\includegraphics[width=0.9\textwidth,keepaspectratio=true]{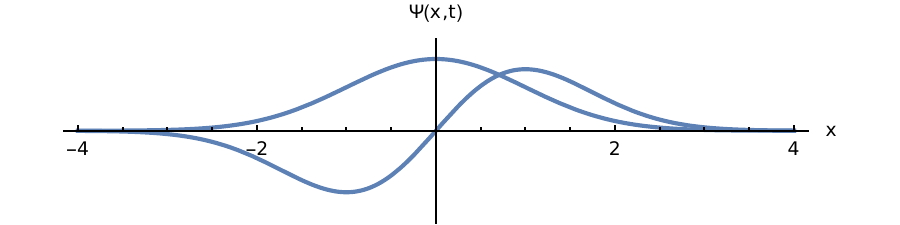}}
\centerline{\includegraphics[width=1.03\textwidth,keepaspectratio=true]{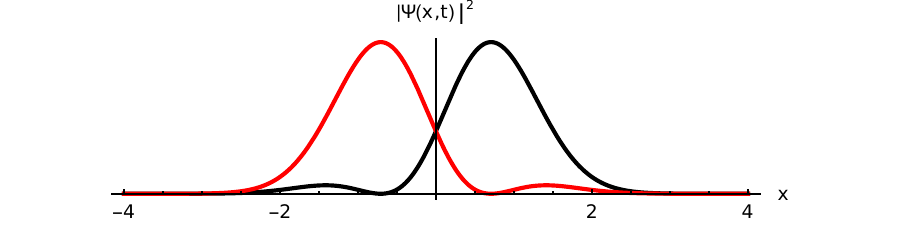}}
\centerline{\includegraphics[width=1.03\textwidth,keepaspectratio=true]{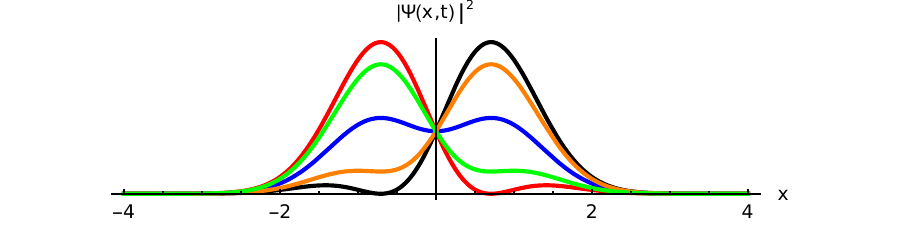}}
\caption{ The ground state wave function and the first excited wave function of a harmonic
oscillator. The ground state wave function $\Psi_0$ has a maximum at $x = 0$, the first excited wave
function $\Psi_1$ has a zero there. The top image shows these two wave functions. The
middle image shows the  probabilities for finding the oscillator for the
superpositions $\Psi_0 + \Psi_1$ (black) und $\Psi_0 - \Psi_1$
(red). In addition to these two superpositions with
phase differences 0$^{\circ}$ and 180$^{\circ}$, the bottom image also shows the  probability densities for the
phase differences 45$^{\circ}$ (orange), 90$^{\circ}$ (blue) and 135$^{\circ}$ (green). 
\tiny{(Graphic: Stolze}) 
}
\label{oszillatorfunktionen}
\end{figure}

\subsection*{Stationary states and transitions between them}

In Section \ref{wie_bewegt} , we learned about \textit{stationary states} and recognized them as particularly
useful: A stationary state has a certain fixed energy and its  probability density does
not change over time, hence the name. In certain situations (when the state is bound,
i.e. the particle is restricted to some finite region in space), only certain values of the energy are
allowed, i.e. the energy is \textit{quantized}. We used simple examples to show how this
comes about and thus understood a very important aspect of quantum physics.

Since the  probability density of a stationary state does not change over time, such a
state cannot ultimately describe any motion. Why stationary states are nevertheless
useful for describing motion is explained below.
The \zitateng{trick} is that you have to superimpose \textit{several} stationary states to observe
motion. (Remember Feynman: the only mystery is the superposition principle...)

In a stationary state, the absolute value $|\Psi(x,t)|$ of the wave
 function
 does not depend on
the time $t$ at all, but only on the position $x$. If the absolute value does not depend on the
time, only the phase can depend on it, the only question is how. It could change
arbitrarily in time, and differently at each position $x$. Fortunately, it does not; this is
what the Schrödinger equation says and what experiments also confirm.

You do not have to bother with the details of the mathematical manipulations
involved in solving the Schrödinger equation; we will simply reveal the solution
here: The phase of the wave function $\Psi(x,t)$, which belongs to a stationary state with
energy $E$, does not depend on the position and changes proportionally to time,
according to the law
$$
    \frac{\varphi}{360^{\circ}} = -\frac Eh t  ,
$$
where $h$ is Planck's constant.

To understand what this means, let us recall Figure \ref{zahlenebene} and ask ourselves what a
complex number does whose phase behaves as the equation above says. Assume that
the energy $E$ is positive; then the angle $\varphi$ continues to decrease and thus the complex
number describes a uniform circular motion in a clockwise direction. As the angles $\varphi = 0^{\circ}, \pm360^{\circ}, \pm720^{\circ},...$ are equivalent, the rotation 
 continues periodically
around the circle, even if the phase becomes ever smaller or larger. We note:
The value $\Psi(x,t)$  of the wave function of a stationary state at a fixed point $x$ in space
rotates on a circle in the complex plane at a constant angular velocity. The number of
revolutions per second is the \textit{frequency} of this periodic movement: $f= \frac Eh$. The radius
of the circle is constant and the square of the radius indicates the probability of finding the particle at point x. That probability therefore remains constant.

 So far, so uninteresting. But now we consider a \textit{superposition} of two stationary states
with \textit{different} energies $E_1$ and $E_2$. For the sake of simplicity, we superimpose the
two states at time $t = 0$ with the same amplitude and the same phase, just like the two
oscillator states in Figure \ref{oszillatorfunktionen}. However, the phases of the two states do not remain
the same, as their energies are different. As a result, the two states rotate at different
speeds in the complete number plane. This creates a phase difference between the two
states, which increases at a speed proportional to the energy difference. Then exactly
the situation occurs that is shown in the lower part of Figure \ref{oszillatorfunktionen} : As the phase difference
increases, the shape of the wave function changes and the probability of the particle's
position shifts, i.e. the particle moves.

Figure \ref{oszillatorfunktionen} (bottom image) shows how the probability of finding the particle changes when the
phase difference increases from $0^{\circ}$ to $180^{\circ}$; for the
further increase from $180^{\circ}$ to $360^{\circ}$ (i.e. effectively $0^{\circ}$), the probability  in
this example \zitateng{sloshes back} in exactly the opposite direction. 
This is due to the particularly simple structure of the wave functions chosen here; in
general, the temporal behavior of the probability density can look more
complicated, but whenever the phase difference has reached $360^{\circ}$ (or a multiple
thereof), the system is back in the initial state and the motion starts all over again.
This means: A superposition of two stationary states with energies $E_1$ and $E_2$, with $E_2 <E_1$, undergoes a periodic evolution with the frequency
$$
f=\frac 1h (E_1-E_2).
$$

We already know similar relationships between energies and frequencies from
Section \ref{wellenfunktion} : Planck and Einstein came to the conclusion that an electromagnetic
field can only absorb or release energy in the form of quanta of size $E = hf$. In Section \ref{wie_bewegt} we mentioned that electromagnetic radiation of frequency $f=\frac{\Delta E}h$
is connected to the transition between two stationary states with energy difference $\Delta E$.
For this process, we can now imagine a simple
\zitateng{classical picture}: In a superposition state,
 the probability of finding a particle  moves periodically with the
frequency $f$, and so does the charge, because the
particles in an atom are charged. However, a charge oscillating back and forth with
the frequency $f$ generates electromagnetic waves with exactly this frequency; this is
known from classical electrodynamics. However, this classical idea is too simple,
because it immediately leads to contradictions if you pursue it further: 
A \zitateng{classically oscillating} charge emits energy continuously (and not quantized), and
this leads to the negatively charged electrons ultimately colliding with the positively charged
atomic nucleus and thus all atoms are unstable. However, as shown in Section \ref{unschaerfe} , quantum
mechanics prevents this form of doomsday with the help of the uncertainty principle.

Stationary states therefore do not possess any physically meaningful dynamics
\textit{individually}, since the associated probability density does not change over time.
However, we have seen that superpositions of several stationary states enable
motion in space, since the phase differences between the states are constantly
changing and thus lead to time-dependent interferences (as in Figure \ref{oszillatorfunktionen}). The totality
of all stationary states for a given system forms a basis in the space of all possible
states of this system; that is, \textit{every} possible state of the system can be described as a
superposition of the stationary states. Depending on how many states are involved in
such a superposition, further calculations then become more or less laborious.

\subsection*{Wave packets also fit into the picture}

In this section, we have explained the motion of a particle in space with the
superposition of stationary states. In Section \ref{wie_bewegt} , we have already dealt with motions
described by \textit{wave packets}. How do these wave packets fit to the superpositions of
stationary states?

In Section \ref{wie_bewegt} , we dealt with \textit{free} particles, i.e. particles that are not subject to any
forces and can move freely throughout space. They therefore have no potential energy
and the stationary states are relatively easy to determine from the Schrödinger
equation. Since the free particles are not spatially confined, they can have any
(kinetic) energy $E$. This means that for given energy, the associated momentum $p$ of the particle is fixed
in terms of magnitude, since the kinetic energy is $E=\frac{p^2}{2m}$.
 The direction of the momentum has (in one dimension, as always with
us) only two possibilities. The wave function is a wave \footnote{ To be more precise, the wave function $\Psi(x,t)$ is  complex, with a constant absolute value, because the
probability of a free particle being present is the same everywhere. The phase $\varphi$ determines the real part
and the imaginary part of the wave function according to 
$e(\varphi)=\cos(\varphi) + i \sin(\varphi)$
We are therefore
actually dealing with \textit{two} waves, one each for the real and imaginary parts. In previous chapters, we only considered
the real part for the sake of simplicity.
} whose wavelength $\lambda$ is related to the momentum: $p=\frac h{\lambda}$. We have already learned about all these
relationships in earlier sections. The phase of the wave function changes according
to the energy, as just discussed, and the entire state undergoes a periodic movement
at each location $x$ with the frequency $f= \frac Eh$. Every second, $f$ wave crests (or troughs) therefore
pass by a fixed position, and thus the velocity of the wave is $v_{\textsf{wave}}=\lambda f$.
If we now take into account how $\lambda$ is related to $p$ and $f$ is related to $E$ and thus
also to $p$, then we obtain for the velocity
$$
v_{\textsf{wave}}= \frac p{2m}.
$$
We had already used this relationship in Section \ref{wie_bewegt} . Since the velocity of a wave
depends on the momentum $p$ and thus on the wavelength $\lambda$, wave packets change
their shape over time, as discussed in Section \ref{wie_bewegt} . The wave packets are therefore also
superpositions of stationary states. Here, too, a single stationary state cannot describe
any motion, because its probability density is constant in both space and time.

\chapter{Spins, 
quantum computers, cats, and spies}
\label{VI} 

This chapter will deal with further aspects of quantum physics that are already of great importance for technical applications and also for our daily lives or could become so in the future. 

One quantity that is very important in this context is the \textit{spin} of a particle, sometimes also called the intrinsic angular momentum. In Section \ref{was_ist_spin} we will first get to know this quantity.

Section \ref{magnetresonanz} explains how spins can be used for insights into the human body (\textit{magnetic resonance imaging}) and also for the elucidation of molecular structures in chemistry
and biology.

In Chapter \ref{IV} we saw how important superpositions of several states are in quantum mechanics, and how many phenomena can be traced back to them. This also applies to the different states that a particle with spin can assume. If several spins are connected
together, there are extremely many possibilities for superposition states that can be used to build \textit{quantum computers}. Section \ref{quantencomputer} explains why a quantum computer can
do more than a current computer based on classical physics. For example, it can \textit{break} the usual encryptions used to transmit confidential data (about credit cards, bank accounts,
...) on the internet. Fortunately, quantum mechanics can also be used to \textit{generate} secure encryption. This application of quantum mechanics is called \textit{quantum cryptography}; an experiment is currently under construction at \textit{Treffpunkt Quantenmechanik}. The simple procedure for \zitateng{key distribution} demonstrated in that experiment is explained in Section \ref{quantencomputer} .

In the last section, \ref{verschraenkung} , things become even more difficult in terms of clarity, because it deals with superpositions that lead to \textit{entanglement} between two (or more) systems.
We will then see what this is and what it has to do with Mr. Schrödinger's famous (or infamous) cat.
Several Nobel laureates in physics will appear and we will explore the question of whether quantum mechanics makes something like teleportation possible.

\section{What is a spin?}
\label{was_ist_spin} 

You may have heard of a  ball \textit{with a spin} in tennis or table tennis. The player gives the ball a rotation around its axis
which changes the trajectory and the impact on the table or floor
and thus confuses the opponent. One variant is known as \textit{top spin}; the English word \textit{spin} generally means \textit{turn} or \textit{rotation}. Physics measures how much \textit{spin} such a
ball has by its \textbf{angular momentum}; this depends on how large and how heavy the ball is, how its mass is distributed and, of course, how fast it rotates.
Systems consisting of several physical objects can also have angular
momentum, e.g. the earth and the sun in their orbit around the common center of gravity\footnote{ Since the mass of the Sun is much greater than that of the Earth, the common center of gravity lies deep
inside the Sun , and simplifyingly, one can also say that the Earth moves around the Sun.
} . Kepler's laws of planetary motion state (among other things) that the angular momentum in this system remains constant, i.e. does not change over time.

The rotational movement of the earth around its own axis also has an angular momentum that is conserved\footnote{At least, if you don't look too closely: The tides of the oceans and of the Earth's body, which is not completely rigid, cause the rotation to slow down very slowly.
} . The angular momentum not only has a magnitude, but also
a direction; it is therefore a vector. When the Earth rotates, the vector of angular
momentum points along the Earth's axis, namely from the South Pole to the North
Pole.

Just like the earth orbiting the sun, a quantum mechanical particle, e.g. an electron in an atom, can also have angular momentum due to its motion around the atomic nucleus. This is referred to as \textbf{orbital angular momentum}, even though the concept
of orbit in quantum mechanics is actually very dubious due to the uncertainty principle. Just like the earth, however, the electron also has an \textbf{intrinsic angular momentum} or \textbf{spin}, which is analogous to the earth's rotation around its own axis.
However, this descriptive image should not be overstretched, as the electron is \textit{point-like} within the limits of the measurement accuracy achievable today, and the rotation of a point \zitateng{around its own axis} is conceptually extremely problematic. 

In classical physics, the angular momentum of a particle can assume any size and point in any direction. In quantum mechanics, in contrast, the angular momentum of a particle is \textit{quantized}. The natural unit for all quantum mechanical angular momenta
is $\hbar$ (read: \zitateng{h-bar}),  the Planck constant $h$ divided by $2\pi$:
$$
\hbar = \frac h{2\pi} = 1,055 \cdot 10^{-34} \text{ Js}.
$$
This order of magnitude is so extremely much smaller than the accuracy of any everyday measuring device that, at first glance, the quantization of angular momentum is completely irrelevant for everyday life. For example, the angular momentum of the Earth's rotation is (roughly) $10^{67}$ times as large as $\hbar$, while the
angular momentum of the slow hand of a very small ladies' wristwatch is still $10^{20}$ times as large. However, we will see that the quantization of angular momentum is nevertheless important for the health of many people.

The quantization of angular momentum obeys certain rules, which we will now explain and which result from the mathematical properties of the \textit{operators} used to describe angular momentum. The angular momentum (more precisely: the orbital angular momentum) can be traced back to the physical quantities of position and momentum, and from their mathematical properties (as operators) follow those of the angular momentum. Since position and momentum are in an uncertainty relation with each other, we expect uncertainty effects for the angular momentum as well.

\begin{mdframed}[style=tpq]

\begin{center} 
\textsf{Operator}
\end{center}
Operators mathematically describe manipulations of a quantum mechanical state, for example in connection with measurements of a physical quantity. Because states admit different mathematical representations, operators can also appear in different forms. If a state is written as a wave function $\Psi(x)$,
an operator can be, for example, a derivative with respect to the variable $x$, or the multiplication with another function $f (x)$. If a state is represented as a vector, the operators are matrices. More information on vectors and matrices can be found in Section \ref{vektoren}.

\end{mdframed}

The \textit{orbital angular momentum} of a particle with respect to a center of motion can only be integer multiples of $\hbar$, i.e. $0, \hbar, 2\hbar, 3\hbar,$ etc. With spin, on the other hand,
there are two possibilities. One type of particle has integer spin (in units of $\hbar$, of course), another type has half-integer spin, i.e.$ \frac 12 \hbar, \frac 32 \hbar, \frac 52 \hbar,$ etc.
The particles with integer spin are called \textit{bosons}, the others, \textit{fermions}, as already
briefly mentioned in Section \ref{unschaerfe} .

In addition to these conditions for the magnitude of the angular momentum, there are further restrictions for the directions or the individual components of the angular momentum, which is a vector after all. One can measure \textit{one} component of the spin (or orbital angular momentum); traditionally this direction is called $z$ (and the other two $x$ and $y$, but these are usually uninteresting anyway). For a particle with integer spin, e.g. spin $2\hbar$, the z-component can have the values from $-2\hbar$ to $+2\hbar$ in $\hbar$ steps, i.e. the values $-2\hbar, -\hbar, 0, \hbar, 2\hbar$. For particles  with half-integer spin, the same rule applies, so for a fermion with spin $\frac 32 \hbar$, the $z$ component can have the values $-\frac 32 \hbar, -\frac 12 \hbar, \frac 12 \hbar, \frac 32 \hbar$.
 If the $z$ component of the spin has been measured (and thus
has \textit{exactly} the measured value), the other two components can only be measured to within a few $\hbar$  and are therefore usually uninteresting. Of course, all this also applies to
angular momenta in our classical everyday world, but because of the extreme smallness of $\hbar$, this uncertainty need not worry us any more than the position-momentum uncertainty from
Section \ref{unschaerfe}.

In the following, we will only deal with \textbf{spin-1/2 particles}. Since they have the spin $\frac 12 \hbar$ the $z$ component of the spin can assume the values $+\frac 12 \hbar$ and  $-\frac 12 \hbar$. These values define also the only two \zitateng{genuinely different} quantum mechanical states of such a particle; \footnote{This only refers to the spin of the particle; of course, apart from the spin, the particle can still move at will in space, just like the ping-pong ball with a \textit{spin}.
} all other states are superpositions of these two. Because of the either positive or negative $z$ component of the spin, the two states are labeled with arrow symbols:$\ket{\uparrow}$ and $\ket{\downarrow}$.  The $\ket{\; \;}$ 
symbol reminds us that quantum mechanical states themselves can be represented as vectors in an abstract space, as discussed
 in Section \ref{vektoren} .

The reason for this restriction to particles with spin $\frac 12 \hbar$ is that electrons as well as protons and neutrons have this property. Another reason is convenience: A quantum mechanical state space with only two dimensions is on the one hand particularly simple, but on the other hand already complicated enough to exhibit all the effects of interest to us here. All three particles, electron, proton and neutron, have a \textit{magnetic moment}, i.e. they behave like tiny magnets\footnote{We do not go into the origin of the magnetic moment in detail, as \zitateng{descriptive} classical explanations are more confusing than clarifying.
} each with a north and south pole. In an external magnetic field, these mini magnets align themselves like compass needles. The external magnetic field determines the
direction along which the spins align, i.e. the $z$ direction. The particles will try, for example\footnote{Not all of the particles will  be in this state at the same time, as the thermal motion keeps disturbing them.
} to bring their spin into the state $\ket{\uparrow}$. Anyone who has ever played with a compass needle knows that you have to apply a small force to reverse its direction; you therefore have to do work and that means adding energy. Translated into quantum mechanics, this means that one of the states $\ket{\uparrow}$ and $\ket{\downarrow}$ is energetically raised in an external magnetic
 field and the other is lowered. The energy difference is proportional to the strength of the magnetic field. This is precisely the basis of magnetic resonance, to be discussed in the next section.

\section{Magnetic resonance}
\label{magnetresonanz}

In Section \ref{wie_bewegt} we had seen that a particle can only have very specific values of energy if its motion is restricted in all directions by external forces. For the sake of simplicity, we had only considered a one-dimensional motion, but even these
considerations were not exactly easy. For a \textit{fixed} particle with spin $\frac 12 \hbar$ and a magnetic moment in a magnetic field, however, the situation is quite simple: the particle can only exist in the two states $\ket{\uparrow}$ and $\ket{\downarrow}$.  In these two states, the component of the magnetic moment in the field direction has opposite values  and the other components are undetermined; however,
they are also unimportant for the energy of the particle. There are therefore only two possible energy values. Transitions between the two states with energy difference $\Delta E$ are excited by electromagnetic oscillations with the frequency $f$ when $\Delta E = hf$\footnote{We assume here that no other forms of energy are involved, i.e. that the particle is not moving in space
and that no forces other than the magnetic field are acting.
} . If the frequency fulfills this condition, the particle can both absorb energy from the electromagnetic field and emit energy to the field. In doing so, it jumps back and forth between the states $\ket{\uparrow}$ and $\ket{\downarrow}$ and generates a measurable signal. This process is known as \textbf{magnetic resonance}.

\subsection*{The belly in the tube}

The signal of the resonance is of course the stronger the more particles are present in the measurement apparatus, and this fact is utilized in magnetic resonance imaging (MRI) to obtain images from inside the human body. To do this, the patient is placed inside a
large coil that generates the magnetic field that is responsible for the energy difference between the states $\ket{\uparrow}$ and $\ket{\downarrow}$. The particles whose spin is manipulated here are usually\footnote{ For very special purposes, other atoms are also used, which are either present in the body or are
introduced as  contrast agents.
} protons,  the atomic nuclei of the simplest of all atoms: hydrogen. Why hydrogen? The body consists largely of water, which contains hydrogen, as the formula H$_2$O says, and fatty tissue also contains a lot of hydrogen.

So now the person lies in the \zitateng{tube} that encloses the coil generating a magnetic field of e.g. 2 T (Tesla, unit of magnetic flux density). This is a fairly strong field\footnote{The stronger the field, the stronger the measured signal.}; the Earth's magnetic field (in Central Europe) is around 40,000 times weaker. At this field strength, the protons have a resonant frequency of
85.16 MHz (roughly in the range of FM radio stations, which is why we also speak of radio frequency fields). The strength of the measured resonance signal then provides information about how many protons this person contains in total - but that was not actually the aim of the investigation! You have to make sure that only the protons in a small area of the body are measured. This is achieved by not making the magnetic field equally strong everywhere, but by allowing it to increase slowly along the body from the feet to the head, for example, using additional coils. Then the magnetic field
may have exactly the right value for the resonance at 85.16 MHz in the abdominal area, and the number of protons in a thin disk in the patient's abdomen can be measured. Further metrological and mathematical tricks can be used to spatially resolve the density of the protons within the \zitateng{belly disk} and thus obtain a cross-sectional image. By changing the magnetic field the patient's entire body is scanned bit by bit. A computer compiles a three-
dimensional image of the examined area from the measured values, which is later evaluated by the examining doctor and discussed with the patient.

The examination$\ket{\uparrow}$ therefore consists of a large number of individual measurements, each with different settings of the magnetic fields involved. The rapid switching of the magnetic coils to different field strengths causes the unpleasant noise, which every patient is warned about before they are put \zitateng{into the tube}. In addition to this method, which \zitateng{only} provides a detailed image of the hydrogen distribution in the body, there are other possibilities. The proton spins can be brought \zitateng{out of equilibrium}  by short pulses of additional radio frequency fields and then one can track how quickly the spins return to equilibrium. That
enables more precise distinctions to be made between different types of tissue. However, we want to leave these details to the physicians and medical physicists.You can easily find impressive examples of MRI images on the internet, for example in the Wikipedia article on magnetic resonance imaging.

Incidentally, quantum mechanics also plays an important role in MRI  in a completely different way. The coil that generates the constant strong magnetic field (of 2 T in our example) is made of \textit{superconducting} wire. Superconductivity is the name given to the fact that some materials conduct electrical current at very low temperatures completely without resistance. A coil made of such material is energized and then the ends of the coil are connected together (superconductingly). The current then continues to flow in a circle without resistance (and therefore without energy supply) as long as the temperature remains low enough. The phenomenon of
superconductivity can only be understood with the help of quantum mechanics, but only after a lengthy journey (2-3 years of study) through the various fields of physics.

\subsection*{Is the chemistry right?}

Another important field of application for magnetic resonance is chemistry, where it is a standard method for analyzing and determining molecular structures. In addition to hydrogen, many other atoms whose atomic nuclei have a spin are of interest for chemistry. The resonance frequency of an atomic nucleus depends on which other
atoms are in its neighborhood and how they are arranged. This is known as a \textit{chemical shift}.\newpage
\begin{wrapfigure}{r}{3cm}
  \includegraphics[width=3cm]{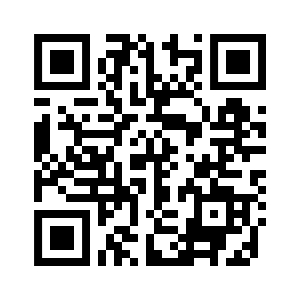}
 \end{wrapfigure}
 If two or more identical atoms are close to each other, they
can interact with each other and a change in the resonance
behavior in the sense of a splitting of the original resonance
can occur. This results in two or more resonances with
slightly different frequencies. This phenomenon is also
known from mechanics: Two pendulums of the same length
oscillate at the same frequency, but if you couple them (e.g.
with a weak rubber band) you observe two different
oscillation shapes with slightly increased/decreased
frequency. 
You can find a two minute video displaying the phenomenon at \url{https://www.youtube.com/watch?v=Wk7YSEJIGDs}.

 The usual unit for the chemical shift is ppm (parts per million). The original resonance therefore only shifts very slightly and the measurements must be correspondingly accurate. An absolute specification of resonance shifts, for example in the unit Hz, would be very impractical, by the way, because all frequencies in magnetic resonance are proportional to the strength of the magnetic field, and there are many different devices with different magnetic fields.
Magnetic resonance spectroscopy of course has been used to analyze many substances already, and their \zitateng{fingerprints} are stored in databases so that a spectrometer connected to the 
computer immediately provides the information as to whether your sample contains water, alcohol or a specific fatty acid. Less automatic is the determination of a molecular structure. Here, in the interesting cases (that are not already in the database) you have
to deal with many resonances, and you have to e.g. specifically separate certain parts of the molecule, replace them or add other parts and see how the spectra change in order to get an idea of which atoms are in close interaction with which others.

\section{Quantum computers and codes}

\label{quantencomputer}

Magnetic resonance utilizes the fact that the states $\ket{\uparrow}$ and $\ket{\downarrow}$ of a particle with spin $\frac 12 \hbar$ have
different energies  in a magnetic field and that
transitions between these states are associated with the absorption or emission of energy. Systems with only two possible states are also important in a completely different field of knowledge, namely in \textit{computer science}. All common computers work with binary digits, called \textit{bits} for short, which can exist in two states $\ket 0$ and $\ket 1$. 
(We use the $\ket{\; \;}$ symbol again here to remind you that we are not talking about the numbers 0 and 1 of something, but about two \textit{states}.) These two states can be technically realized in different ways: A switch is open or closed, a capacitor is charged or uncharged and so on. It is important that the two possible states of a bit can be easily distinguished so that the state of each individual bit is unambiguously defined at all times.

Of course, one can come up with the idea that a particle with spin $\frac 12 \hbar$ can also be used as a bit, because it also has the two states $\ket{\uparrow}$ and $\ket{\downarrow}$. Such a particle
 would  then  be a quantum mechanical bit or, in short, a \textbf{qubit}. However, there is a crucial difference
between the switch or capacitor in a conventional computer and the state of a qubit:
As we discussed especially in Chapter \ref{IV} , a quantum mechanical system can exist in a \textit{superposition} of two or more different states, unlike a conventional (classical) bit. Is this good or bad? We will see in this section that a computer with qubits offers
spectacular possibilities. There are already many different proposals for building such a \textbf{quantum computer}, all with their own peculiar technical advantages and disadvantages, which we cannot discuss here. However, the \textit{basic advantages} of a quantum computer
can also be understood without going into the technical details.

In Chapter \ref{IV} we saw that quantum mechanical states can be represented as \textit{vectors} because they obey the superposition principle: A superposition of two vectors is again a vector. For simplicity we worked with two-dimensional vectors $\vecz{x}{y} $ which could be written as superpositions of two \textit{basis vectors}:
$$
\vecz{x}{y} = x \vecz{1}{0} + y \vecz{0}{1} .
$$
Admittedly, this all looked a bit abstract, but now we have a concrete physical example of a system with two basis states, namely a particle with spin $\frac 12 \hbar$, and we identify its basis
states with the abstract two-dimensional vectors:
$$
\ket{\uparrow} = \vecz 1 0 \qquad \ket{\downarrow} = \vecz 01.
$$
This also shows us what the abstract operator defined in Chapter \ref{IV}
$$
\hat{\mathbf{Z}} = \matz 100{-1} 
$$
really does:
$$
\hat{\mathbf{Z}} \ket{\uparrow} = \matz 100{-1} \vecz 10 = +1 \vecz 10 = +1 \ket{\uparrow} \quad \mbox{and} \quad \hat{\mathbf{Z}} \ket{\downarrow} = -1 \ket{\downarrow} .
$$
The operator $\hat{\mathbf{Z}}$ therefore provides the $z$ component
 of the spin except for a factor $\frac 12 \hbar$, or in other words:
$\frac 12 \hbar \hat{\mathbf{Z}}$ is the operator for the $z$ component of the spin. It was precisely for this purpose that Wolfgang Pauli invented the matrices or operators $\hat{\mathbf{Z}}$ und $\hat{\mathbf{X}}$ and  about a century ago. You can now guess what physical quantity the operator $\hat{\mathbf{X}}$
corresponds to.

\begin{sloppypar}
Back to the qubits: If we identify our two basis states for a particle with spin $\frac 12 \hbar$ with the two basis states of a quantum mechanical bit, i.e. if we write
$$
\ket 0 = \ket{\uparrow} = \vecz 10  \quad \mbox{ and } \quad \ket 1 = \ket{\downarrow} = \vecz 01 
$$
then we can start to compare classical and quantum mechanical computers. A classical bit can \textit{either} be in state $\ket 0$ \textit{or} in state $\ket 1$. In contrast, a qubit can be in any 
superposition state
$$
\ket{\Psi} = \alpha \ket 0 + \beta \ket 1.
$$
Here, $\alpha$ and $\beta$ are complex numbers that must satisfy the normalization condition ${|\alpha|^2+|\beta|^2=1}$.
(For the mathematical terms, please refer to Chapter \ref{IV} again if necessary). There are infinitely more possibilities than the classical bit has!
\end{sloppypar}

So can a qubit store infinitely more information than a classical bit? At first glance, it can, because the two complex numbers $\alpha$ and $\beta$ contain the relative weighting between
the two parts of the state and also a phase difference between them. This can be expressed by two real numbers, and a single real number already contains an infinite amount of information in principle, as can be seen from a very well-known real number, for example:
\\ \scriptsize $\pi=3,1415926535897932384626433832795028841971693993751058209749445923078164062862...$\\ 
\normalsize
Unfortunately, the information contained in the complex numbers $\alpha$ and $\beta$  cannot be utilized completely, because you have to carry out a \textit{measurement} on the state $\ket{\Psi}$, and measurements change states and (usually) have several possible results. In view of these uncertainties, one can indeed wonder whether a qubit can reliably store
information \textit{at all}.
However, it can be mathematically proven that a single qubit can
store twice as much information as a classical bit. We will not go into the not-so-simple details here.

What is more interesting about the superposition
state $\ket{\Psi}$ is the fact that the qubit is in \zitateng{two states at the same time}. If you manipulate such a state in a specific way, for example, by applying certain external forces, you basically do not do just \textit{one} experiment, but \textit{two} experiments in parallel, starting from the two qubit states $\ket 0$ and $\ket 1$, respectively. But this can still be greatly improved: If you consider two  qubits together, then there are already four possibilities for their state, namely $\ket 0 \ket 0$, $\ket 1 \ket 0$, $\ket 0 \ket 1$, and $\ket 1 \ket 1$, with three qubits there are eight possibilities,... and with 10 qubits there are  1024 possibilities which can then be treated in parallel. This is known as \textbf{quantum parallelism}, which enables  a quantum computer to
solve certain problems (not all!) much faster than a classical computer.

At first glance, however, it seems very complicated to create a superposition of very many states. In fact, it does not take 1024 steps to produce a superposition of the 1024
possible states of 10 qubits, but \textit{much fewer}:\\
$\bullet$ Steps 1 to 10: Bring each of the 10 qubits into the state $\ket 0$, respectively $\ket{\uparrow}$. Result:
$$
\ket{\uparrow} \ket{\uparrow} \ket{\uparrow}...\ket{\uparrow}.
$$
$\bullet$ Steps 11 to 20:
To each of the 10 qubits, apply the operation 
$$\frac 1{\sqrt 2} \left(\hat{\mathbf{Z}}+  \hat{\mathbf{X}}\right) = \frac 1{\sqrt 2} 
\left( \matz 100{-1} + \matz 0110 \right) = \frac 1{\sqrt 2} \matz 111{-1}$$ 
and the result is then for each qubit
$$
\frac 1{\sqrt 2} \left(\hat{\mathbf{Z}}+  \hat{\mathbf{X}} \right) \ket 0 = \frac 1{\sqrt 2} \left( \ket 0 + \ket 1\right)
$$
(please do the math!) and in total we have
$$
\left( \frac 1{\sqrt 2} \right)^{10} \left( \ket 0 + \ket 1\right)\left( \ket 0 + \ket 1\right)\left( \ket 0 + \ket 1\right)...\left( \ket 0 + \ket 1\right).
$$
$\bullet$ Done! If you multiply and expand the 10 bracket expressions $\left( \ket 0 + \ket 1\right)$ (which you neither
have to nor should do), you get the superposition of the 1024
 different possible  combinations of the basis states of 10 qubits; the factors $\frac 1{\sqrt 2}$ take care of
the normalization of the state.

With this state, the quantum computer can then perform 1024 different calculations at once. However, it must be remembered that the 1024 possibilities are stored here \textit{in a single state}, and that one or more \textit{measurements} must be performed at the end in order to find out the result of the calculation. However, since measurements usually change the state, only relatively little information can be extracted from the state without destroying it. This makes it clear what kind of problems are particularly suitable for quantum computers: \textit{Many possibilities must}
be investigated in parallel, but \textit{only a few results} are required.

An example of this is the search in an \zitateng{unstructured database}. To make it a little more concrete, imagine you want to use a telephone directory with two million entries to find out who lives at \textit{42 Douglas Adams Street}. The phone book is obviously not properly structured for this kind of search, and you will have to look at
an average of one million entries before you find the right one. With a quantum computer, you can use the \textit{Grover algorithm} invented in 1996 which requires only about 1000 steps instead of a million. (More generally, the algorithm requires approximately  $\sqrt N$ steps for $N$ entries). Unfortunately we cannot discuss the technical details
of the algorithm here.

A second example concerns global properties of a mathematical function
$f (x)$, where $x$ is a natural number (i.e. 1,2,3,. ....) and the result $f(x)$ is a natural number too. Then, for example, you can ask the question: \zitateng{Is f (x) always positive for even x?} Another conceivable situation is the following: It is known that $f (x)$ is \textit{periodic}, i.e. that the function values start repeating at some point: $f(N+1)=f(1)$, $f(N+2)=f(2)$, $f(N+3)=f(3)$,
and so on and so forth, but the number $N$ (the period) is \textit{not known} and is to be searched for. This is another case for the quantum computer and is part of the \textit{Shor algorithm} from 1994, which has caused a great deal of excitement, and has also prompted secret services and the military to
support research into quantum computers. We will see why that is so in the next section.

\subsection*{Quanta, bankers, and spies}

If someone orders some product or books a trip on the Internet, reputable providers will point out that the data transmitted (address, bank details, credit card number,...) will be \zitateng{securely encrypted}. In the usual procedure the key is encrypted with a \textit{public key} that is available to everyone,
who wants to transmit their data. However, this public key can only \textit{encrypt} data; \textit{decryption} requires the
associated \textit{private} key, which only the recipient has.

The crucial point here is the use of \textbf{prime numbers}. As a reminder: A prime number is a natural number that is only divisible by 1 and by itself. The first prime numbers are 2,3,5,7,11,13,17,19,23,29. You can already see that this sequence looks irregular, and you cannot calculate prime numbers systematically. You can break down any natural number into prime factors, for example $42=2\cdot3\cdot7$, but this  becomes 
 difficult if the numbers are larger and do not have several easily determinable prime factors, such as 2, 3 or 5.
As an example, you may determine the prime factors of 29083, but please do so without electronic aids! 

The public keys use a very long number, namely a product of two very large prime numbers, to encrypt the data, but decryption requires the two prime factors, which remain secret because only the recipient of the data knows them. The security lies in the fact that all known methods for prime factorization require exponentially increasing
time as the numbers become longer. Therefore, if a credit card fraudster captures the public key, he   can start one of the known prime factor decomposition methods on his computer, but it is more than doubtful whether he will get the result in his lifetime.

However, this only applies to classical computers. The quantum algorithm for searching for prime factors published by Peter Shor in 1994 uses some mathematical methods that we cannot go into in detail here; however, the core piece is in fact a certain mathematical function $f (x)$ ($x$ a natural number) that depends on the (very large) number to be decomposed and has a (very large) period that can be determined comparatively quickly by a quantum computer. Formulas can be used to indicate how the number of steps of the algorithm increases with the length of the number to be decomposed; however, these formulas are rather opaque for laypersons. We will therefore do a numerical example and compare the Shor algorithm with the best
known classical algorithm. Let's assume that the classical method takes one second for 50 digit numbers and the quantum method takes one hour, because classical computers are very advanced today and quantum computers are not yet. For 300 digits, both
methods take about 60 hours, and for 2000 digits, the quantum computer is done within a year, while the classical computer takes about as long as the universe has existed so far: just under 14 billion years.

The fact that the Shor algorithm works was demonstrated in 2001. The example was
$15 = 3 \cdot 5$. Of course, that leaves some room for improvement. But at the latest when there are quantum computers that can deal with really large numbers, bankers and merchants will have to come up with something to replace the encryptions that have been used up to now.

In fact, there are already new encryption methods based on quantum physics. A simple method for a special task, which was developed back in 1984 by Charles Bennett and Gilles Brassard and became known as the \zitateng{BB84 protocol} will be briefly described here because it is particularly simple. The \textit{Quantum cryptography} experiment currently under construction at \textit{Treffpunkt Quantenmechanik} demonstrates this process.

The task is \textbf{key distribution}, which must first be explained. A demonstrably secure method of communication or data transmission is the use of a \textbf{one-time key} (or one-time pad). A key is a list of translation rules for the symbols to be transmitted, such as
\begin{center}
$a \to 6$\\
$f \to g$\\
$k \to m$\\
$w \to q$\\
$0 \to z$\\
.....\\
\end{center}
These translation rules are processed one after the other, i.e. the letter \zitateng{a} is replaced by the number \zitateng{6} and so on, \textit{but only once}; the next \zitateng{a} is replaced by a different symbol (from the next line of the key starting with \zitateng{a}). The key must therefore be at least as long as the
message to be transmitted. Security is guaranteed by the fact that each line of the key
is used only once; otherwise you could search for the most frequently used symbol and assume that it belongs to the most frequent letter, i.e. the \zitateng{e} in German texts. In this procedure, both communication partners (and no one else) must of course possess the key.
In earlier times, secret agents were therefore given code books to take with them, which they were not supposed to lose under any circumstances.

Today, two partners can establish a common key using quantum mechanics, provided they can exchange \textit{photons}. As in all discussions on communication by quantum mechanical means, the partners shall be called \textit{Alice} and \textit{Bob}, and sometimes there is also \textit{Eve}, who takes her name from the English word \zitateng{to eavesdrop}.
Alice and Bob use single photons, and in 
particular their polarization, to transmit information. 
If you no longer know exactly what polarization is all about, you can read again in Section \ref{was_bei_messung} what it says about the experiment \textit{Polarization of light} in the \textit{Treffpunkt Quantenmechanik}; however, we will briefly repeat the most important facts here.

\begin{mdframed}[style=tpq]

\begin{center} 
\textsf{Polarization of light}
\end{center}
Electromagnetic waves consist of coupled electric and magnetic fields that change in space and time. In the case of a monochromatic (only one frequency or wavelength) plane wave in free space, the electric and magnetic fields are perpendicular to each other and both are perpendicular to the direction in which the light propagates. If the electric field only oscillates back and forth in one direction, it is referred to as a \textbf{linearly polarized} wave. If you look in the direction of wave propagation, you can
distinguish between two fundamentally different polarizations (directions of oscillation): \zitateng{right and left} ($x$ polarization) and \zitateng{up and down} ($y$ polarization). All
more complicated waveforms can be composed from these basic forms.

\end{mdframed}

The info box describes the \textit{classical} view of polarization: 
If light is polarized in an \zitateng{oblique} direction, then the electric field is a superposition of fields in the $x$ direction ($\rightarrow$) and $y$-direction ($\uparrow$). A polarization
filter in the $\rightarrow$ direction then blocks the field component in the $\uparrow$ direction, and only
the field component in the $\rightarrow$ direction is allowed through.

In the \textit{quantum mechanical} view, photons have a two-dimensional state space whose basis states correspond to the directions $\rightarrow$ and $\uparrow$. The state of a photon
  can  be an arbitrary superposition
of these two basis states. The amplitudes (\zitateng{weights}) of the two basis states determine the probabilities with which the photon is allowed to pass by polarization filters in the directions $\rightarrow$ and $\uparrow$.
For example, a photon with polarization $\nearrow$ is in an
\zitateng{equal weight} superposition state of $\uparrow$  and $\rightarrow$, and is therefore transmitted by
a $\rightarrow$ polarization filter with 50\% probability.

Alternatively, one can also use the states with the polarizations $\nearrow$ and $\nwarrow$ as basis  states;  then a $\rightarrow$ photon is an equally weighted superposition of these two alternative basis states and is allowed through by a $\nearrow$ polarization filter with 50\% probability.

At a previously agreed time and in a previously agreed rhythm, Alice sends photons towards Bob, randomly changing polarizations
between the variants \zitateng{straight} and \zitateng{skewed}.
 With \zitateng{straight} she (randomly) sends a photon with polarization $\uparrow$ or $\rightarrow$, with \zitateng{skewed} she uses the polarizations $\nearrow$ or $\nwarrow$.
 For each photon, Alice notes down the polarization with which she sent it. Bob does not yet know at this point which polarization Alice is currently using and randomly changes the polarization filter in front of his detector between $\uparrow$ and $\nearrow$.
 He logs the changing settings of his filter and his measurement results. If Alice sends a $\uparrow$  photon and Bob uses the $\uparrow$ filter, then he is sure to measure a photon. If she sends a $\rightarrow$ photon, then with this filter setting he will equally certainly  measure no photon.
However, if she sends a \zitateng{skewed} polarized photon, then Bob is equally likely to measure a photon or no photon\footnote{At this point, the \textit{Quantum cryptography} demonstration experiment (under construction) in the \textit{Treffpunkt Quantenmechanik} \zitateng{cheats}, since for technical reasons (\euro \euro \euro) it does not really work with individual photons, but with short light pulses, each of which contains many photons. In the version described, half of the photons are allowed through and a random generator then \zitateng{corrects} the result to \zitateng{all allowed through} or \zitateng{nothing allowed through}. The quantum mechanical randomness is therefore only simulated here.
}. So if Alice's polarization directions do not match Bob's filter,
Bob will not get a clear result and these measurement results are therefore useless.However, Bob does not yet know which results are unambiguous and which are not; this will only become clear in the next step.

After finishing the transmission, Alice and Bob communicate, e.g. by telephone, about their settings ( \zitateng{straight} or \zitateng{skewed}) for each individual photon and discard all
measurements that cannot have an unambiguous result, because the used
settings did not match. For the remaining measurements, Bob knows for sure what Alice has sent, and these measurement results form the key. Of course, Alice also knows what she has sent and what could be clearly measured by Bob. Both therefore have a sequence of zeros (no photon measured) and ones, and such a bit sequence can
be used to encrypt another bit sequence (the actual message).\footnote{To do this, you can \zitateng{add the two bit sequences bit by bit without carryover}, using the symbol $\oplus$, according to the rules $0 \oplus 0=0, \; \; 0 \oplus 1 = 1$ and $1 \oplus 1=0$. 
For the message $00111$ and the key $01001$ that yields the encrypted text  $00111 \oplus 01001 = 01110$. If you add to this text the key  
$01001$ once again (using the $\oplus$ operation) you obtain the message  $00111$ back, as you can work out for yourself.
}

It is no big deal if the spy Eve taps the phone line and finds out that photon no. 137 has been measured unambiguously; this information is of no use to her, because she does not know the result of the measurement and the photon no longer exists. However, during the transmission of photons, Eve
can try to measure each of Alice's photons in a similar way to Bob
and for the photon destroyed during the measurement to pass on a \zitateng{faked} photon to Bob, which corresponds to the result of her measurement. However, since Eve in half of the cases uses 
the wrong filter, i.e. not the one used by Alice, her \zitateng{fakes} will end up not matching Alice's original in
a quarter of the cases. Alice and Bob can easily notice that
by sacrificing a small part of their secure key in order to compare their data (again, e.g. via the telephone line, which does not have to be tap-proof). If then about a
quarter of the data does not match, they can be sure that Eve has been eavesdropping and have to make a new key.

\subsection*{What's the point?}

Quantum computers, or more generally, quantum information and quantum
communication, therefore offer interesting possibilities. However, quantum computers only offer an advantage for very specific problems, which in those cases is enormous. For word processing or games, on the other hand, they are rather useless. Quantum information (whether  stored in qubits or exchanged via quantum communication) has several \zitateng{faces}. It is very \textit{sensitive}, as it is easily damaged by uncontrolled interaction with the environment of the qubit or can be lost in the event of inappropriate measurements. However, it is also very \textit{secure}, as it cannot be copied. (One example was Eve's inability to falsify the photons sent by Alice). It is also very \textit{powerful}, as quantum computers can solve problems that no classical computer can handle in any reasonable amount of time.

Many groups of physicists, engineers, and computer scientists are working on the technical realization of quantum computers (which cannot be covered here) in academic institutions and in the research and development departments of large companies such as IBM, Google, and Microsoft. There are many very different approaches to what a quantum processor could look like; particles with spin $\frac 12 \hbar$ are by no means the only suitable \zitateng{hardware}. On the other hand, many scientists are working on quantum computer software for a wide variety of applications.
Areas of application range from solving large systems of equations and calculations for 
chemistry to risk analysis in finance.

\section{Entanglement, cats, and teleportation}

\label{verschraenkung}

Some words of  warning are appropriate at the beginning of this section. If you gradually get the feeling while reading that you are a little confused and have not fully understood everything, then rest assured: the majority of physicists were also confused for
several decades as soon as the topics covered here were discussed. That's why these topics were often put aside
as they were \zitateng{not too important for practical  purposes}. However, this changed at the latest when practical applications emerged. We will start here with a drastic example, which is also discussed again and again (but not always correctly):
Schrödinger's cat. Since we have familiarized ourselves with spins, state vectors and other non-everyday concepts in the previous chapters, we can also explain what the serious background for Schrödinger's strange example was. The concept of \textit{entanglement}, which the cat example is intended to explain, can be treated mathematically in a simple way, and with the help of the mathematical description, interesting applications can be analyzed. The future will show just how important or
\zitateng{practical} these applications are; quantum mechanics is still a key topic and 
still good for surprises after more than a hundred years! Two Nobel Prizes in physics (2012 and 2022) are closely related to the content of this section.

\subsection*{Schrödinger's cat}

In 1935, Erwin Schrödinger published a series of articles under the title \zitateng{The current situation in quantum mechanics}, in which he described a thought experiment.
The example was intended to illustrate the strange properties of the then still new theory. A radioactive atom has two possible states: decayed or not decayed. Imagine a small amount of a radioactive element, so small that a decay occurs every hour with a
50\% probability, which is registered by a Geiger counter. The Geiger counter then triggers a mechanism that kills a cat. The cat is enclosed in a locked box together with the entire apparatus, and the box is to be opened again after one hour.

At that time either an atom has decayed and the cat is dead, or no atom has decayed and the cat is still alive. That much is clear; but what about the cat as long as the box is still closed? Is there a superposition of two states here, namely \zitateng{cat alive, atom still whole}, and on the other hand \zitateng{cat dead, atom decayed}? At the end, does the measurement (opening the box) randomly decide what \zitateng{really} happens? 
Schrödinger was an experienced university lecturer and was certainly aware of this drastic example being certain to encourage his readers to reflect and discuss. And they did, and continue to do so today.

Unfortunately, we cannot discuss the example to the end here, as it is far too complicated in its original form. Even the radioactive substance consists of many atoms, and each of them can be intact or decayed. The cat is, of course, even more complex when viewed as a quantum physical system, as it has astronomically many more and vastly more complicated states than the simple alternatives \zitateng{alive} or \zitateng{dead}. All these states occur in a superposition, in
combination with all possible states of the radioactive substance.

The thought experiment with the cat is quite often described by saying, for example, that the cat is both dead and alive at the same time. However, this is wrong already, because if you take quantum mechanics seriously (and you have to, otherwise you can leave this
discussion altogether), then the cat does not have a defined state \textit{on its own}. The state of the cat is inextricably linked to the state of the radioactive substance. Schrödinger
used the term \textbf{entanglement} (in German: Verschränkung) for this. The total system \zitateng{cat plus radioactive substance} has a very specific quantum mechanical state, and this is a superposition of \zitateng{cat alive, all atoms still whole} and \zitateng{cat dead, one atom decayed}.

In contrast to thought experiments with complicated systems (such as cats), one can try to carry out real experiments with simple systems; this is still complicated enough. In 1996, a group led by David Wineland in Boulder (Colorado) succeeded in exactly recreating the Schrödinger cat scenario using a single beryllium ion.\footnote{ An ion is an atom with different numbers of electrons and protons; here there are 4 protons in the atomic nucleus, surrounded by only 3 electrons.
}
 The ion was held in place with electromagnetic fields and could either rest or oscillate back and forth around the resting position. We will call these two possible states $R$ (for rest) and $S$ (for oscillation; German: Schwingung). Irrespective of this, the electrons of the ion could assume many different states. The experimenters  chose two of them, let's call them $A$ and $B$. By skillful irradiation with lasers (i.e. also electromagnetic
fields), it was possible to \textit{entangle} the \zitateng{internal} electron state of the ion  with its motion in space, i.e. to form a superposition of $(A, R)$ with $(B, S)$. (Important: The two other possible combinations $(A, S)$ and $(B, R)$ are \textit{not} present). In this experimental set-up, the state of motion (resting
or vibrating) of the ion alone was undetermined, as was its electron state; however, the overall state of the combined system was uniquely defined.

An even more convincing demonstration was also shown in 1996 by a group led by Serge Haroche in Paris. There, an atom was placed in a superposition of two different excited states before passing through a chamber containing an electromagnetic field. The atom changed the electromagnetic field as it passed through the chamber, depending on its state, which, however did not change in the process. When the atom had left the chamber, the overall system of atom plus field was in a state in which atom and field were  in an entangled state of coexistence, but neither of them was in a definite state \textit{by itself}. This demonstration was particularly convincing because it entangled two physically different and spatially separate subsystems. David Wineland and Serge Haroche shared the 2012 Nobel Prize for Physics. Conclusion: Schrödinger's cat really does
exist, but it doesn't look quite the way Schrödinger imagined it at the time. This also pleases the animal rights activists.

\subsection*{Vectors for cats}

Of course, Schrödinger did not come up with the pitiful cat just for fun, but because he wanted to respond to a contribution to the theoretical discussion by three colleagues: Einstein, Podolsky and Rosen, hereinafter referred to as EPR. The three of them were of the
opinion that quantum mechanics is not a \textit{complete} theory and explained this in an article in a scientific journal using a thought experiment, which we are treating here in a somewhat simplified form, and in which two particles with spin $\frac 12 \hbar$ occur. These two particles, let's call them $A$ and $B$, are prepared in a specific (entangled) state, which is a superposition of two different possibilities:
$$
\ket{\Psi}_{\text{EPR}} = \frac 1{\sqrt 2} \left( \vphantom{\frac 12} \ket{\uparrow}_A \ket{\downarrow}_B +  \ket{\downarrow}_A \ket{\uparrow}_B \right).
$$
It is therefore clear: \textit{If} particle $A$ is in state $\ket{\uparrow}$, \textit{then} particle $B$ is in state $\ket{\downarrow}$, and \textit{if}  $B$ is in state $\ket{\uparrow}$, \textit{then}  $A$ is in state $\ket{\downarrow}$. Neither of the two particles is in a particular state by itself,
but the \textit{overall state} of both particles is uniquely determined. If a measurement of the state is made on one of the two particles, the result is not unique, but the result of the measurement on one particle determines the result of a later measurement on the other particle: If the measurement for particle $A$ determines the state $\ket{\uparrow}$ , then all that
remains of the initial state $\ket{\Psi}_{\text{EPR}}$ of the overall system is the part
$$
\ket{\uparrow}_A \ket{\downarrow}_B,
$$
 according to what we discussed in Section \ref{was_bei_messung} . But then it is clear that a
measurement on particle $B$ will \textit{with certainty} yield the state $\ket{\downarrow}$.

Now imagine that particle $A$ is sent to physicist Alice \textit{before the measurement}, i.e. when the two particles are still in the  initial state $\ket{\Psi}_{\text{EPR}}$, and particle $B$ is sent to physicist Bob, with the two being very far apart. Neither of them knows what a measurement will yield on the particle that they now have in their respective labs, but once Alice has measured her particle, she knows for sure what state Bob's particle is in, no matter how far away she is from him!

Einstein referred to this consequence of the EPR thought experiment as \textit{spooky action at a distance} and was not prepared to accept a theory that would have allowed such a phenomenon.
Schrödinger used the word \zitateng{Verschränkung}  for the way in which the two particles are joined together. The term comes from furniture construction and refers to the connection of two pieces of wood where the teeth of one piece engage in slots in the other piece. In English, this phenomenon of quantum mechanics is called \textit{entanglement}. To show \textit{how} strange this consequence of quantum mechanics is, Schrödinger then came up with the example of the cat. He himself described entanglement as \textit{the} central feature in which classical physics and quantum physics differ. Neither Schrödinger nor Einstein were able to come to terms with this phenomenon, which clearly results from the mathematical structure of quantum mechanics.

\subsection*{John Bell and the hidden variables}

One conceivable way out, which many tried to use to calm themselves down, was to think that they simply hadn't looked closely enough yet, that there existed mechanisms that \zitateng{in reality} control the apparent randomness and already determine at the beginning of the experiment that Alice will measure the state $\ket{\uparrow}$  and
Bob then just the state $\ket{\downarrow}$, and that 
 the physicists involved are simply too clumsy to
discover and measure this mechanism. This is the idea of so-called \zitateng{hidden variables}.

A non-quantum mechanical example of a hidden variable acting would be the following situation: You randomly select two people from a large crowd as test subjects and are stunned to discover that both are tall, slim, red-haired and have green eyes. The hidden
variables were then the identical genes, because you picked two identical twins...

The idea of hidden variables served to calm people's minds, even if nobody had any idea what these variables might be. For about 30 years, basic physics research had plenty of other problems to solve and the questions surrounding entanglement were considered unimportant for practical purposes, until John Bell took up the matter
again in the 1960s. Bell took the idea of hidden variables seriously and investigated their practical consequences by calculating how certain measurement quantities are statistically distributed when a large number of measurements are performed on pairs of particles in the EPR state. The assumption was that there is indeed \textit{some} quantity that determines what Alice and Bob will measure for each such EPR pair, and that this quantity obeys \textit{some} probability distribution. Bell was then able to prove that certain combinations of measurements could not exceed certain numerical values, regardless
of the probability distribution of the hidden variables; for example, a certain quantity can have a maximum value of 2. This statement is referred to as \textit{Bell's inequality}. However, if quantum mechanics is correct, then the same quantity can be up to $2 \sqrt 2=2,828..$. This is a significant effect and not just a deviation in the fourth decimal place. If Bell's inequality is clearly violated, i.e. the quantity investigated by Bell is clearly greater than 2, then this shows two things: entanglement is a real phenomenon, and it cannot be explained by hidden variables.

When Bell published his ideas, the experiments he proposed were not yet technically feasible. It took several years before precise measurements became possible. In these experiments, however, no pairs of particles with spin $\frac 12 \hbar$  were used, but pairs of photons,  which can also assume two fundamentally different states ($\uparrow$ and $\rightarrow$) due to their polarization. In certain physical processes,  photons  are always produced in pairs and with entangled polarizations, which is ideal for these experiments. In the 1970s, John Clauser and coworkers (USA) showed that Bell's inequality is clearly violated, but there were still a few \zitateng{loopholes} through which the theory of hidden variables could  have \zitateng{saved} itself. However, these loopholes were closed by the experiments of Alain Aspect and coworkers (France), so that the phenomenon of entanglement has been proven beyond doubt and the existence of hidden variables is at the same time excluded beyond doubt. A series of \textit{tricks} can be performed with entangled phonons, which also have practical applications; Anton Zeilinger and coworkers ( Austria) have demonstrated this in many experiments. Aspect, Clauser and
Zeilinger were jointly awarded the Nobel Prize\footnote{www.nobelprize.org provides explanations of the Nobel Prizes on both elementary and more advanced levels.} for physics in 2022.

\subsection*{Tricks with entanglement: Key distribution}

Entanglement can be used to encrypt messages; more specifically, to create a one-time key that can be used later to send and receive actual encrypted messages.

We have already seen what a one-time key is in Section \ref{quantencomputer} : A translation table
between plaintext letters and encrypted symbols, from which each line may be used only once to prevent decryption. The construction of such a table was possible by Alice sending a number of single photons with randomly chosen polarization to Bob, which Bob could measure with randomly rotated polarization filters. Alice and Bob could then compare for which photons the transmitter polarization and the receiver polarization matched. In these cases, Bob's measurement results were unambiguous and could be used as a key, in the other cases they were random, i.e. not useful. It was possible to determine whether the spy Eve had listened in.

With pairs of entangled photons, a somewhat trickier system for quantum key distribution can be constructed, which has additional safety features. Again the polarization of the photons is used, and again different polarization directions are measured. As a reminder: If a photon is polarized in the $x$ direction, it is passed
through a polarization filter aligned in the $x$ direction and registered by a detector behind it. If, on the other hand, the photon is polarized in the $y$ direction, it is definitely not transmitted and therefore not registered. In both cases, the result is
unambiguous. If, on the other hand, the polarization filter 
is rotated by 45$^\circ$, i.e. aligned along the bisector between the $x$ and $y$ axes, it will be registered or not, with 50\% probability each, irrespective of whether it was polarized in  the $x$ direction or in the $y$ direction.

In Section \ref{quantencomputer}, Alice and Bob both used either the \zitateng{0$^\circ$ rotated} (i.e. the original, non-rotated) axis system for polarization, or the one rotated by 45$^\circ$. Both acted 
 randomly and independently of each other. In 50\% of the cases, both
had used the same system and thus obtained useful results. The quantum key distribution with single photons that we had discussed was based on this. In the system employing \textit{pairs} of photons, things get a bit more complicated. Both Alice and Bob
act as receivers here and measure single photons with polarization filters. Alice randomly uses one of three axis systems with rotation angles 0$^\circ$ , 45$^\circ$ and 22.5$^\circ$ ; Bob randomly
chooses 0$^\circ$ , +22.5$^\circ$ and -22.5$^\circ$.

A source emits photon pairs in a very specific state, namely
$$
\frac 1{\sqrt 2} \left( \vphantom{\frac 12} \ket{\rightarrow}_A \ket{\uparrow}_B -  \ket{\uparrow}_A \ket{\rightarrow}_B \right).
$$
One of the photons is sent to Alice, the other one to Bob. In the formula above $A$ or $B$ denote who was sent the photon, the arrows denote the polarization and the symbol $\ket{\phantom{x}}$ reminds us that we are dealing with quantum mechanical states here. This state has a property which needs some getting used to but which is useful: It \zitateng{looks the same in every system of axes},i.e. it is \textit{identical} to the state
$$
\frac 1{\sqrt 2} \left( \vphantom{\frac 12} \ket{\nearrow}_A \ket{\nwarrow}_B -  \ket{\nwarrow}_A \ket{\nearrow}_B \right).
$$

This means the following: If Bob measures with a polarization filter rotated by 45$^\circ$ and registers a photon, then this photon \textit{has} the polarization  $\nearrow$ and Alice's photon  \textit{must} have the polarization $\nwarrow$. If, on the other
hand, Bob does not measure a photon, then it was obviously polarized in the direction $\nwarrow$ and therefore Alice's photon \textit{must} have the polarization $\nearrow$. This works not only for the angle 45$^\circ$ , but for any other angle\footnote{Well-practiced quantum mechanists can do the math to verify this in a few minutes; we'll save ourselves the trouble here. The important thing is: This only works with exactly this state, with the minus sign between the two
parts.
}. If you look at it closely, the information only \zitateng{arises} as soon as Alice or Bob have measured their respective photon; before that time the photon pair contains no information.

After Alice and Bob have measured a previously agreed number of photons, they inform each other which axis system they used for the measurement. They can do this publicly, because the photons are gone and can no longer be measured. In the cases in which both have used the same axis system, they have unambiguous measurement results (which, of course, they do not disclose publicly), and these measurement results form the key. Since both of them choose 3 different axis systems at random, there are 9 combinations, 2 of which are useful (both 0$^\circ$ or both 22.5$^\circ$). This means that only 2/9 = 22.22\% of the photon pairs can be used for the key, whereas with the
method with single photons it was 50\%; the additional security of the new method comes at a price. The three combinations in which the two axis systems form an angle of
45$^\circ$, provide completely random measurement results and therefore help no one. From the four other combinations, with angles $\pm 22.5^\circ$ or 67.5$^\circ$ it is possible to determine precisely  that combination of measured quantities, which for the state used reaches the value $2\sqrt 2$ which \zitateng{maximally violates} Bell's inequality. These measurements can be used to check security:
 If they show a significantly smaller value
than $2\sqrt 2$, then Eve has intercepted photons, passed on faked photons, and thus falsified the original state. \textit{Small} deviations from the maximum value of $2\sqrt 2$ can, however, also occur without espionage due to inaccuracies in the
transmission or measurement errors.

This method, which was proposed by Artur Ekert in 1991, has since been tested several times in practice, with pioneering work being carried out by Anton Zeilinger's group, e.g. during the transmission between the Canary Islands of Tenerife and La Palma via 144km distance (2006).

\subsection*{Tricks with entanglement: entanglement swapping}

In order to transmit photons over 144 km, high mountains (due to the curvature of the earth's surface) and very clear weather conditions are required, which unfortunately cannot be relied upon.
 Fiber optic cables, which are already available for \textit{classical} data transmission in the Internet and telephone system, offer an alternative. 
However, optical fibers also suffer from losses, so that a photon may not even reach the end of the cable. This is not a problem with classic data transmission. There, each signal consists of a large number of photons, so that one more or less does not matter.
If, after a longer distance, the signal no longer contains many photons, it is measured in an amplifier (\zitateng{repeater}) and an amplified copy is forwarded. However, this is \textit{not possible} for individual photons: You cannot copy the state of a photon (or qubit)\footnote{ It is often said that a qubit is not copied but cloned, and the impossibility of producing a copy is called
the \textit{no-cloning theorem}.
} unless you know that it is \textit{either} in
the state $\ket{\uparrow}$ \textit{or} in the state $\ket{\rightarrow}$; but in that case  the situation is exactly the same as with a classical bit and such a restricted qubit offers no advantages.

However, it is possible to entangle two photons (or other qubits) that do \textit{not} originate from the same source. This requires another entangled pair of photons, whose entanglement can then be transferred to the other two photons. This process is referred to as \textbf{entanglement swapping}. This idea offers a possibility to build \zitateng{quantum repeaters} that allow the transmission of quantum states over longer distances. The first experiments on entanglement swapping were carried out in 1998 in Anton Zeilinger's group; we explain the simplest example here.

To do this, we need a little mathematical preparation. We remember that a single photon has a two-dimensional state space, with the basis vectors\footnote{ If you can't remember what that was, please refer to Section \ref{vektoren} .
} $\ket{\rightarrow}$ and $\ket{\uparrow}$.
 A general  state  is then a superposition of these two basis states. Here we are dealing with \textit{pairs} of photons and have to be able to describe their states. To this end, we write e.g.$\ket{\uparrow \rightarrow}_{12}$ and mean \zitateng{photon 1 is
in the state $\ket{\uparrow}$ and photon 2 is in the state $\ket{\rightarrow}$}. The photon pair has the four basis states
$\ket{\uparrow \uparrow}_{12}, \ket{\rightarrow \rightarrow}_{12}, \ket{\uparrow \rightarrow}_{12}, \ket{\rightarrow \uparrow}_{12}$, and a general state is a superposition of these four basis states.

For our example it is, however, more appropriate to choose a different set of four basis vectors, namely the two states
$$
\ket{= \pm}_{12} = \frac 1{\sqrt 2} \left( \vphantom{\frac 12} \ket{\rightarrow \rightarrow}_{12} \pm \ket{\uparrow \uparrow}_{12} \right),
$$
where the states of the two photons are 
\textit{equal} ($=$), and the two states 
$$
\ket{\neq \pm}_{12} = \frac 1{\sqrt 2} \left( \vphantom{\frac 12} \ket{\rightarrow \uparrow}_{12} \pm \ket{\uparrow \rightarrow}_{12} \right),
$$
in which the states of the two photons are \textit{different} ($\neq$). In the discussion of the EPR experiment and Bell's inequality, we already encountered the state $\ket{\neq +}$, for the situation with two spin $\frac 12$ particles  instead  of two photons. The state $\ket{\neq -}$ was already used in the the quantum key distribution scheme with photon pairs. These four basis states are often referred to as \textbf{Bell states} or 
EPR states; a device that can generate such pairs is called an \zitateng{EPR source}.

For the entanglement swapping we now need two entangled photon pairs,
 (1,2) and (3,4), both in the state $\ket{\neq -}$. These pairs are created by two EPR sources. Alice and Bob are very far away from each other; between them is the \zitateng{intermediary} Charlie. One EPR source is located between Alice and Charlie, the second one between Charlie and Bob. Photon 1 is sent to Alice, photons 2 and 3
go to Charlie, Photon 4 to Bob.

Who is now entangled with whom? As before, 1 and 2 as well as 3 and 4 are entangled with each other, so that, for example, a measurement on 1 has an influence on 2, but measurements on 1 or 2 have no influence on the state of 3 or 4; there is therefore no entanglement between the two pairs. However, this changes when Charlie performs a certain type of measurement on the two photons that arrive at his place.
This measurement enforces an entanglement between photons 2 and 3 (which are destroyed during the measurement) and thus between the two photons 1, for Alice, and 4, for Bob, which previously had nothing to do with each other.

Charlie has a measuring device for photon pairs that can determine in which of the four Bell states $\ket{= \pm}_{23}$ and $\ket{\neq \pm}_{23}$ is the photon pair that he receives. The pair
(1,4) that Alice and Bob share is, after Charlie's measurement, \textit{in exactly the state that Charlie measured on his pair (2,3)}. When Charlie tells Alice and Bob what he has measured, they know which of the four Bell states they share.

The fact that the states of the two pairs (1,4) and (2,3) (the latter no longer exists) are \textit{the same} is, of course, an astounding statement that you can either simply believe or, much better, prove, i.e. do the math. If you find this inconvenient, you can simply skip the following calculation (about one page). Anyone who hangs in there will at least have understood a small piece of the work that was awarded the Nobel Prize.

As already mentioned, we start with the photon pairs (1,2) and (3,4), both in the state $\ket{\neq -}$ and write this state in a little more detailed way by multiplying the bracket expressions:
\begin{multline*}
\ket{\neq -}_{12} \ket{\neq -}_{34} = \frac 12 \left( \vphantom{\frac 12} 
 \ket{\rightarrow \uparrow}_{12} - \ket{\uparrow \rightarrow}_{12} \right)
 \left( \vphantom{\frac 12} 
 \ket{\rightarrow \uparrow}_{34} - \ket{\uparrow \rightarrow}_{34} \right) =\\
  \frac 12 \left[ \vphantom{\frac 12} \ket{\rightarrow \uparrow}_{12} \ket{\rightarrow \uparrow}_{34}
 -\ket{\rightarrow \uparrow}_{12} \ket{\uparrow \rightarrow}_{34}
 -\ket{\uparrow \rightarrow}_{12} \ket{\rightarrow \uparrow}_{34} 
 +\ket{\uparrow \rightarrow}_{12}\ket{\uparrow \rightarrow}_{34} \right] .
\end{multline*}
Now we rewrite the last expression so that the two photons at Charlie's are combined into one pair and the other two (at Alice's and Bob's) into a second pair. This is just a slight rearrangement; \textit{nothing} changes physically or mathematically:
\begin{multline*}
\ket{\neq -}_{12} \ket{\neq -}_{34} = \\ \frac 12 \left[ \vphantom{\frac 12}
\ket{\rightarrow \uparrow}_{14} \ket{\uparrow \rightarrow}_{23} 
   -\ket{\rightarrow \rightarrow}_{14} \ket{\uparrow \uparrow}_{23}
   -\ket{\uparrow \uparrow}_{14} \ket{\rightarrow \rightarrow}_{23}
   +\ket{\uparrow \rightarrow}_{14} \ket{\rightarrow \uparrow}_{23} \right].
\end{multline*}
The statement we wanted to prove was the following: If Charlie enforces one of the four Bell states by his measurement on his photon pair (2,3), then the photon pair (1,4) is also in this Bell state. It therefore makes sense to look at states in which those two photon pairs are in the same Bell state and to check whether there is anything similar to the state we have just constructed. We start with
\begin{multline*} 
\ket{\neq \pm}_{14} \ket{\neq \pm}_{23} = \\
\frac 12 
\left( \vphantom{\frac 12} \ket{\rightarrow \uparrow}_{14} \pm \ket{\uparrow \rightarrow}_{14} \right)
\left( \vphantom{\frac 12} \ket{\rightarrow \uparrow}_{23} \pm \ket{\uparrow \rightarrow}_{23} \right) =\\
\frac 12 \left[ \vphantom{\frac 12}
\ket{\rightarrow \uparrow}_{14} \ket{\rightarrow \uparrow}_{23} \pm
\ket{\rightarrow \uparrow}_{14} \ket{\uparrow \rightarrow}_{23} 
\pm \ket{\uparrow \rightarrow}_{14} \ket{\rightarrow \uparrow}_{23}
+ \ket{\uparrow \rightarrow}_{14} \ket{\uparrow \rightarrow}_{23}
\right].
\end{multline*}
We recognize the two terms with the $\pm$ sign in our rewritten
initial state $\ket{\neq -}_{12} \ket{\neq -}_{34}$.
 We subtract from each other the two states just calculated
so that only the terms with the $\pm$ sign remain and are left with:
$$
\ket{\neq +}_{14} \ket{\neq +}_{23}- \ket{\neq -}_{14} \ket{\neq -}_{23}
\quad = \quad \ket{\rightarrow \uparrow}_{14} \ket{\uparrow \rightarrow}_{23} 
+\ket{\uparrow \rightarrow}_{14} \ket{\rightarrow \uparrow}_{23}.
$$
We have thus already checked one half of the initial state; the \textit{other half} must have some connection to the Bell states in which the two photons are equally polarized; we therefore write
\begin{multline*} 
\ket{= \pm}_{14} \ket{= \pm}_{23} = \\
\frac 12 
\left( \vphantom{\frac 12} \ket{\rightarrow \rightarrow}_{14} \pm \ket{\uparrow \uparrow}_{14} \right)
\left( \vphantom{\frac 12} \ket{\rightarrow \rightarrow}_{23} \pm \ket{\uparrow \uparrow}_{23} \right) =\\
\frac 12 \left[ \vphantom{\frac 12}
\ket{\rightarrow \rightarrow}_{14} \ket{\rightarrow \rightarrow}_{23}
  \pm \ket{\rightarrow \rightarrow}_{14} \ket{\uparrow \uparrow}_{23}
  \pm \ket{\uparrow \uparrow}_{14} \ket{\rightarrow \rightarrow}_{23}
  + \ket{\uparrow \uparrow}_{14} \ket{\uparrow \uparrow}_{23}
\right].
\end{multline*}
Again, it is the two terms with the $\pm$ sign that we also see in the initial state, both
with the sign \zitateng{$-$}. So we have
$$
\ket{=-}_{14}\ket{=-}_{23} - \ket{=+}_{14}\ket{=+}_{23} \quad = \quad - \ket{\rightarrow \rightarrow}_{14} \ket{\uparrow \uparrow}_{23}
  - \ket{\uparrow \uparrow}_{14} \ket{\rightarrow \rightarrow}_{23}.
$$
\textit{That's it}: The initial state of our two photon pairs can in total be written as follows:
\begin{multline*}
\ket{\neq -}_{12} \ket{\neq -}_{34} =\\ \frac 12 \left[ \vphantom{\frac 12} 
\ket{\neq +}_{14} \ket{\neq +}_{23}- \ket{\neq -}_{14} \ket{\neq -}_{23}
+ \ket{=-}_{14}\ket{=-}_{23} - \ket{=+}_{14}\ket{=+}_{23} \right].
\end{multline*}
So the expression in the large square brackets is actually an equal-weight superposition of states in which the two photon pairs (1,4) and (2,3) are both in the same Bell state. If Charlie now measures \textit{which} of the four possible Bell states his pair (2,3) is in, he will get a result; let's say he measures $\ket{=+}_{23}$.
The state of the overall system then collapses to $\ket{=+}_{14} \ket{=+}_{23}$, and Alice and Bob share the photon pair (1,4) in the state $\ket{=+}_{14}$. However, Charlie has
to tell them the result of his measurement so that they know this.

In this way, an entangled pair can be distributed over a greater distance than the range of a single EPR source, but a second EPR source and Charlie as an intermediate station are required. One can imagine continuing the procedure with a suitable network of further EPR sources and intermediate stations and thus distributing an
entangled pair over even greater distances. It is also possible to use other initial conditions than in our example; additional steps in the procedure may then be necessary.

At first glance, however, the procedure does not appear to be suitable for a reliable key distribution, because Charlie (or one of the many Charlies at a greater distance) could be a spy, and Charlie also transmits information to Alice and Bob that should actually remain secret, namely which Bell state the two share. If this information is intercepted, the whole operation makes no sense.

\subsection*{Tricks with entanglement: Teleportation}

When we hear the word \zitateng{teleportation}, everyone probably thinks of scenes from science fiction films in which someone climbs into a glass cabin, slowly \zitateng{fades away} and reappears in another solar system at the next moment. We are
more modest here and understand teleportation as follows: The quantum mechanical state of a single qubit is transferred\footnote{Not \textit{copied}, this is not possible according to the no-cloning theorem.}
to another qubit without one of the two
qubits itself being moved.

Once we have understood how this works, we can consider how much more difficult it is to turn the science fiction movie scenario into reality. The teleportation of individual qubits was first proposed theoretically in 1993 and confirmed experimentally in 1997,
among others by Anton Zeilinger's research group.

Our usual suspects Alice and Bob enter again. Alice has a qubit in a quantum state\footnote{Where Alice  got this state from is not of interest to us here. What is important is that Alice does not \textit{know} the state, for example because she produced it herself, otherwise she could simply tell Bob the numbers $\alpha$ and $\beta$ as \zitateng{building instructions}.}
$$
\ket{\Psi} = \alpha \ket{0} + \beta \ket 1 ,
$$
furthermore, Alice and Bob have supplied themselves with a qubit pair in one of the Bell states, each of them keeping one of those qubits:
$$
\ket{=+} = \frac 1{\sqrt 2} \left( \vphantom{ \frac 12} \ket{00} + \ket{11} \right) .
$$
Similar to the beginning of Section \ref{quantencomputer}, we here  identify the two basis states for a
particle with spin $\frac 12 \hbar$ with the two basis states of a qubit or the basis states in the $xy$ plane:
$$
\ket 0 = \ket{\uparrow} = \vecz 10  \quad \mbox{ and } \quad \ket 1 = \ket{\downarrow} = \vecz 01 .
$$
Before it gets a bit mathematical again, here is a description of teleportation without math, in three simple steps:\\
1) Alice entangles her two qubits using certain mathematical operations, whereby Bob's qubit is also influenced.\\
2) Alice carries out a measurement on her two qubits, as a result of which the overall state of all three qubits \zitateng{collapses}, i.e. is no longer a superposition of several possibilities,
but a unique single state that is related to Alice's initial state $\ket{\Psi}$, but Bob doesn't know the relation exactly yet.\\
 3) Alice tells Bob the result of her measurement; Bob needs this information to
reproduce the state $\ket{\Psi}$ \textit{exactly} at his place.\\
If you don't want to know more, you can stop reading now.

We number the qubits: The state $\ket{\Psi}$, which is to be teleported, is in qubit 1, qubit 2 is owned by Alice and qubit 3 by Bob; 2 and 3 are in the Bell state $\ket{=+}$. To follow what is going on we write down the initial state in its full glory:
$$
\ket{\Psi} \ket{=+} = 
 \frac 1{\sqrt 2} \left[\alpha \ket 0 \left( \vphantom{ \frac 12}  \ket{00} + \ket{11} \right)
 + \beta \ket 1 \left( \vphantom{ \frac 12}  \ket{00} + \ket{11} \right) \right].
$$
The first thing Alice does with her two qubits is to perform an operation that is usually called CNOT(1,2), which means \zitateng{controlled not} (conditional negation). This
operation acts on qubit 2 depending on the state of qubit 1: If 1 is in state $\ket 0$, 2 is not changed; however, if 1 is in state $\ket 1$,  2 is inverted, i.e. brought from $\ket 0$ to $\ket 1$, and vice versa. That yields the following result:
$$
\text{CNOT}(1,2) \ket{\Psi} \ket{=+} = 
 \frac 1{\sqrt 2} \left[\alpha \ket 0 \left( \vphantom{ \frac 12}  \ket{00} + \ket{11} \right)
 + \beta \ket 1 \left( \vphantom{ \frac 12}  \ket{10} + \ket{01} \right) \right].
$$
We have already briefly familiarized ourselves with Alice's next operation in Section \ref{quantencomputer} ,but we will repeat the individual components once again. The operation $\hat{\mathbf X}$ is
the negation; it transforms $\ket{\uparrow}$ into $\ket{\downarrow}$ and vice versa, or $\ket 0$ into $\ket 1$ and vice versa.
The operation $\hat{\mathbf Z}$ adds a minus sign to $\ket 1$ or $\ket{\downarrow}$ and does nothing else. We combine the two
operations and see that the following happens:
$$
\frac 1{\sqrt 2} \left( \hat{\mathbf X} + \hat{\mathbf Z} \right) \ket 0
= \frac 1{\sqrt 2} \left( \ket 1 + \ket 0 \right)
\qquad
\frac 1{\sqrt 2} \left(  \hat{\mathbf X} + \hat{\mathbf Z} \right) \ket 1
= \frac 1{\sqrt 2} \left( \ket 0 - \ket 1 \right).
$$
Alice now applies this operation to qubit 1, which is indicated by an attached subscript \zitateng{1}:
\begin{multline*}
\frac 1{\sqrt 2} \left( \hat{\mathbf X}_1 + \hat{\mathbf Z}_1 \right) \text{CNOT}(1,2) \ket{\Psi} \ket{=+} = \\
\frac 12 \left[\alpha (\ket 0+\ket 1) \left( \vphantom{ \frac 12}  \ket{00} + \ket{11} \right)
 + \beta (\ket 0 -\ket 1) \left( \vphantom{ \frac 12}  \ket{10} + \ket{01} \right) \right]=\\
 \frac 12 \left[\alpha \ket{000} + \alpha\ket{100}  +\alpha \ket{011} +\alpha \ket {111} + \beta \ket{010} - \beta \ket{110} + \beta \ket{001}  - \beta \ket{101}  \right].
\end{multline*}
In the last step, all brackets were multiplied out and in all of the resulting terms all three qubits were written down, of course in the order 1,2,3. Alice's qubits 1 and 2 appear in all four 
possible combinations: 00, 01, 10 and 11, each time once with an $\alpha$ and once with a $\beta$. This can be rearranged differently again:
\begin{multline*}
\frac 1{\sqrt 2} \left( \hat{\mathbf X}_1 + \hat{\mathbf Z}_1 \right) \text{CNOT}(1,2) \ket{\Psi} \ket{=+} = \\
\frac 12 \left[ \ket{00} \left( \alpha \ket 0 + \beta \ket 1  \right) +
                \ket{01} \left( \alpha \ket 1 + \beta \ket 0  \right) +
                \ket{10} \left( \alpha \ket 0 - \beta \ket 1  \right) +
                \ket{11} \left( \alpha \ket 1 - \beta \ket 0  \right) \right] .
\end{multline*}
Alice can now carry out the second step and measure her two qubits. If she measures the values (0,0), the superposition of four states collapses and only the first part remains.
Bob then has the state $\alpha \ket 0 + \beta \ket 1 = \ket{\Psi}$, exactly the state that Alice started with. The state has been teleported to Bob, and Alice only has  the  rather uninteresting state $\ket{00}$ in her hand.

If Alice measures the combination (0,1), Bob has the state $\ket{\Psi}$  with reversed roles of $\ket 0$ und $\ket 1$.
This can be easily fixed by Bob applying the operation $\hat{\mathbf X}$ to his qubit. With the measurement result (1,0), in Bob's state the component $\ket 1$ has the wrong sign, which can be corrected
 with a $\hat{\mathbf Z}$. For the last possible result (1,1),
Bob must first apply $\hat{\mathbf X}$ and then correct the sign with $\hat{\mathbf Z}$ -- in exactly this order.

To summarize: Alice transmits the two measurement results, i.e. (classic) bits, to Bob. If the second bit has the value 1, Bob must apply $\hat{\mathbf X}$. He then looks at the first
bit, and if this has the value 1, he applies $\hat{\mathbf Z}$. In the end, he has transferred the exact state $\ket{\Psi}$ to his qubit.

 \subsection*{\zitateng{Beam me up, Scotty}?}
 
 It is therefore actually possible, and comparatively simple if you have a well-equipped laboratory, to transfer the state of \textit{a single} qubit at one location to  another qubit at a different location. However, this requires an additional auxiliary pair of qubits in a Bell state, one qubit at the start and one at the destination of the teleportation.
Message transmission between the two locations must be possible, and quantum mechanical operations and measurements must be possible. All this was verified in experiments.

 So can Captain Kirk now simply give his chief engineer Scotty on the \textit{Enterprise} the order
\zitateng{Beam me up, Scotty!} and off we go to the Andromeda Nebula?  -- For this to happen, someone would have to produce a pair of auxiliary qubits in a Bell
state for each of Kirk's qubits and deposit one each on the \textit{Enterprise} and in the Andromeda Nebula. (Providing the same number of qubit pairs again for the return trip would also be a good idea!)

 However, Kirk does not consist of qubits that have exactly two different states, but of atoms, and even a single atom without interaction with other atoms has an \textit{infinite}
number of bound (i.e. stable) states. We have not even considered the fact that Kirk's atoms have to interact with each other in a very finely tuned way for him to feel comfortable and perform heroic deeds. The sheer number of atoms or molecules does not make it any easier: a man of 75kg mass contains about 45kg of water, that is
roughly $1.5 \cdot 10^{27}$ water molecules alone, and we are not even talking about the other components of a human body.

All this looks rather difficult...

\appendix

\chapter{Glossary}
\begin{mdframed}[style=tpq]

\begin{center} 
\textsf{Amplitude}
\end{center}

The \textbf{amplitude} is the maximum displacement of a wave, e.g. in the case of a
water wave, the height of the wave crest above the calm water surface. In
addition to the amplitude, the \textbf{phase} of a wave is also important, as it
determines, for example, exactly where the wave crests are located.
When waves are superimposed, both the amplitudes and the phases of the
partial waves are important, keyword interference.

In the case of a superposition of several quantum mechanical wave
functions, one also speaks of the amplitudes of the different wave functions
and thus means the respective components (expressed as complex
numbers).

\end{mdframed}
\begin{mdframed}[style=tpq]

\begin{center} 
\textsf{Classical physics}
\end{center}

The state of knowledge of physics around 1900 is termed classical: mechanics obeys the principles found by Newton, extended by Einstein's special theory of relativity. Electrical and magnetic phenomena are fully described by Maxwell's electrodynamics. Even today, classical physics still describes very well all physical phenomena that occur in everyday life and in many areas of technology. Around 1900, only a few phenomena were not
fully described by classical physics.
The spectral lines of the chemical elements and the spectrum of thermal radiation, for example, could \textit{not} be explained in terms of classical physics.
Only the development of quantum mechanics in 
 the 1920s provided  the theory of these \zitateng{non-classical} phenomena.
\end{mdframed}
\begin{mdframed}[style=tpq]

\begin{center} 
\textsf{Complex number}
\end{center}

A \textbf{complex number} $z$ is made up from two real numbers $x$ and $y$ according to the prescription $$ z= x+iy.$$ Here $i$  is the \textbf{imaginary unit} with the property $$i^2=-1.$$
$x$ is called the \textbf{real part} and $y$ the \textbf{imaginary part} of $z$. Complex numbers
can be represented in a plane with $x$ and $y$ as coordinates in the usual way (Gaussian number plane). The quantity $$|z|^2=x^2+y^2$$ is called \textbf{absolute value squared} of $z$. The absolute value $|z|$ of the number $z$ then indicates the distance between $z$ and the zero point of the Gaussian number plane.

To warm up, simply calculate the product of $3 + 4i$ and
$3 - 4i$. Hint: The result is real and a square.

\end{mdframed}
\begin{mdframed}[style=tpq]

\begin{center} 
\textsf{De-Broglie wavelength}
\end{center}

A particle with mass $m$ moving at speed $v$ has momentum $p = mv$. In
quantum mechanics, this particle then has a wavelength
$$
\lambda = \frac hp,
$$
where $h$ is Planck's constant. The heavier the particle and the faster it
moves, the shorter the wave length.

\end{mdframed}
\begin{mdframed}[style=tpq]

\begin{center} 
\textsf{Equations of motion}
\end{center}

In physics, equations of motion are always differential equations,
i.e. they indicate at a certain point in time $t$ how the state changes over
time (mathematically: derivative with respect to time). From this, the state
at a later point in time can be calculated.

For a \textit{classical} particle, the equation of motion is given by Newton's laws, e.g. in the form \zitateng{force = mass times acceleration} If the force is known, the acceleration can be used to
calculate how the velocity changes, and from this, how the position
changes. Then you can follow the path of the particle in space and time, e.g. for a
satellite orbiting the earth.

The equation of motion for a \textit{quantum mechanical} particle is the Schrödinger
equation, a differential equation for the wave function. Due to Heisenberg's
uncertainty principle, the concept of orbit does not exist for a quantum
mechanical particle (e.g. an electron in an atom).

\end{mdframed}
\begin{mdframed}[style=tpq]

\begin{center} 
\textsf{Exponential function}
\end{center}

The exponential function is given by
$$f(x)=e^x,$$
where $e=2.71828...$ is Euler's number. You can see that this function 
increases very strongly when $x$ becomes larger: 
$e^0=1$, $e^1=e$, $e^2=e\cdot e=7.389$, ... $e^{10}=22026.5$, $e^{20}=485,165,195$.
 Each time $x$ grows by 1, $e^x$ is multiplied by $e$.
 This is called an exponential \textit{growth}.

The function $e^{-x}=\frac 1{e^x}$ shows an exponential \textit{decay}. 
Each time $x$ increases by 1, it decreases by a factor $\frac 1e=0.3679$, so that $e^{-10}=0.000045$.

Radioactive decay provides an example of exponential decay (in time $t$).
The number $N (t)$ of radioactive particles present at time $t$ is
$$
N(t)=N_0 e^{-t/\tau},
$$
where $N_0$ is the number at the initial time $t = 0$. $\tau$ (Greek letter tau) is the average \zitateng{lifetime} of a radioactive particle until its decay. $t_{1/2}=0.6931 \tau$ is the half-life after which half of the particles
present at the beginning are still there. After ten half-lives, only 1/1024 of
the radioactive particles remain.

\end{mdframed}
\begin{mdframed}[style=tpq]

\begin{center} 
\textsf{Kinetic and potential energy, classical}
\end{center}

Kinetic energy is the energy of motion. If a particle of mass $m$ moves at velocity $v$, it has the energy
$$
E_{\textrm{kin}} = \frac 12 m v^2.
$$
If the particle transfers its energy to a spring, for example, which is tensioned by the impact, then the energy is stored in the spring as potential energy $E_\textrm{pot}$. The total energy, i.e. the sum of kinetic and potential energies, is constant. This law of conservation of energy  is one of the most important principles of physics.

$E_\textrm{pot}$ often depends on a spatial coordinate $x$, in our example on the position at
which the particle is stopped by the spring. The faster it was at the
beginning, the further the spring is compressed and the greater the potential
energy. In physics, such a potential energy dependent on the location $x$ is
usually referred to as $V (x)$ for short and is also briefly called potential.
(A somewhat misleading term, as in electricity theory the potential energy
\textit{per charge} is referred to as electric potential).

The kinetic energy is often expressed by the momentum $p = mv$; then one has the relationship
$$
E_{\textrm{kin}} = \frac{p^2}{2m}.
$$
However, this relationship only applies if the particle is moving at speeds that are very small compared to the speed of light $c$. If this is not the case, the theory of
relativity must be employed, with the result
$$
E_{\textrm{kin}} = \sqrt{m^2 c^4 + p^2 c^2} - m c^2.
$$
In this text, however, the theory of relativity only becomes important at one
point, namely in the Compton effect.

\end{mdframed}

\begin{mdframed}[style=tpq]

\begin{center} 
\textsf{Measurements and probabilities}
\end{center}

The state of a particle in one dimension at time $t$ is \textit{completely} determined by the wave function $\Psi(x,t)$. Nevertheless, the result of a \textbf{single measurement} of the position $x$ (or another measurable quantity) is not predictable. However, if the following operations are carried out very often one after the other:
\begin{enumerate}
\item Prepare the state $\Psi(x,t)$,
\item Measure the position $x$ and
\item Register the result of the position measurement,
\end{enumerate}
then the distribution of the measurement results of the position, which is given by the probability density $|\Psi(x,t)|^2$, is gradually obtained.

In contrast to a single measurement, this is a statistical measurement; experts often speak of \textbf{ensemble measurements}. It is important that the system is always prepared anew, i.e. the initial state $\Psi(x,t)$ is created.
Alternatively, one could also imagine to prepare a  large number of particles (an ensemble of particles) all in the desired state $\Psi(x,t)$ and measure each particle once.

After a single measurement, the state of the particle is usually \textit{no longer} $\Psi(x,t)$, but a different one, depending on which physical quantity was measured and which value this measurement produced.

\textit{A measurement changes the state of a system.}

\end{mdframed}

\begin{mdframed}[style=tpq]

\begin{center} 
\textsf{Momentum}
\end{center}
The momentum $p$ of a classical particle is simply the product of mass
$m$ and speed $v$:
$$
p=m v .
$$
The momentum of a force-free quantum mechanical particle is related to the
wavelength $\lambda$ of the wave function:
$$
p=\frac h{\lambda},
$$
where $h$ is the Planck constant.

If forces act on the quantum mechanical particle, $p$ is somewhat more
complicated to calculate.

A photon (light quantum) also has momentum; this is defined by the
wavelength $\lambda$
of light: $$p=\frac h{\lambda}.$$ 

The momentum of the photon is related to its energy $E$
by $$p=\frac Ec,$$ with $c$ the speed of light.

\end{mdframed}
\begin{mdframed}[style=tpq]

\begin{center} 
\textsf{Operator}
\end{center}
Operators mathematically describe manipulations of a quantum mechanical state, for example in connection with measurements of a physical quantity. Because states admit different mathematical representations, operators can also appear in different forms. If a state is written as a wave function $\Psi(x)$,
an operator can be, for example, a derivative with respect to the variable $x$, or the multiplication with another function $f (x)$. If a state is represented as a vector, the operators are matrices. More information on vectors and matrices can be found in Section \ref{vektoren}.

\end{mdframed}
\begin{mdframed}[style=tpq]

\begin{center} 
\textsf{Phase relation, phase difference}
\end{center}

Two waves with the same wavelength $\lambda$ can have different amplitudes and they can be shifted relative to each other so that the wave crests are at different positions. For mathematical reasons, this shift is expressed as an angle. Half a wave length corresponds to an angle of 180$^\circ$, for example. This angle is
called the \textbf{phase difference}, and if the phase difference between two
waves remains constant, they are referred to as having a fixed \textbf{phase relation}.
Important special cases are 0$^\circ$ phase difference (in-phase) and 180$^\circ$ (out-of-phase). An angle of 360$^\circ$ is, as always, completely equivalent to an angle of 0$^\circ$, because a shift of one entire wavelength has no effect, as does a shift
of two wavelengths, i.e. 720$^\circ$, etc.

\end{mdframed}
\begin{mdframed}[style=tpq]

\begin{center} 
\textsf{Polarization of light}
\end{center}
Electromagnetic waves consist of coupled electric and magnetic fields that change in space and time. In the case of a monochromatic (only one frequency or wavelength) plane wave in free space, the electric and magnetic fields are perpendicular to each other and both are perpendicular to the direction in which the light propagates. If the electric field only oscillates back and forth in one direction, it is referred to as a \textbf{linearly polarized} wave. If you look in the direction of wave propagation, you can
distinguish between two fundamentally different polarizations (directions of oscillation): \zitateng{right and left} ($x$ polarization) and \zitateng{up and down} ($y$ polarization). All
more complicated waveforms can be composed from these basic forms.

\end{mdframed}

\begin{mdframed}[style=tpq]

\begin{center} 
\textsf{Spectrum (plural: spectra), spectroscopy
}
\end{center}
A \textbf{spectrum} is the set of the wavelengths (or frequencies) of
electromagnetic radiation that can be observed in a range, together with
their respective intensities. 
If all wavelengths can occur within some range, the spectrum is called
\textbf{continuous}, e.g. in the case of thermal radiation from a hot object. If only
very specific wavelengths can occur, the spectrum is called \textbf{discrete}. This is
the case for pure chemical elements, for example. Discrete spectra are also
called \textbf{line spectra}, named after the image of colored lines on a dark
background that appears in an optical spectrometer. By examining such line
spectra in \textbf{spectral analysis}, it is possible to determine which chemical
elements the substance under investigation contains. The study of
spectra is called \textbf{spectroscopy}.

In more general terms, the possible values of the energy of a quantum system
are also referred to as the \textbf{energy spectrum}.

\end{mdframed}
\begin{mdframed}[style=tpq]

\begin{center} 
\textsf{State of a particle}

\end{center}
The state of a classical particle moving in the $x$ direction is determined at time $t$ by its position $x$ and its momentum $p$ (or alternatively its velocity $v$). The 
values of $x$ and $p$ are precisely determined at every instant.

The state of a quantum mechanical particle, on the other hand, is
determined by a function that depends on $x$ and $t$, the wave function. The wave function allows statements to be made about the position and momentum of the particle, but only within the limits of the Heisenberg uncertainty principle.

Vectors are an alternative way of representing a quantum mechanical state
mathematically. More about this can be found in section \ref{vektoren} .

\end{mdframed}
\begin{mdframed}[style=tpq]

\begin{center} 
\textsf{Uncertainty relation}
\end{center}
Heisenberg's uncertainty relation, often also called the indeterminacy
relation, states the accuracy $\Delta x$  with which the position $x$ of a particle
moving along the $x$ axis can be determined if the momentum $p$ of the
particle is also determined with accuracy $\Delta p$ in the same state:
$$
\Delta x \Delta p \geq \frac{\hbar}2 .
$$
The \textit{uncertainties} $\Delta x$ and $\Delta p$ are therefore related: If you want to determine $x$ very precisely, $p$ becomes very imprecise, and vice versa.

Not only the position and momentum, but also other pairs of \zitateng{incompatible} quantities of quantum mechanics are linked by indeterminacy relations.

$\hbar$ (read: \zitateng{h-bar})   is the Planck constant divided by $2\pi$; $$\hbar = 1,055 \cdot 10^{-34} \textrm{Js}.$$
Since this number is so tiny, we do not notice the uncertainty principle in
everyday life, but the atoms feel it very clearly.

\end{mdframed}
\begin{mdframed}[style=tpq]

\begin{center} 
\textsf{Wave function}
\small
Caution: Contains higher mathematics.

\normalsize
\end{center}
The \textbf{wave function} of a quantum mechanical particle moving on the $x$ axis depends on the position $x$ and the time $t$; it is usually denoted by $\Psi(x,t)$
(read: \zitateng{Psi of $x$ and $t$}).

$\Psi(x,t)$ ) is a \textit{complex} function, i.e. it consists of two parts, $ \textsf{Re}\Psi(x,t)$ and $\textsf{Im}\Psi(x,t)$
(read: \zitateng{real part of Psi and imaginary part of Psi}). 
The parts play a similar role to the coordinates $x$ and $y$ of a point in a plane, where, for example, the distance $r$ of the point $(x, y)$ from the zero point can be calculated using Pythagoras' theorem:
$$
r^2=x^2+y^2 .
$$
In a similar way, the wave function yields the important quantity 
$$
| \Psi(x,t)|^2 = ( \textsf{Re}\Psi(x,t))^2 + (\textsf{Im}\Psi(x,t))^2
$$
(read: \zitateng{Psi absolute value squared}), which indicates how likely it is to find the particle at position $x$ at  time $t$. (More 
 precisely, $| \Psi(x,t)|^2$ is the
probability density of finding the particle).

\end{mdframed}
\chapter{List of experiments}


\begin{itemize}
\item Diffraction of laser light (Sections \ref{interferenz} and \ref{interferenz_quanten})
\phantom{xxxxxx}
\item Detection of single photons  (Section \ref{einzelphotonen})
\phantom{xxxxxx}
\item Double slit with single photons (Section \ref{interferenz_quanten})
\phantom{xxxxxx}
\item Electron diffraction  (Section \ref{interferenz_quanten})
\phantom{xxxxxx}
\item Franck-Hertz experiment (Section \ref{wie_bewegt})
\phantom{xxxxxx}
\item Mach-Zehnder interferometer (Section \ref{interferenz_quanten})
\phantom{xxxxxx}
\item Millikan experiment  (Section \ref{millikan})
\phantom{xxxxxx}
\item Photoelectric effect (Section \ref{photoeffekt})
\phantom{xxxxxx}
\item Polarization of light  (Section \ref{was_bei_messung}
\phantom{xxxxxx}
\item Quantum cryptography (under construction; Section \ref{quantencomputer})
\item Blackbody radiation (Section \ref{schwarzkoerper})
\phantom{xxxxxx}
\item Spectroscopy (Sections \ref{wie_bewegt} and \ref{spektroskopie})
\phantom{xxxxxx}
\item Wave trough (Sections \ref{interferenz} and \ref{unschaerfe})
\phantom{xxxxxx}
\end{itemize}

\Huge

\tre{Important note:}
\\

\normalsize

This is the internet version of the handbook for the \textit{Treffpunkt Quantenmechanik} laboratory. It contains only the
overview part in six chapters.

For copyright reasons, we cannot make the detailed instructions for the individual
experiments in \textit{Treffpunkt Quantenmechanik} available on the internet.

Those instructions are only available in the printed version of the handbook or as
individual printouts in the rooms of the \textit{Treffpunkt Quantenmechanik}.

\bibliography{general}
\printindex
\end{document}